%% file: paper_main.tex
\documentclass[twocolumn,english,aps,prb,10pt,superscriptaddress,floatfix,longbibliography]{revtex4-2}
\usepackage[colorlinks,bookmarks=false,citecolor=blue,linkcolor=blue,urlcolor=blue]{hyperref}
\usepackage{graphicx}
\usepackage{dcolumn}
\usepackage{bm}
\usepackage[T1]{fontenc}
\usepackage[latin9]{inputenc}
\usepackage{mathrsfs}
\usepackage{amsmath}
\usepackage{amssymb}
\usepackage{wasysym}
\usepackage{esint}
\usepackage{babel}
\usepackage{xcolor}
\usepackage{braket}
\usepackage[export]{adjustbox}
\usepackage[normalem]{ulem}
\usepackage{verbatim}
\usepackage{booktabs}

\usepackage{nicematrix}
\usepackage{tikz}

\usepackage{listings}

\newcommand{\pdag}{{\phantom{\dagger}}}
\newcommand{\tGD}{\widetilde\Gamma(\omega_n;D)}
\newcommand{\tG}{\widetilde\Gamma(\omega_n)}
\newcommand{\wh}{\widetilde{h}}
\newcommand{\wip}{\widetilde{p}}
\newcommand{\wiq}{\widetilde{q}}
\newcommand{\wD}{\widetilde{\Delta}}
\newcommand{\lowplus}{\mathbin{\lower1.5ex\hbox{$+$}}}

\definecolor{mygreen}{rgb}{0,0.6,0}
\definecolor{mygray}{rgb}{0.5,0.5,0.5}
\definecolor{mymauve}{rgb}{0.58,0,0.82}

\begin{document}

\title{Scalable Effective Models for Superconducting Nanostructures:\\ Applications to Double, Triple, and Quadruple Quantum Dots}

\author{Daniel Bobok}
\affiliation{Department of Condensed Matter Physics, Faculty of Mathematics and Physics, Charles University, Ke Karlovu 5, Praha 2 CZ-121 16, Czech Republic}

\author{Luk\'a\v{s} Frk}
\affiliation{Department of Condensed Matter Physics, Faculty of Mathematics and Physics, Charles University, Ke Karlovu 5, Praha 2 CZ-121 16, Czech Republic}

\author{Vladislav Pokorn\'y}
\affiliation{Institute of Physics (FZU), Czech Academy of Sciences, Na Slovance 2, 182 00 Prague 8, Czech Republic}

\author{Martin \v{Z}onda}
\affiliation{Department of Condensed Matter Physics, Faculty of Mathematics and Physics, Charles University, Ke Karlovu 5, Praha 2 CZ-121 16, Czech Republic}

\begin{abstract}
We introduce a versatile and scalable framework for constructing effective models of superconducting (SC) nanostructures described by the generalized SC Anderson impurity model with multiple quantum dots and leads. Our Chain Expansion (ChE) method maps each SC lead onto a finite tight-binding chain with parameters obtained from \emph{Pad\'e} approximants of the tunneling self-energy. We provide an explicit algorithm for the general case as well as simple analytical expressions for the chain parameters in the wide-band and infinite-chain limits. This mapping preserves low-energy physics while enabling efficient simulations: short chains are tractable using exact diagonalization, and longer ones are handled with density matrix renormalization group methods. The approach remains reliable and computationally efficient across diverse geometries, both in and out of equilibrium. We use ChE to map the ground-state phase diagrams of double, triple, and quadruple quantum dots coupled to a single SC lead. While half-filled symmetric systems show similar overall diagrams, the particular phases differ substantially with the dot number. Here, large parameter regions are entirely missed by the widely used zero-bandwidth approximation but are captured by ChE. Away from half-filling, additional dots markedly increase diagram complexity, producing a rich variety of stable phases. These results demonstrate ChE as a fast, accurate, and systematically improvable tool for exploring complex SC nanostructures.
\end{abstract}

\maketitle

\section{Introduction}

In the last decades, significant progress has been achieved in
experimental techniques that allow for coupling nanodevices with only a few active orbitals, here called quantum dots (QDs) for simplicity, with superconducting (SC) reservoirs~\cite{DeFranceschi2010hybrid,Heinrich2018single,Benito2020hybrid}. Consequently, the overall complexity of these experimental systems increases steadily. Junctions, where SC leads are bridged by effective QDs, have been prepared in multi-lead arrangements~\cite{Draelos2019supercurrent,Pankratova2020multiterminal}. Tunable double quantum dots (DQDs) have been constructed in serial~\cite{Saldana2018supercurrent,Saldana2020two,Rasmussen2018yu-shiba-rusinov} as well as parallel~\cite{Vekris2021josephson,Steffensen2022direct} configurations. The scanning tunneling microscopy (STM) and spectroscopy (STS) experiments, where QDs are placed on SC surface are already probing dimers~\cite{Ruby2018wave,Choi2018influence, Kamlapure2018engineering, Kezilebieke2018coupled, Ding2021tuning, Beck2021spin, Kuster2021long} and more complicated setups~\cite{Howon2018toward,Liebhaber2022quantum,Li2025individual,Liu2025spin,Li2025negative,Trivini_Mn_imps}. They promise future applications in superconducting electronics, spintronics, and other computational devices, and represent tunable platforms for basic research~\cite{DeFranceschi2010hybrid,Heinrich2018single,Benito2020hybrid,Steffensen2025YSR, Zaldivar2025, Lakic2025quantum}.

Naturally, this progress also motivates advances in the theoretical study of complex QD-SC systems. The paradigmatic framework for describing such setups is the SC Anderson impurity model (SC-AIM), which involves one or more impurity levels~\cite{Meden2019the}. Interestingly, even the simplest case of a single QD coupled to a single SC lead remains an active area of research~\cite{Ilicic2025variational, keliri2025slaves}. For instance, it has only recently been shown that a Kondo cloud screens the impurity spin even in the presence of an SC gap~\cite{Moca2021kondo}, and methods for treating non-equilibrium phenomena are still under active development~\cite{Keliri2023driven,Cheng2024quasiparticle}.  More complex configurations -- such as DQDs or multiple leads -- have already revealed unexpected features, including previously unaccounted-for phases~\cite{Zitko2015numerical,Zonda2023generalized,zalom2024double} and overlooked symmetries~\cite{Kadlecova2017quantum,Kadlecova2019practical,Zalom2023hidden}, and serve as a basis for recent qubit proposals~\cite{Geier2024fermion,Steffensen2025YSR}.       

As one proceeds to even more complex structures, now commonly addressed in experiments, reliable numerical methods such as numerical renormalization group (NRG)~\cite{Bulla2008numerical,Yoshioka2000numerical,Bauer2007spectral,Yao2014phase,Zitko2015numerical,Zitko2016spectral,Zalom2021spectral,Zalom2021tunable, Zalom2023rigorous} or various types of quantum Monte Carlo 
~\cite{Siano2004josephson,Luitz2010weak,Luitz-2012,Pokorny-2018,Pokorny2021footprints} 
becomes too expensive even in equilibrium. 
In certain regimes, a feasible alternative are analytic approximations, e.g., various mean-field approaches~\cite{Yeyati1997resonant,Rodero1999general,Yoshioka2000numerical,Rodero2011josephson,Ptok-2017-NbSe2}, perturbation expansions~\cite{Alastalo1998the,Vecino2003josephson,Meng2009self,Zonda2015perturbation,Zonda2016perturbation}, and functional renormalization group techniques~\cite{Karrasch2008josephson,Wentzell2016magnetoelectric}. Another strategy is to use simple effective models, such as the zero-bandwidth approximation (ZBW)~\cite{Vecino2003josephson,Rasmussen2018yu-shiba-rusinov} and the SC atomic limit (AL)~\cite{Meden2019the}. Their great advantage is that they can be exactly solvable even in complex scenarios. However, ZBW and AL are limited to qualitative description~\cite{Meden2019the,Zitko2015numerical}. For AL, this problem can be mitigated to a large extent by generalized AL (GAL)~\cite{Zonda2023generalized,Pokorny2023effective}. Yet, it was recently shown that ZBW and (G)AL can fail even qualitatively in some regimes of setups as simple as a DQD coupled in parallel to a single SC lead~\cite{zalom2024double}. An alternative method that allows one to overcome these limitations is the recently introduced surrogate model~\cite{Baran2023surrogate,Baran2024BCS,Souto2025majorana}, which relies on exact diagonalization (ED) of SC-AIM with a discretized reservoir using only a small number of effective levels. The parameters of the approximate model are determined by minimizing a cost function that measures the difference between the surrogate and the exact tunneling self-energy.

Here, we present a conceptually different approach. We show that by directly employing a tight-binding chain representation of the hybridization function, one can construct optimal and scalable effective models for the low-energy physics of SC-AIM with arbitrary numbers of leads and QDs in a straightforward and systematic way. The idea of expanding the tunneling self-energy into a chain is not new. Exact unitary transformations that map a system coupled to a continuum onto a one-dimensional chain with only nearest-neighbor interactions have been derived for various models~\cite{Chin2010exact}. Similar mappings are widely used in methods such as NRG~\cite{Bulla2008numerical}, the density matrix renormalization group (DMRG)~\cite{Moca2021kondo}, 
auxiliary quantum master equation approaches~\cite{Arigoni2013nonequilibrium,Dorda2014auxiliari,Schwarz2016Lindblad}, and even machine learning techniques for extracting effective quantum impurity models~\cite{Rigo2024unsupervised,Rigo2020machine}. 

However, in this paper, we show that a properly constructed chain expansion (ChE) model naturally yields a \emph{Pad\'e} approximant of the original tunneling self-energy of the SC-AIM. Consequently, for many relevant problems and geometries, finite chains of only a few sites long are already sufficient to capture the effects of the SC leads. Moreover, the model parameters can be obtained explicitly and reduced to simple analytic formulas in both the wide-band and infinite-chain limits. To complement the paper, we provide a Jupyter notebook~\cite{ChECode} with examples illustrating how the ChE models introduced here can be applied to different system geometries.
It is worth noting that \emph{Pad\'e} approximants are also widely used in other contexts related to impurity problems, for example, as analytical continuation methods~\cite{Beach2000reliable,Osolin-Zitko-2013Pade,Schott2016analytic,Fei2021Nevanlinna} and to decompose the bath's Fermi function~\cite{Hu2010Pade,Hu2011Pade,Smorka2024dynamics}.

We first demonstrate the advantages of the ChE approach by revisiting selected problems that were previously studied using other methods, both in and out of equilibrium. We then focus on SC systems with multiple QDs arranged in parallel with respect to the SC lead. We show that small ChE models can be efficiently solved using ED, while longer chains, when needed, can be treated using DMRG. We also discuss how the method can be incorporated into other equilibrium and non-equilibrium techniques. This allows for the construction of fast, reliable, and systematically improvable effective models that are well-suited for a broad range of problems in both theoretical and experimental contexts.

The remainder of the paper is structured as follows. In section~\ref{sec:MandM}, we first introduce the SC-AIM in its general form (Sec.~\ref{sec:AIM}). Then we construct its low-energy effective models using the ChE approach and derive their parameters (Sec.~\ref{sec:ChE}), with further technical details provided in Appendices~\ref{app:CF} and ~\ref{app:multileat}. Section~\ref{sec:Bench} illustrates both the strengths and limitations of the ChE models by reexamining systems and problems previously addressed using accurate numerical methods and alternative effective models. It also outlines general strategies for applying ChE models to various types of systems in several physical regimes. Section~\ref{sec:Results} begins with an analysis of a DQD in a parallel configuration (Sec.~\ref{sec:DQD}), focusing on the emergence of the triplet state and competing singlet phases. We then extend the analysis to the ground-state properties of triple and quadruple QD systems (Sec.~\ref{sec:TQQD}). The paper concludes with a summary of the main findings (Sec.~\ref{sec:Conc}). Certain technical details are deferred to the Appendices~\ref{app:CF}-\ref{app:RGmethods}.    

\section{Models and Methods~\label{sec:MandM}}

\subsection{Superconducting Anderson impurity model with multiple impurities ~\label{sec:AIM}}
\begin{figure}[!ht]
    \centering
    \includegraphics[width=1\linewidth]{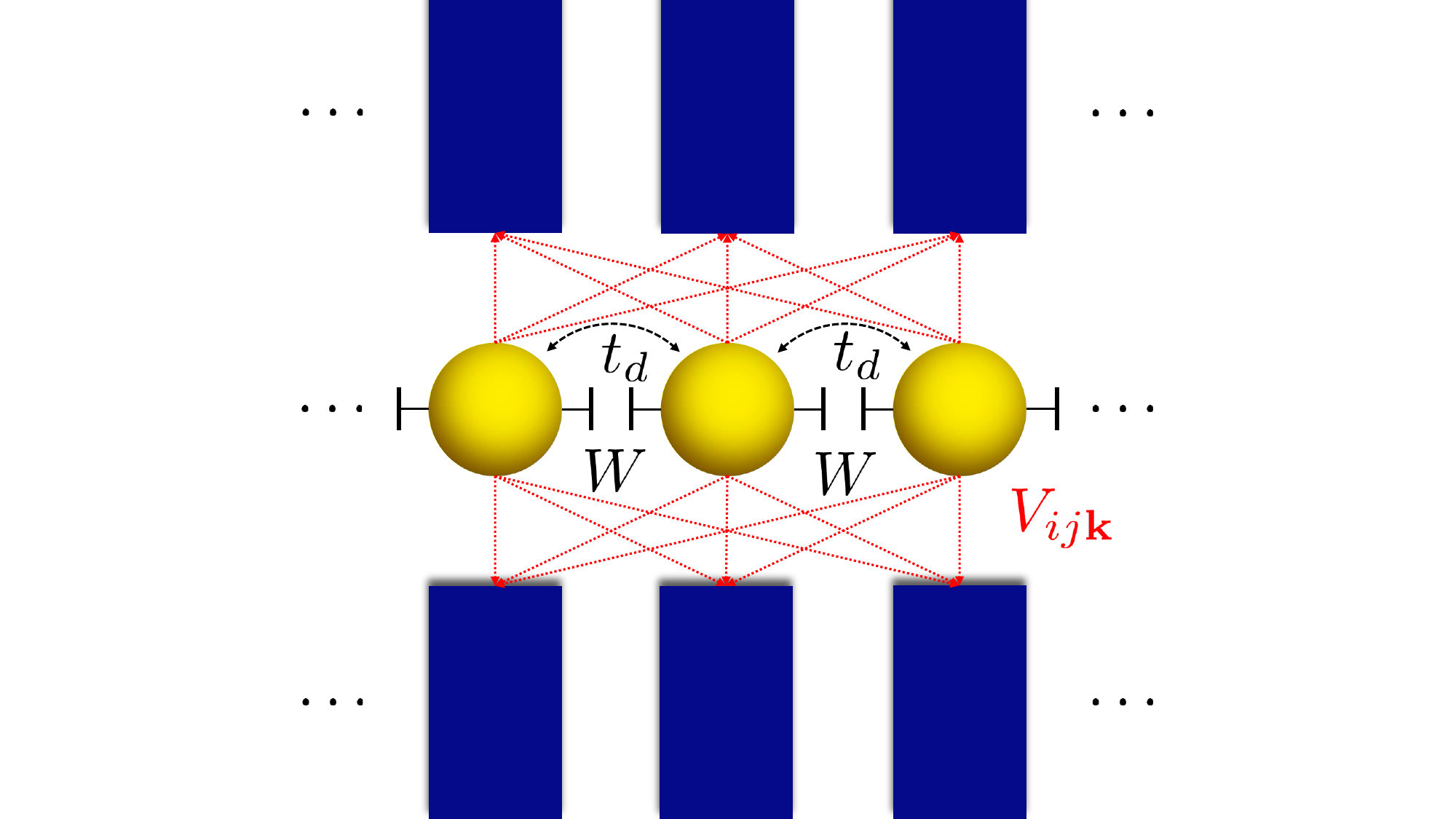}
    \caption{Illustration of $N_d$ QDs coupled to $N_l$ SC leads. Here, only nearest neighbor QDs interact with each other through hopping $t_d$ and capacitance coupling $W$. Dots' interactions with leads are described by parameters $V_{ij\mathbf{k}}$.}
    \label{fig:SCAIMmulti}
\end{figure}

We consider a general model of interacting QDs that are coupled to SC reservoirs treated within the BCS theory (Fig.~\ref{fig:SCAIMmulti}). Hamiltonian of such a hybrid structure with $N_d$ QDs and $N_l$ SC leads consists of three parts,
\begin{equation}
    \mathcal{H} = \mathcal{H}^{\text{en}} + \sum_{l=1}^{N_l}\mathcal{H}^{\text{le}}_l + \sum_{j=1}^{N_d}\sum_{l=1}^{N_l}\mathcal{H}^{\text{hy}}_{jl}.
    \label{eq:NDNLeq}
\end{equation}
The first part describes an ensemble of $N_d$ interacting QDs,
\begin{align}
    \mathcal{H}^{\text{en}} 
    &= \sum_{j=1}^{N_d}\sum_\sigma\epsilon_j n_{j\sigma} +
    \sum_{j=1}^{N_d}U_j n_{j\uparrow}n_{j\downarrow}\\ 
    &-\sum_{i=1}^{N_d}\sum_{j=1}^{N_d}\sum_{\sigma}t_{ij} d_{i\sigma}^\dagger d_{j\sigma}^\pdag \nonumber\\
    &+\sum_{i=1}^{N_d-1}\sum_{j>i}^{N_d}W_{ij}
    \left( n_{i\uparrow} + n_{i\downarrow}\right)\left( n_{j\uparrow} + n_{j\downarrow}\right)\nonumber,
\end{align}
where $d_{j\sigma}^\dag$ ($d_{j\sigma}^\pdag$) creates (annihilates) an electron with spin $\sigma$ on the impurity $j$ with energy $\epsilon_j$ and $n_{j\sigma}=d_{j\sigma}^\dagger d_{j\sigma}^\pdag$. The second term stands for the local Coulomb interaction with strength (charging energy) $U_j$ on the dot $j$. The third term describes direct hopping between the dots with $t_{jj}=0$ for all $j$, and the last term describes the inter-dot capacitive coupling. Matrix elements $t_{ij}$ and $W_{ij}$ encode the geometry of the ensemble. For simplicity, in the following, whenever applied, we assume only nearest-neighbor hopping and capacitive coupling, with equal hopping amplitude $t_d$ and coupling $W$ for all connections. For convenience, we present results using shifted energy levels
$\varepsilon_j = \epsilon_j + U_j/2 + \sum_{i \neq j} W_{ji}$ such that $\varepsilon_j = 0$ corresponds to the half-filled case on a bipartite lattice.

The second part represents $N_l$ SC leads, each described by a standard BCS Hamiltonian,
\begin{equation}
    \mathcal{H}^{\text{le}}_l = \sum_{\mathbf{k}\sigma} \epsilon^{\pdag}_{l\mathbf{k}} c^{\dagger}_{l\mathbf{k} \sigma} c^{\pdag}_{l\mathbf{k}\sigma} - \Delta\sum_{\mathbf{k}} \left(e^{i\varphi_l}c^{\dagger}_{l\mathbf{k} \uparrow}
    c^{\dagger}_{l,-\mathbf{k} \downarrow} + \mathrm{H.c.}\right),
    \label{eq:H_le}
\end{equation}
where $c_{l\mathbf{k}\sigma}^\dagger$ ($c_{l\mathbf{k}\sigma}^\pdag$) creates (annihilates) an electron in lead $l$ with spin $\sigma$ and energy $\epsilon^{\pdag}_{l\mathbf{k}}$, $\Delta$ is the SC gap and $\varphi_l$ is the SC phase of lead $l$. We assume that all leads are made from the same material, as is common in experiments, and we use $\Delta$ as the energy unit within the paper and natural units for Planck constant $\hbar=1$ and elementary charge $e=1$. 

The third part is the hybridization between the QD ensemble and the SC leads, which reads
\begin{equation}
    \mathcal{H}^{\text{hy}}_{jl} = \sum_{\mathbf{k}\sigma} 
    \left(V_{jl\mathbf{k}}  c^{\dagger}_{l\mathbf{k} \sigma} d^{\pdag}_{j\sigma} + \mathrm{H.c.}\right).
    \label{eq:H_hyjl}
\end{equation}
The coupling to the leads is characterized by the tunneling rate matrix 
$\Gamma_{ij,l}(\epsilon)=\pi \sum_{\mathbf{k}}V_{il\mathbf{k}}V_{jl\mathbf{k}}\delta(\epsilon-\epsilon_{l\mathbf{k}})$, where the cross terms $i\neq j$ arise if multiple QDs are coupled to the same lead $l$. We assume a flat density of states with half-bandwidth $D$, $\rho(\epsilon) = \Theta(D-|\epsilon|)/(2D)$ and real $V_{il\mathbf{k}}$ which leads to  
$\Gamma_{ij,l}(\epsilon)=\Gamma_{ij,l}\Theta(|D-\epsilon|)$ with constant tunneling rates 
$\Gamma_{ij,l}$. In the paper, we use the convention of omitting subscripts to parameters whenever the same magnitude is assumed for all dots or leads.

\subsection{Chain expansion and effective models \label{sec:ChE}}
\begin{figure}[ht!]
    \centering    \includegraphics[width=1.\linewidth]{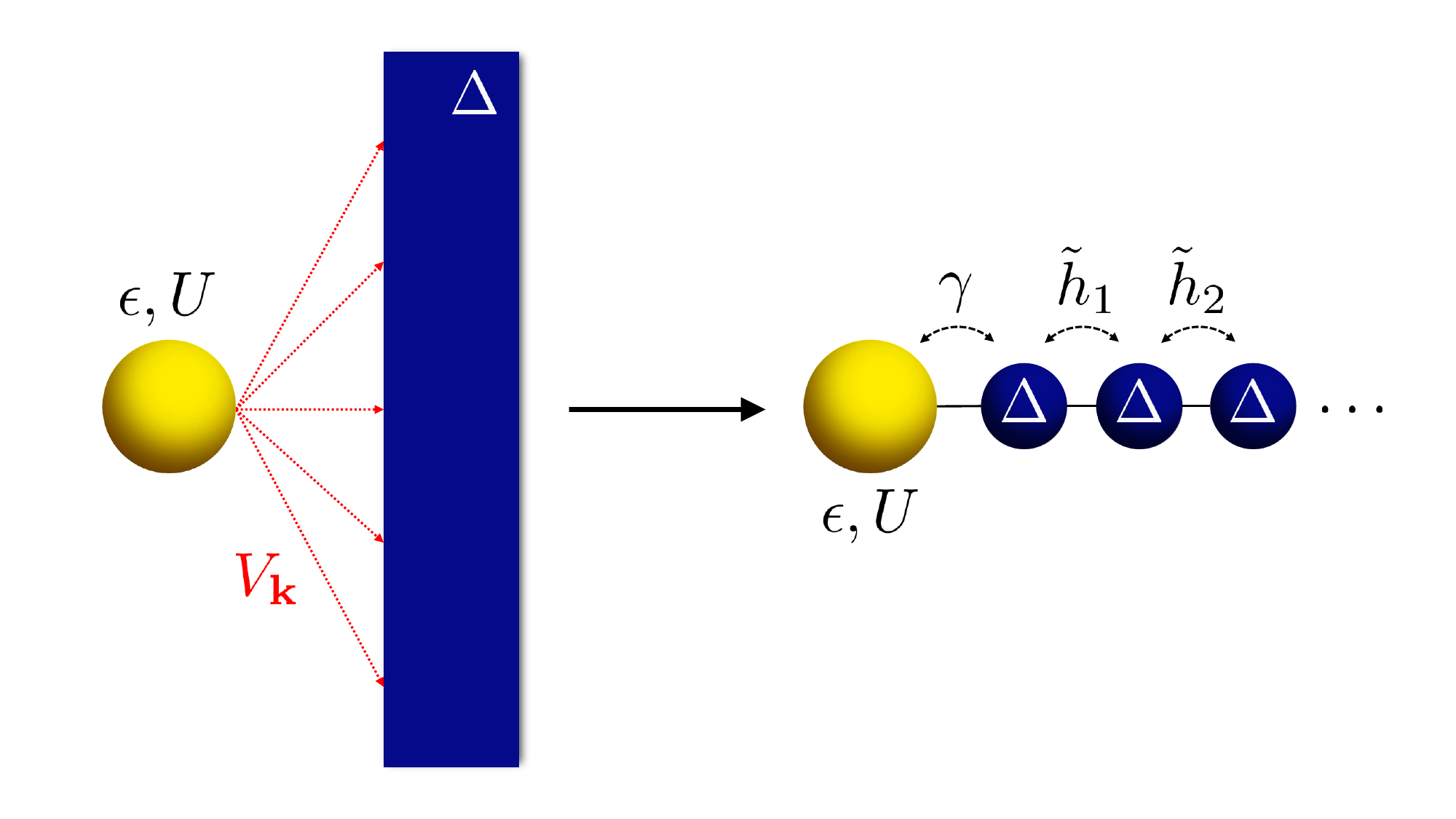}
    \caption{
    Mapping of an SC lead onto a finite tight-binding chain with site-dependent hopping amplitudes $\wh_\ell$ and local SC pairing. In the derivation of the ChE models, it is convenient to introduce the substitutions $\gamma=\sqrt{\Gamma\Delta h_0}$ and $\widetilde{h}_\ell=\sqrt{h_\ell}\Delta$ with dimensionless coefficients $h_\ell$.
    \label{fig:mapping}}
\end{figure}
We demonstrate how to systematically construct increasingly accurate low-energy effective models using ChE of the tunneling self-energy. To illustrate the central idea, it suffices to consider the simplest case of a single QD coupled to a single lead, starting with the noninteracting limit $U = 0$.

To avoid the complex analytic structure of real-frequency Green's functions, we adopt the Matsubara (imaginary-frequency) formalism.
This eliminates the need to treat the behavior inside and outside the superconducting gap separately. Employing the Nambu spinor notation $\mathscr{D}^\dagger = (d^\dagger_{\uparrow}, d_{\downarrow})$ the impurity (QD) Green's function becomes~\cite{Meden2019the}:
\begin{align}
    &G_0(i\omega_n)^{-1}=\nonumber\\
    &\begin{pmatrix}
    i\omega_n\left[1+\frac{\Gamma}{\Delta}\tGD\right]-\epsilon, & 
    \Gamma\,\tGD\\
    \Gamma\,\tGD & 
    i\omega_n\left[1+\frac{\Gamma}{\Delta}\tGD\right]+\epsilon
    \end{pmatrix}.
    \label{eq:G0}
\end{align}
Here, $\omega_n=(2n+1)\pi k_BT$ denotes the $n$th fermionic Matsubara frequency at temperature $T$, and the dimensionless hybridization function $\tGD$ is given by
\begin{equation}
    \tGD=\frac{1}{\sqrt{1 + \frac{\omega_n^2}{\Delta^2}}}\frac{2}{\pi}
    \arctan\left[\frac{D/\Delta}{\sqrt{1+\frac{\omega_n^2}{\Delta^2}}}\right]
    \label{eq:GomegaD}
\end{equation}
and
\begin{equation}
    \tG\equiv\widetilde\Gamma(\omega_n; D\rightarrow\infty)=
    \frac{1}{\sqrt{1 + \frac{\omega_n^2}{\Delta^2}}},
    \label{eq:GomegaWBL}
\end{equation}
with Eq.~\eqref{eq:GomegaWBL} corresponding to the wide band limit (WBL). 
In what follows, we consider $\omega_n$ to be a continuous variable, consistent with the limit of $T=0$.
The hybridization function plays a central role in our approach. We show next that due to the structure of Eq.~\eqref{eq:G0}, the additive nature of the tunneling self-energy, and the form of ChE, we are actually seeking the best approximations of 
Eq.~\eqref{eq:GomegaD} or Eq.~\eqref{eq:GomegaWBL}, respectively.

Our main aim is to describe low-energy phenomena such as quantum phase transitions (QPTs), Andreev bound states, Josephson current (supercurrent), and related quantities in QD assemblies. Accordingly, our objective is to construct accurate, effective models in this regime. However, we are also interested in time-dependent phenomena, which may require an alternative effective model. Nevertheless, in both cases, we approximate the continuum bath with a finite chain of the same form. Among various chain configurations tested, the simplest and most effective is described by the Hamiltonian
\begin{align}
    \mathcal{H}^{\mathrm{ChE}}_{U=0}&= \sum_\sigma\epsilon_\sigma d_{\sigma}^\dagger d_{\sigma}^\pdag-\sum_{\sigma} \left(\gamma d_{\sigma}^\dagger c^\pdag_{1\sigma} +  \mathrm{H.c.}\right)\nonumber\\
    &-\sum_{\ell=1}^{L-1}\sum_\sigma \left(\wh^\pdag_{\ell} c_{\ell\sigma}^\dagger c^\pdag_{\ell+1,\sigma} +  \mathrm{H.c.}\right)\\ 
    &-\sum_{\ell=1}^L\left(\wD c_{\ell\uparrow}^{\dagger}c_{\ell\downarrow}^\dagger + \mathrm{H.c.}\right),\nonumber
    \label{eq:ChEM}
\end{align}
where the lead is modeled by a tight-binding chain with $L$ sites, as shown in Fig.~\ref{fig:mapping}. Here, $\gamma$ is the hopping between the dot and the first site of the chain, $\wh_\ell$ are the inter-site hopping amplitudes, and $\wD$ is the local SC pairing. 

Using again the Nambu spinor notation 
$\mathscr{D}^\dagger_{\mathrm{ChE}}=(d^\dagger_\uparrow,
d^\pdag_\downarrow,c^\dagger_{1\uparrow},
c^\pdag_{1\downarrow},c^\dagger_{2\uparrow},
c^\pdag_{2\downarrow},\dots,c^\dagger_{L\uparrow},
c^\pdag_{L\downarrow})$ 
we can write the noninteracting Green's function as
\begin{equation}
\begin{aligned}
    \label{eq:ChEGF}
    &G_{U=0}^\text{ChE}\left(i\omega_n\right)^{-1}=\\
    &\begin{pmatrix}
    i\omega_n-\epsilon & 0 & \gamma & 0 & 0 & 0 & \cdots & 0 & 0  \\
    0& i\omega_n+\epsilon & 0 & -\gamma  & 0 & 0 & \cdots & 0 & 0 \\
    \gamma & 0 &  i\omega_n & \wD & \wh_1 & 0 & \cdots & 0 & 0 \\
    0 & -\gamma & \wD & i\omega_n & 0 & -\wh_1 & \cdots & 0 & 0 \\
    0 & 0 & \wh_1 & 0 & i\omega_n & \wD & \cdots & 0 & 0 \\
    0 & 0 & 0 & -\wh_1 & \wD & i\omega_n  & \cdots & 0 & 0\\
    \vdots & \vdots & \vdots & \vdots & \vdots & \vdots & \ddots & \vdots & \vdots\\ 
    0 & 0 & 0 & 0 & 0 & 0 & \cdots & i\omega_n & \wD  \\
    0 & 0 & 0 & 0 & 0 & 0 & \cdots & \wD & i\omega_n 
    \end{pmatrix}. 
\end{aligned}
\end{equation}
It is also convenient to introduce rescaling
\begin{equation}
    \gamma = \sqrt{\frac{\Gamma}{\Delta}h_0}\Delta,\qquad
    \wh_\ell = \sqrt{h_\ell}\Delta,
    \label{eq:resc}
\end{equation}
with dimensionless coefficients $h_\ell$ and $\wD=\Delta$ (see Appendix~\ref{app:CF}). 

As shown in Appendix~\ref{app:CF}, the resulting dot Green's function takes a form analogous to Eq.~\eqref{eq:G0}:
\begin{equation}
\label{eq:G0ChE1}
\begin{aligned}
    &G^\text{ChE}_{d,U=0}(i\omega_n)^{-1} =\\
    &\begin{pmatrix}
    i \omega_n \left[1+\frac{\Gamma}{\Delta}{\cal P}_L(\omega_n)\right]-\epsilon& \Gamma{\cal P}_L(\omega_n) \\
    \Gamma{\cal P}_L(\omega_n) & i \omega_n \left[1+\frac{\Gamma}{\Delta}{\cal P}_L(\omega_n)\right]+\epsilon\\
    \end{pmatrix},
\end{aligned}
\end{equation}
where ${\cal P}_L(\omega_n;\{h_\ell\})$ is a rational function, with the form:
\begin{equation}
    {\cal P}_L(\omega_n;\{h_\ell\})=\frac{\sum_{j=0}^{L/2-1}p_j \omega_n^{2j}}
    {\sum_{j=0}^{L/2}q_j \omega_n^{2j}},
    \label{eq:PadeEven}
\end{equation}
for even $L$ and
\begin{equation}
    {\cal P}_L(\omega_n;\{h_\ell\})=\frac{1}{1+\omega_n^2}\frac{\sum_{j=0}^{(L-1)/2}p_j \omega_n^{2j}}
    {\sum_{j=0}^{(L-1)/2}q_j \omega_n^{2j}},
    \label{eq:PadeOdd}
\end{equation}
for odd $L$.
The coefficients $p_i$ and $q_i$ are functions of $h_\ell$ and follow directly from Eq.~\eqref{eq:PMP}. 

Here, the crucial observation is that Eq.~\eqref{eq:PadeEven} is of the same form as the \emph{Pad\'e} approximant~\cite{Baker1996pade} $[L-2/L]_{\tG}(\omega_n)$ of the total hybridization function $\tGD$ from Eq.~\eqref{eq:GomegaD} and that the second ratio in Eq.~\eqref{eq:PadeOdd} has the form of the scaled \emph{Pad\'e} approximant $[L-2/L-2]_{(1+\omega_n^2)\tG}(\omega_n)$. 
Therefore, to obtain ${\cal P}_L(\omega_n) \approx \tGD$ around $\omega_n^2 = 0$ to order $2L-2$, one simply needs to find coefficients $h_\ell$ that reproduce the respective \emph{Pad\'e} coefficients. This is convenient as \emph{Pad\'e} approximant of function $f(x)$ is a rational fraction $[n/m]_{f}(x)$ (a degree-$n$ polynomial in the numerator over a degree-$m$ polynomial in the denominator) whose Maclaurin series matches that of $f(x)$ up to order $n+m$. However, its radius of convergence can be much larger than that of the Maclaurin series.

\begin{figure}
    \centering
    \includegraphics[width=1.\linewidth]{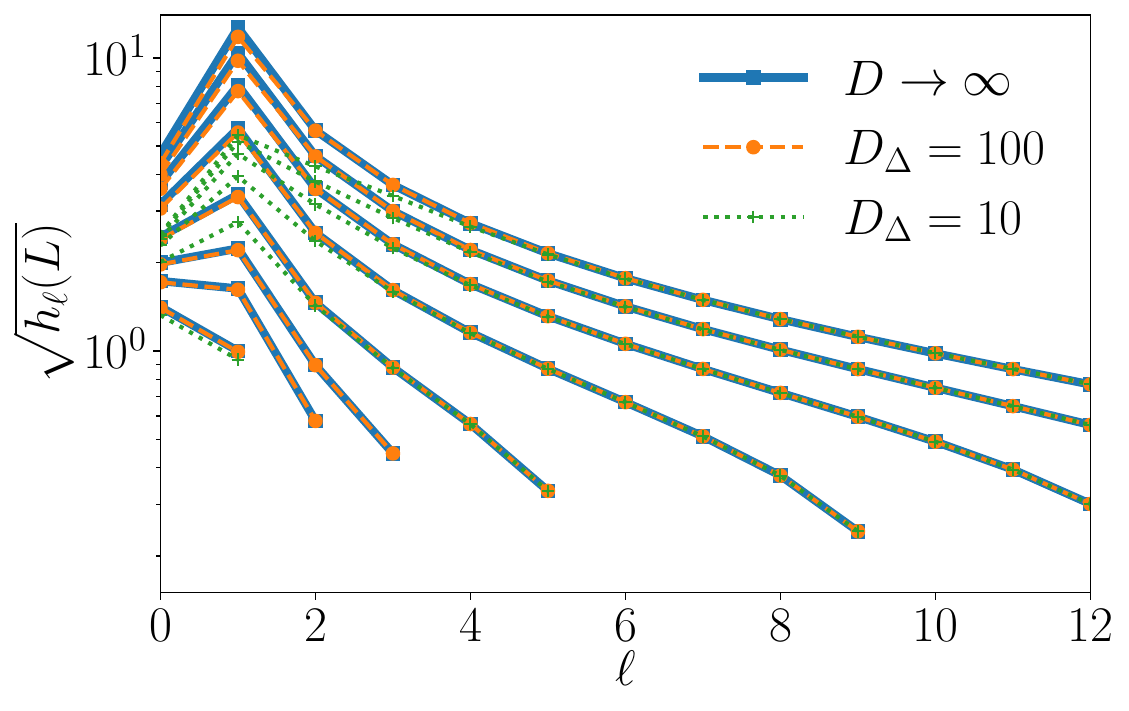}
    \caption{Square roots of first thirteen coefficients ${h_\ell}$ for ChE models with chain lengths ranging for (from left to right) $L = 2$, $3$, $4$, $6$, $10$, $14$, $20$, and $22$. Blue squares represent the ChE($L$) model in WBL, orange circles correspond to ChE($L$, $D_{\Delta} = 100$), and green crosses to ChE($L$, $D_\Delta = 10$). Lines are included as guides to the eye. Note that we plot $\sqrt{h_\ell}$ as this is proportional to the hopping terms in the chain. Coefficients for $\ell>12$ are practically aligned for all three cases. 
    \label{fig:coeffs22}}
\end{figure}

The coefficients $h_\ell$ can be obtained by directly solving the system of equations for all independent parameters in the \emph{Pad\'e} approximant. However, a more convenient approach is to use well-known continued fraction identities~\cite{wall2018analytic}, which provide the $h_\ell$ coefficients directly. We outline this straightforward procedure in Appendix~\ref{app:CF}, which includes illustrative examples, and also provide a Jupyter notebook with the implemented algorithm~\cite{ChECode}.

Given the popularity of the ZBW approximation and the practical relevance of the extended ZBW (eZBW)~\cite{zalom2024double} -- which correspond to ChE with $L=1$ and $L=2$, respectively -- we record the closed-form coefficients below. Unlike typical ZBW/eZBW implementations, where parameters are phenomenologically fitted or tuned to a specific target, the ChE construction fixes them. For $L=1$
\begin{equation}
    h_0 = \frac{2}{\pi} \arctan{( D_{\Delta})}
    \label{eq:h0_expression}
\end{equation}
and $L=2$ case,
\begin{equation}
\begin{aligned}
    h_0 &= \frac{2}{\pi} f(D_{\Delta})\arctan{(D_{\Delta})},\\
    h_1 &= f (D_{\Delta}) - 1,
\end{aligned}
\end{equation}
where
\begin{equation}
    f(D_{\Delta}) = 2\frac{(1+D_{\Delta}^2)\arctan{( D_{\Delta})}}{D_{\Delta} + (1+D_{\Delta}^2)\arctan{( D_{\Delta})}}
\end{equation}
with normalized half-bandwidth $D_{\Delta}=D/\Delta$. 

\begin{figure}
    \centering    
    \includegraphics[width=1.\linewidth]{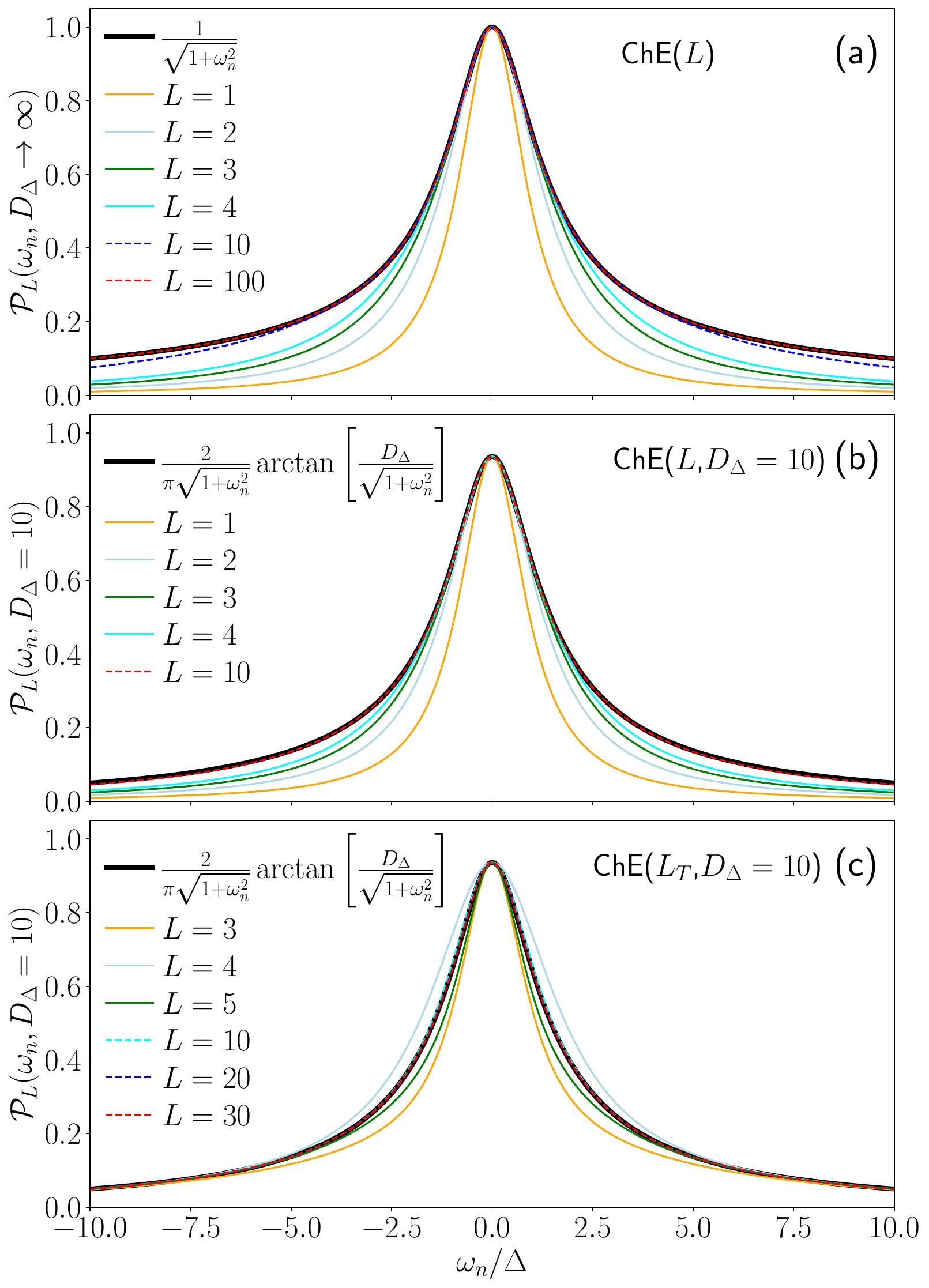}
    \caption{Comparison of the approximate function ${\cal P}_L(\omega_n,{h_\ell})$ with the full hybridization function $\tGD$ for $\Delta = 1$.
    (a) WBL, where the coefficients ${h_\ell}$ take a simple analytic form~\eqref{eq:coefWBL}.
    (b) Finite-bandwidth case with $D_\Delta = 10$, where ${h_\ell}$ are obtained using the method described in the main text and Appendix~\ref{app:CF}, based on matching the \emph{Pad\'e} approximant to the full tunneling self-energy.
    (c) Truncated expansion of the infinite-chain model, where the last coefficient is evaluated using Eq.~\eqref{eq:trancation}, with $D_\Delta = 10$. See also Fig.~\ref{fig:rozvoj_app}.
    \label{fig:rozvoj}}
\end{figure}

Fortunately, in WBL ($D\rightarrow \infty$), the coefficients $h_\ell$ 
take a simple form of rational numbers:
\begin{equation}
    h_0 = L,\qquad
    h_{\ell > 0} = \frac{L^2-\ell^2}{4\ell^2-1}.
    \label{eq:coefWBL}
\end{equation}
For large but finite $D \gg \Delta$, the finite bandwidth introduces only small corrections to the coefficients ${h_\ell}$, as illustrated in Fig.~\ref{fig:coeffs22}. Consequently, the effective models constructed using the above coefficients are sufficient for most realistic systems where $D$ is usually the largest energy scale. Therefore, most of the new results presented here are based on these simpler expressions. 

However, finite-bandwidth corrections offer some distinct advantages. The finite-$D$ approximation converges more rapidly to the exact hybridization function with increasing $L$ than in the WBL [compare Figs.~\ref{fig:rozvoj}(a) and (b)]. Moreover, unlike the WBL expression, the finite-$D$ coefficients do not diverge with $L$. The drawback of using finite $D$ is that the calculation of the coefficients $h_\ell$ for $L \gtrsim 20$ becomes computationally demanding when arbitrary-precision arithmetic is employed, or numerically unstable otherwise, due to the appearance of large numbers in the \emph{Pad\'e} approximant. This is not a problem for most ground-state investigations, where a typical system requires only a few sites in the chain, but it prohibits investigation of non-equilibrium phenomena where long chains are required.   

However, there is a simple remedy for that. As demonstrated in Appendix~\ref{app:CF}, a straightforward continued-fraction expansion of Eq.~\eqref{eq:GomegaD} can be used to derive the exact form of $h_\ell$ in the limit $L \rightarrow \infty$,
\begin{equation}
    h_0=\frac{2}{\pi}D_{\Delta},\qquad
    h_{\ell>0}=\frac{\ell^2 D_{\Delta}^2}{4\ell^2-1}.
\label{eq:coefLinf}
\end{equation}
Although not ideal for short-chain effective models, this form is useful when long chains are needed. It is particularly convenient when $D_{\Delta}$ is not too large (best if $D_{\Delta}< 20$). As a general strategy for the investigation of the non-equilibrium phenomena, we use both Eq.~\eqref{eq:coefWBL} and Eq.~\eqref{eq:coefLinf} to test the convergence. 
Both have their advantages and limitations as discussed in Sec.~\ref{sec:Bench}, but give the same result for long enough chains and large enough $D_{\Delta}$. When using Eq.~\eqref{eq:coefLinf}, it is also convenient to truncate the chain by changing $h_{L-1}$ to the value calculated using Eq.~\eqref{eq:trancation}, which significantly improves the low-energy behavior of shorter chains. 

To distinguish between the methods using different expansion coefficients, we adopt the following notation: ChE($L$) refers to models with coefficients ${h_\ell}$ given by Eq.~\eqref{eq:coefWBL}, corresponding to WBL with $L$ sites in the chain; ChE($L$, $D_{\Delta}$) denotes models with coefficients ${h_\ell}$ derived for a finite bandwidth $D_{\Delta}$ and $L$ sites in the chain; and ChE($L_T$, $D_{\Delta}$) represents models where the coefficients ${h_\ell}$ are taken from the $L \rightarrow \infty$ solution in Eq.\eqref{eq:coefLinf}, but truncated via Eq.~\eqref{eq:trancation} to include only the first $L$ terms. The notation with $L_{\infty}$ means that although the results are calculated with finite $L$, the correction of Eq.~\eqref{eq:trancation} was not used to change the last $h_{{L-1}}$. 

Figure~\ref{fig:rozvoj} shows how the various expansions approximate the tunneling self-energy in the vicinity of $\omega = 0$ on the imaginary-frequency axis. Together with its counterpart in Fig.~\ref{fig:rozvoj_app} in Appendix~\ref{app:CF}, it reveals that the ChE($L$) and ChE($L$, $D_\Delta$) schemes are deliberately tailored to be optimal at $\omega_n = 0$, and that their accuracy systematically extends to a wider imaginary frequency window as $L$ grows. 
Nevertheless, both schemes become inaccurate in the high-frequency limit, $|\omega_n| \to \infty$.
By contrast, the ChE($L_T$, $D_\Delta$) expansion reproduces the high-frequency tails correctly even for short chains, yet a substantially larger number of sites is needed to capture the low-frequency regime around $\omega_n = 0$, particularly for large bandwidths [see Fig.~\ref{fig:rozvoj}(c) and Fig.~\ref{fig:rozvoj_app}(c) in Appendix~\ref{app:CF}].

The generalization of ChE models to several dots and/or leads is straightforward. When a quantum dot $j$ is coupled to multiple leads indexed by $l$, a separate chain can be introduced for each lead, characterized by its own hybridization $\Gamma_{jl}$, bandwidth $D$, superconducting order parameter $\wD_l = \Delta_l e^{i\varphi_l}$, and length $L$.
However, if several leads made of the same material (i.e., with identical $D$ and $\Delta$) are connected only to the same QD, they can be combined into a single effective chain (sChE). This mapping employs the geometric factor $\bm{\chi}$ introduced in Ref.~\cite{Zalom2023hidden}, along with minor adjustments of the chain parameters, as discussed in detail in Appendix~\ref{app:multileat}.

The situation is more involved when several dots are connected to the same lead. Consider, for instance, two dots ($j=1,2$) coupled to a common SC reservoir. The non-interacting Green's function, discussed in Appendix~\ref{app:double dot}, then contains non-local terms such as $V_{1l\mathbf{k}}V_{2l\mathbf{k}}$, which encode, among other things, the spatial separation of the impurities. 
Assuming real $V_{il\mathbf{k}}$, these mixed contributions can again be approximated by energy-independent tunneling rates $\Gamma_{jj',l}$. This is, however, a significant simplification; the actual dependence can be much more complex even in an isotropic environment~\cite{Eickhoff2018effective}. 

By the Cauchy-Schwarz inequality, one may then write $\Gamma_{jj',l}= \zeta\sqrt{\Gamma_{j,l}\Gamma_{j',l}}$ with $\zeta\in\langle 0,1\rangle$, which quantifies the correlations between the dots mediated by the shared lead. The most correlated case $\zeta=1$ requires only a single effective channel, whereas generic $\zeta$ values necessitate two. Within the ChE framework we therefore connect both dots to two identical chains ($l=1,2$) with tunneling rates chosen as $\Gamma_{1,1}=(1-\delta)\Gamma_{1}$, $\Gamma_{12}=\delta\Gamma_{1}$, $\Gamma_{21}=\delta\Gamma_{2}$, and $\Gamma_{22}=(1-\delta)\Gamma_{2}$, where $\delta\in\langle 0,0.5\rangle$ and $\zeta=2\sqrt{\delta(1-\delta)}$. The limit $\zeta=0$ ($\delta=0$) represents two dots that are not correlated through the substrate and thus effectively couple to independent reservoirs, whereas $\zeta=1$ ($\delta=0.5$) corresponds to a local two-level impurity coupled to a single SC bath. The ChE construction for systems with more than two dots follows by straightforward generalization. Some systems not treated in this work require a more careful application of the chain-expansion scheme. An example is a QD Aharonov--Bohm interferometer, where the two leads are directly hybridized~\cite{zalom2025andreev}. 

Throughout this work, we adopt the standard gauge convention~\cite{Meden2019the}, which shifts the SC phases from the chain sites into the hybridization rates. This transformation leads to $\gamma_{jl} = \sqrt{h_0 \Delta \Gamma_{jl}} e^{i\varphi_l/2}$, while keeping the SC pairing term $\wD = \Delta$ and all other chain parameters real. The current operator $J_{jl}$, describing the particle current between dot $j$ and lead $l$ within the ChE framework, can be derived directly from the time evolution of the particle number operator associated with the lead
\begin{equation}
\begin{aligned}
J_{jl} &= -\frac{d N_l}{dt} = i\left[N_l,\mathcal{H}^{\mathrm{ChE}(L)}\right]\nonumber,\\
\label{eq:current1}
\end{aligned}
\end{equation}
where the relevant part is given by the commutator with the dot-lead coupling term:
\begin{equation}
    J_{jl} = -i \sum_\sigma \left( \gamma^*  c^\dagger_{1l\sigma}   d^{\pdag}_{j\sigma} - \gamma d_{j \sigma}^{\dagger} c^{\pdag}_{1l\sigma} \right).
\end{equation}

Before presenting our results, we emphasize that the ChE framework is quite general and naturally encompasses several approaches previously employed in the literature. For example, the widely used ZBW method is equivalent to ChE with $L = 1$, although the ZBW parameters are usually chosen arbitrarily. Our analysis reveals that ChE($L=1$) fails to approximate the tunneling self-energy even at leading order in $\omega_n$. In light of this, it is perhaps unsurprising that the ZBW method can yield qualitatively incorrect results, as recently demonstrated by Zalom \emph{et~al.}~\cite{zalom2024double}. This observation is particularly significant given that ZBW is frequently used not only for qualitative understanding but also in the interpretation of experimental data.

In contrast, the slightly more sophisticated ChE($L = 2$), which corresponds to the eZBW method~\cite{zalom2024double}, often resolves these shortcomings. 
Unlike earlier eZBW implementations, our formulation fixes the coefficients in the low-energy expansion exactly, yielding a consistent and systematically controlled approximation. The surrogate models proposed by Baran \emph{et~al.}~\cite{Baran2023surrogate}, although formulated differently, can likewise be interpreted as particular instances of the ChE framework. 
In their approach, the discretized reservoir parameters are adjusted to reproduce the tunneling self-energy across the full bandwidth for finite $D$, though the fitting procedure is intrinsically weighted toward low frequencies.
However, to gain more insight into the surrogate-model eigen-states, the chain representation of discretized lead is also employed~\cite{Baran2023surrogate}. In this sense, their construction resembles ChE($L_T$, $D_\Delta$), whereas in our approach, the coefficients are provided analytically.

\section{Benchmarking, Usage Strategies, and Ancillary Results \label{sec:Bench}}
\begin{figure}
    \centering
\includegraphics[width=1.\linewidth]{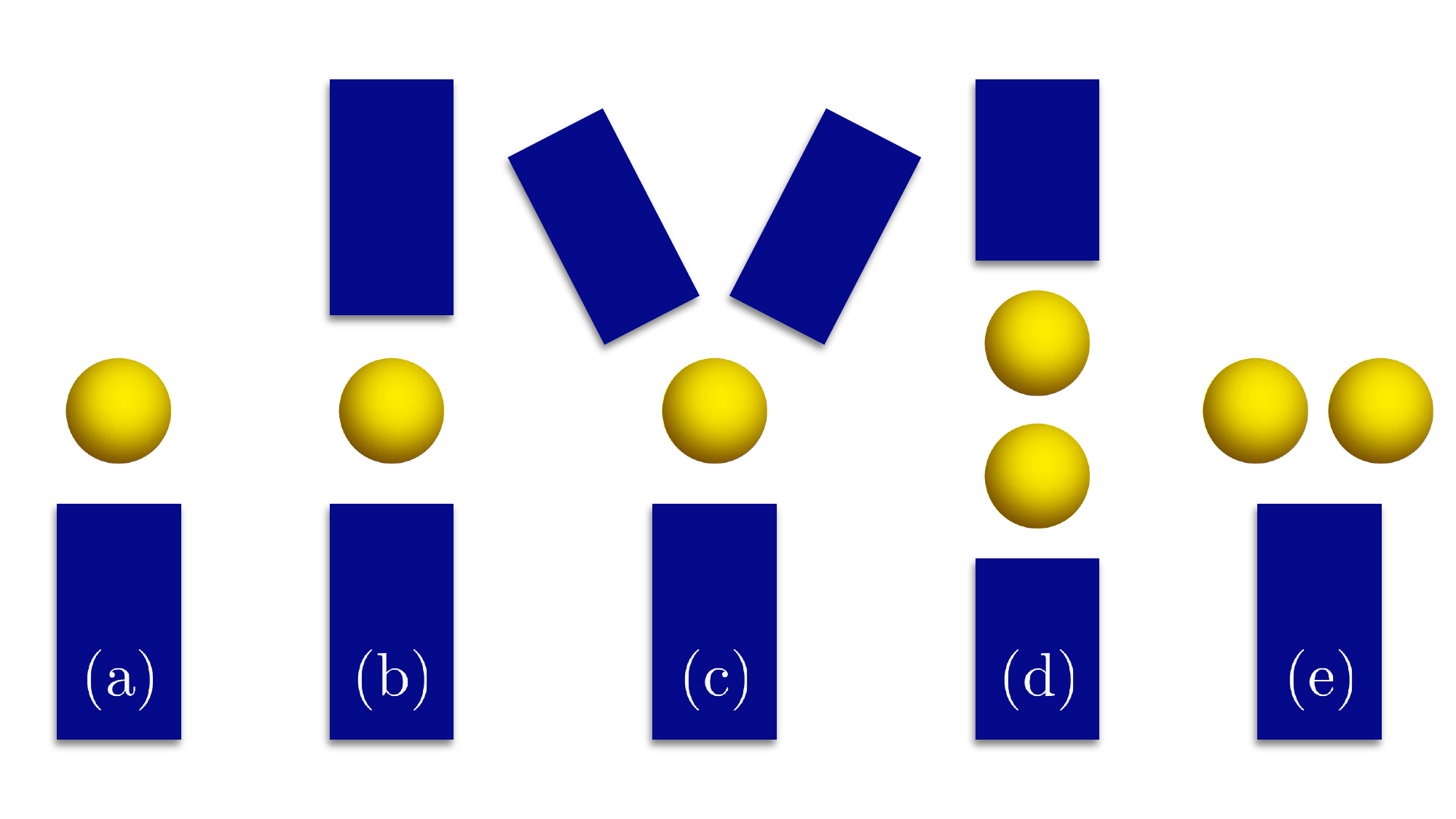}
    \caption{Illustration of systems used to benchmark the ChE models against previous results. (a) Single QD coupled to one lead. (b) Junction bridged by a QD. (c) Single QD coupled to three leads. (d) DQD in serial configuration. (e) DQD in parallel configuration.
    \label{fig:bench}}
\end{figure}

To illustrate both the strengths and limitations of effective ChE models, we reexamine a set of systems and problems that have previously been studied in the literature using precise methods and other effective models. Accordingly, we do not dwell on detailed recapitulations of established results. We instead concentrate on introducing some general strategies for employing ChE for different types of problems and testing its various versions. In this respect, the section also contains new ancillary results. For ground-state properties, we focus on setups that are amenable to accurate solutions via NRG. For time-dependent dynamics, we compare our results to those obtained using non-equilibrium Green's function (NEGF) techniques and briefly discuss the relation to other methods. To show the versatility of the ChE scheme, we address several system geometries illustrated in Fig.~\ref{fig:bench}.

\subsection{Single QD coupled to one SC lead}
\begin{figure}
    \centering
    \includegraphics[width=1\linewidth]{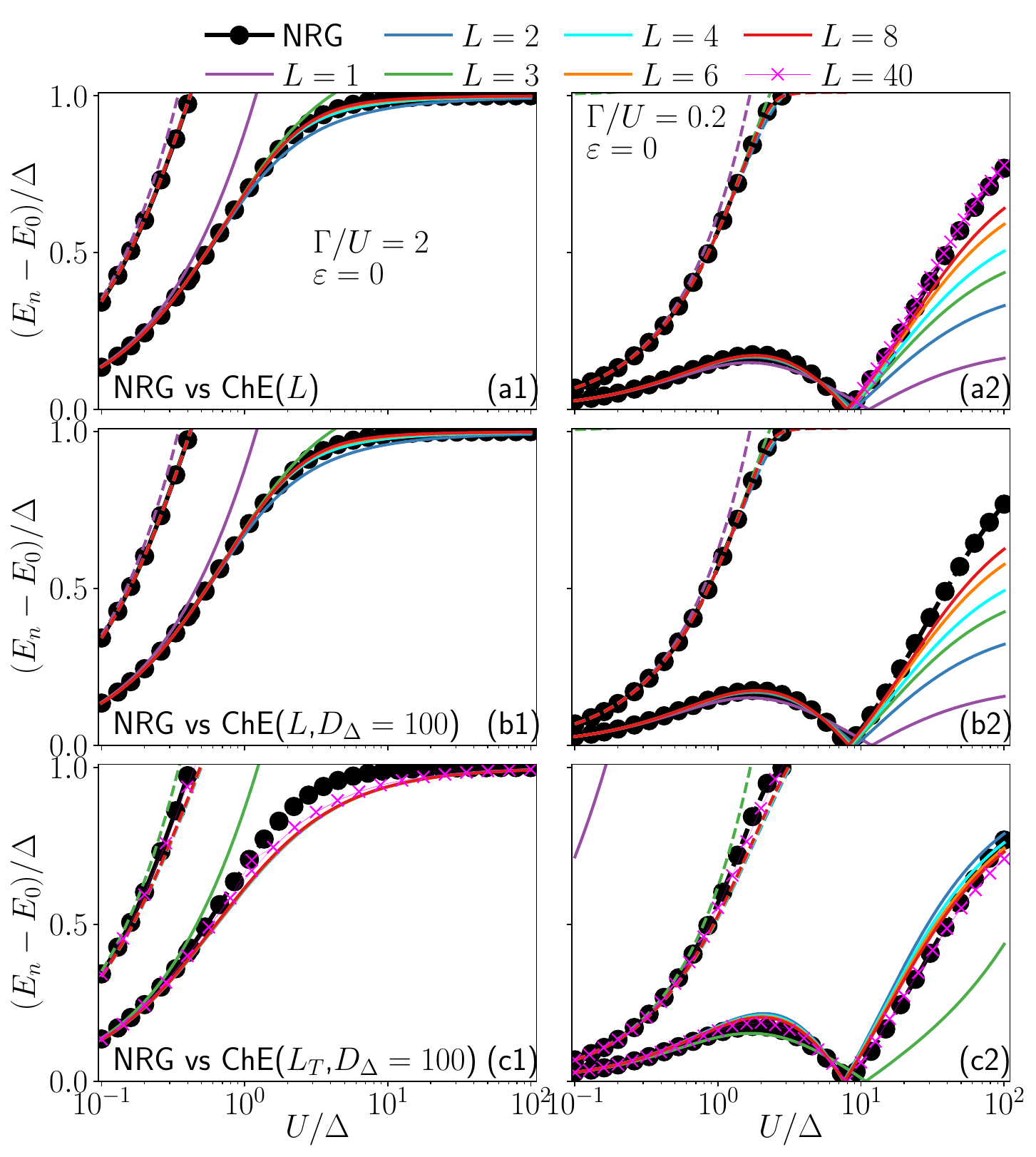}
    \caption{Single QD coupled to a single SC lead, as illustrated in Fig.~\ref{fig:bench}(a). Comparison of the excited in-gap states calculated with various versions of ChE effective models to the NRG data over three orders of magnitude of $U$ for fixed ratios $\Gamma/U=2$ (left column) and $\Gamma/U=0.2$ (right column). The NRG data (black circles) have been calculated for half-bandwidth $D=100\Delta$. Panels (a1) and (a2) show infinite band ChE($L$), panels (b1) and (b2) the finite band ChE($L$,$D_\Delta=100$), and panels (c1) and (c2) the truncated ChE($L_T,D_\Delta=100$). All results are for the half-filled case $\varepsilon=\epsilon + U/2=0$.} 
    \label{fig:1d1lChe}
\end{figure}

The equilibrium properties of a single QD coupled to one SC lead [Fig.~\ref{fig:bench}(a)] can be efficiently addressed even by computationally demanding methods such as NRG and CT-HYB quantum Monte Carlo. This makes it an ideal benchmark system for testing the validity and convergence of ChE-based effective models. Figure~\ref{fig:1d1lChe} shows the evolution of excited in-gap many-body states over three orders of magnitude in $U$, for fixed ratios of $\Gamma/U = 2$ (left column) and $\Gamma/U = 0.2$ (right column). The black circles indicate NRG data calculated for a half-bandwidth of $D = 100\Delta$. Consequently, results in the upper range correspond to values of $U$ and $\Gamma$ comparable to $D$, placing the system effectively in a small-gap regime where our approximations, relying on the SC character of the leads, are expected to break down.

We test three variants of ChE: the infinite-band approximation ChE($L$) in (a1) and (a2); finite-band models ChE($L$,~$D_{\Delta}=100$) in (b1) and (b2); and the truncated expansion ChE($L_T$,~$D_\Delta=100$) in (c1) and (c2). Panels (a1)-(b2) clearly demonstrate that both ChE($L$) and ChE($L$,~$D_\Delta$) with two to four sites in the chain suffice to accurately reproduce the NRG results over more than two orders of magnitude in $U/\Delta$. Furthermore, as $L$ increases, the ChE models show systematic improvement even in the regime $U \gg \Delta$. We present results up to $L = 8$, which can be readily solved via Lanczos ED within minutes on a standard PC. To illustrate convergence, we also include DMRG results for ChE with $L = 40$ [purple crosses in panel (a2)], which are in very good agreement with NRG, despite the use of the infinite-$D$ approximation in a regime where $U$ is comparable with $D$.
 
Panels (c1) and (c2) show that, although the truncated expansion does not match the precision of ChE($L$) at small $L$, it still yields robust effective models, particularly for small $\Gamma/U$, and its accuracy improves with increasing $L$. Nevertheless, the true usefulness of this approach will become apparent in problems that require long chains.
\begin{figure}[ht]
    \centering
    \includegraphics[width=1\linewidth]{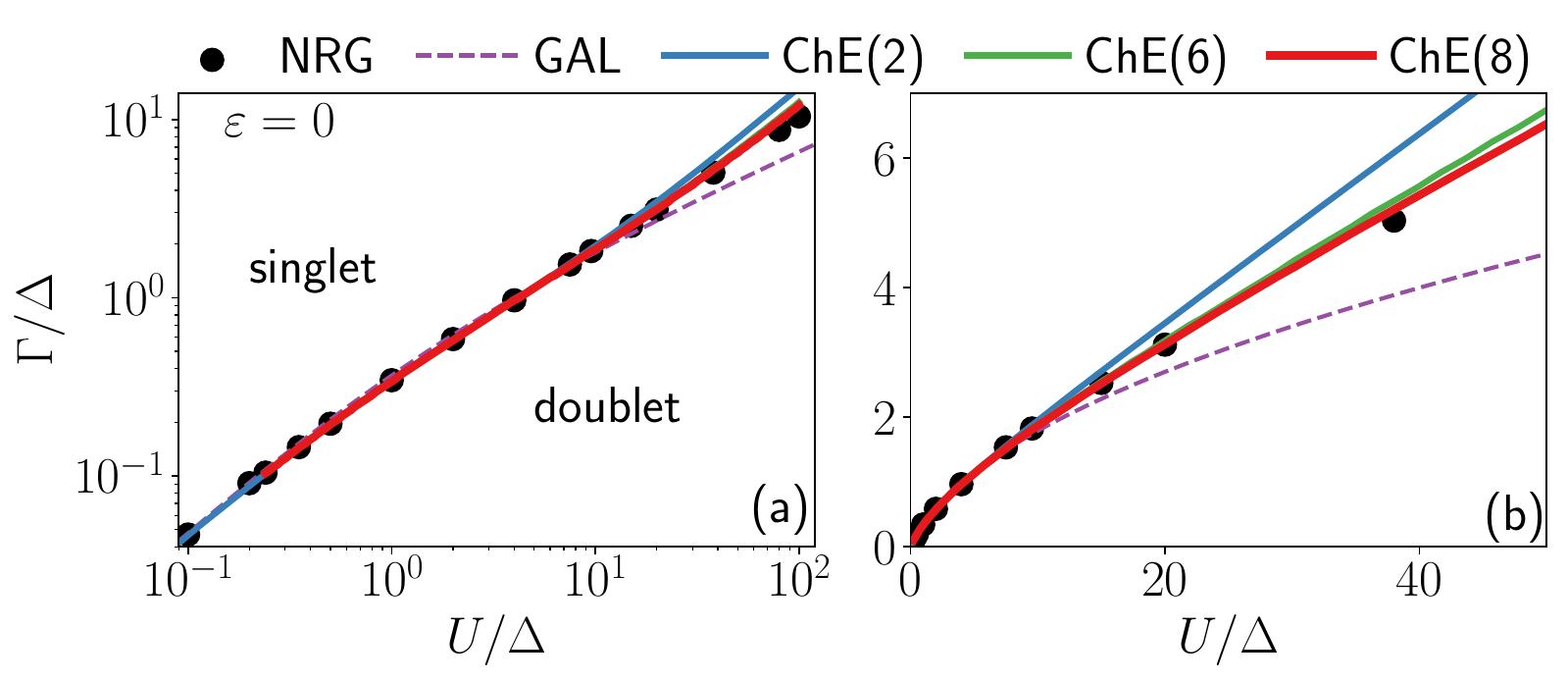}
    \caption{Single QD coupled to a single SC lead as illustrated in Fig.~\ref{fig:bench}(a). Phase boundary between the singlet and doublet ground state at half-filling estimated from NRG (black circles), GAL (dashed purple line), and ChE(L) (solid lines) for different $L$. Panels (a) and (b) show the same results using different scales.} 
    \label{fig:1d1lChe2}
\end{figure}

Figure~\ref{fig:1d1lChe2} shows the phase diagram of the half-filled ($\varepsilon = \epsilon + U/2 =0$) SC-AIM, with the phase boundary separating the singlet and doublet ground states. The NRG results (black circles) are compared with GAL (purple dashed line) and ChE($L$) for $L = 2$, $6$, and $8$ (solid lines). Panels (a) and (b) present the same data on different scales to better emphasize the differences across regimes. Notably, even ChE($L = 2$) shows better agreement with NRG (being virtually exact for $U < 10\Delta$) than GAL. Moreover, increasing $L$ extends this agreement to larger values of $U/\Delta$.

However, it is important not to overstate these results. Although computationally trivial, ChE($L = 2$) still represents an interacting model with three sites. In contrast, the GAL result is based on a compact analytical formula~\cite{Zonda2015perturbation,Zonda2023generalized}:
\begin{equation}
    \left(\frac{U}{2\left(1+\frac{\Gamma_T}{\Delta}\right)}\right)^2
    =\varepsilon^2 + \Gamma_T^2 \cos^2\left(\frac{\phi}{2}\right),
    \label{eq:GAL}
\end{equation}
shown here for the more general case of two equally coupled leads with $\Gamma_1 = \Gamma_2 = \Gamma_T/2$ and a Josephson phase $\varphi = \varphi_1 - \varphi_2$. For a single lead, we have $\varphi = 0$ and $\Gamma_T = \Gamma$, which further simplifies the expression. Despite its simplicity, GAL captures the phase boundary with experimental-level precision in the weak to intermediate interaction regime.

The true strength of the ChE approach, however, becomes evident away from half-filling, where GAL requires phenomenological corrections, and in more complex systems where GAL and other commonly used effective models may even fail qualitatively.

\subsection{Single QD coupled to multiple SC leads}
The system of two SC leads connected via QD [Fig.~\ref{fig:bench}(b)] is of central importance. Depending on the parameter regime, it represents a generalized Josephson junction or an impurity on the SC surface probed by an SC STM tip. Moreover, it has recently been shown that any single-dot multiterminal system without direct inter-lead coupling can be exactly mapped to a QD symmetrically coupled to only two leads~\cite{Zalom2023hidden}. This is of particular importance, as three-terminal QD devices can be used to construct SC diodes and transistors, and even greater functionality is expected in more complex configurations. In addition, considerable effort has been devoted to studying out-of-equilibrium phenomena, including current at finite voltages and the microwave response of such systems. These offer enhanced tunability and access to dynamical information via the supercurrent response. Consequently, there is significant value in developing effective models that enable fast and reliable parameter scans in equilibrium as well as to address long-time dynamics.
\paragraph{Equilibrium properties:}
The central quantity for the investigation of the junction is the supercurrent, which, for a finite Josephson phase $\varphi$, flows through the QD even in equilibrium. A sudden reversal of its direction upon tuning a parameter signals a QPT between singlet and doublet phase, i.e., crossing of the many-body ground state with the first excited state of different character.

\begin{figure}[ht]
    \centering
   \includegraphics[width=1\linewidth]{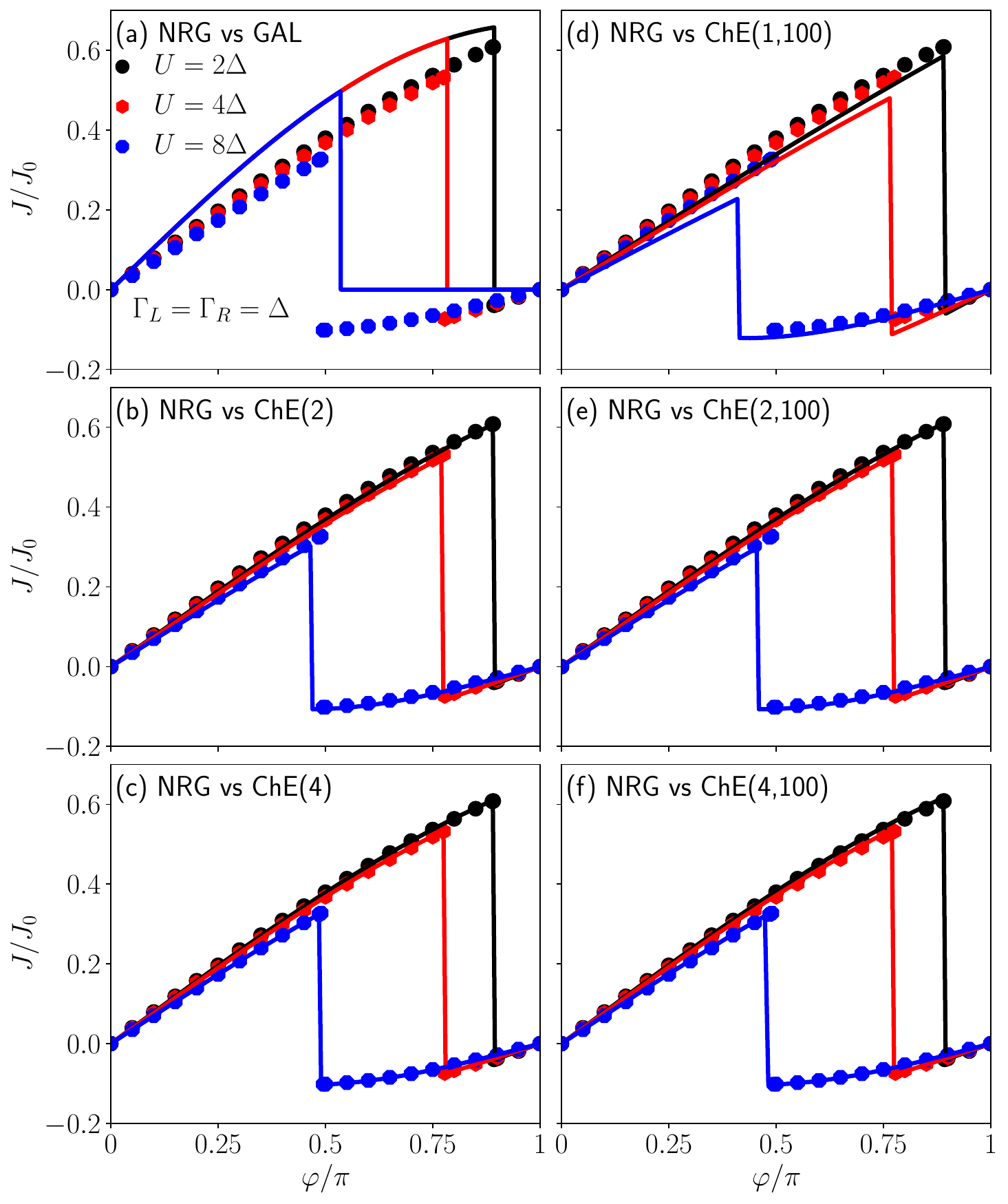}
    \caption{Single QD between two SC leads as illustrated in Fig.~\ref{fig:bench}(b). Comparison of the supercurrent as a function of the Josephson phase for different $U$ at half-filling ($\varepsilon=0$) between NRG (black circles) and GAL (a), infinite band ChE($L=2$) (b), ChE($L=4$) (c), finite band ChE($L=1$,$D_{\Delta}=100$) (d), ChE($L=2$,$D_{\Delta}=100$) (e), and ChE($L=4$,$D_{\Delta}=100$) (f). The NRG data were calculated for $D=100\Delta$ and $J_0=e\Delta/\hbar$. 
} 
    \label{fig:1d2lChe_Cur}
\end{figure}

In Fig.~\ref{fig:1d2lChe_Cur} we compare the NRG results for the supercurrent in a system with $\Gamma_1=\Gamma_2\equiv\Gamma=\Delta$ and three interaction strengths $U=2$ (black), $U=4$ (blue) and $U=8$ (red circles) with the predictions of GAL (a), ChE($L=2,4$) (b,c) and ChE($L=1,2,4$,$D_{\Delta}=100$) (d-f). 
As shown in panel (a) and discussed in detail in Ref.~\cite{Zonda2023generalized}, for the geometry of Fig.~\ref{fig:bench}(b), the GAL model yields a \emph{qualitatively} incorrect Josephson current $J$ as it vanishes in the doublet phase and is independent of $U$ in the singlet phase. A band correction, denoted GAL+C in Ref.~\cite{Pokorny2023effective}, is needed to tame these shortcomings. In contrast, ChE-based models do not suffer from these limitations because the band influence is encoded directly in the chain representation. Remarkably, even for $L=1$, equivalent to ZBW [Fig.~\ref{fig:1d2lChe_Cur}(d)], ChE($L=1$,$D_{\Delta}=100$) already provides a reasonable estimate of $J$, and the agreement improves systematically with increasing $L$. Panels (b) and (c) further demonstrate that for short chains and large $U$, the simpler ChE($L$), which uses analytical chain coefficients, can match  ChE($L$,$D_{\Delta}$), justifying its use for systems with sufficiently wide band.

\begin{figure}[ht]
    \centering
\includegraphics[width=1\linewidth]{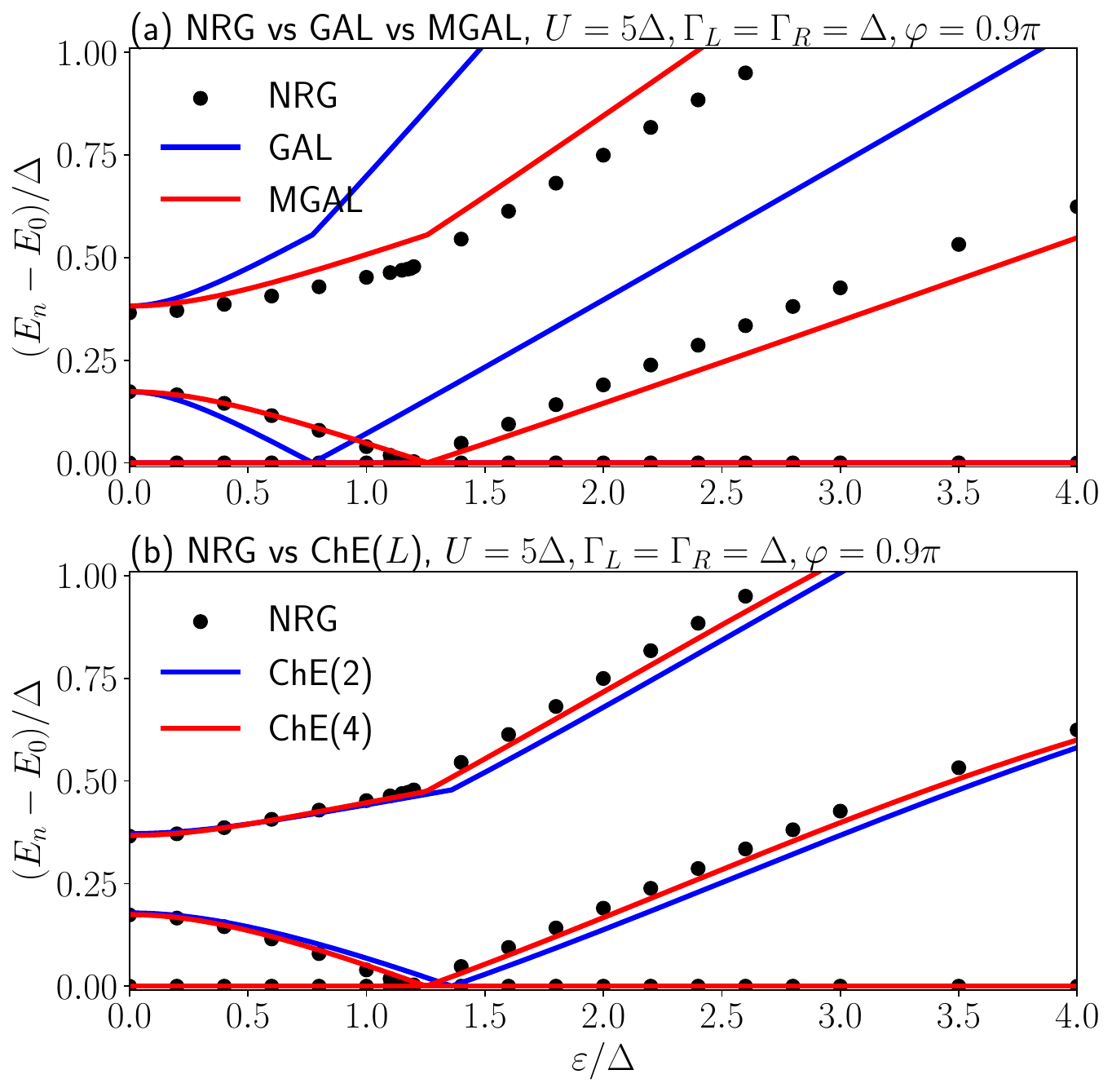}
    \caption{Evolution of the in-gap states away from half filling for a single QD between two SC leads as illustrated in Fig.~\ref{fig:bench}(b). 
    (a) Comparison between the NRG (black circles), GAL (blue lines), and MGAL (red lines) results. 
    (b) Comparison of NRG data (black circles) with ChE($L=2$) (blue lines) and ChE($L=4$) (red lines).
}   
    \label{fig:1d2lCheABS}
\end{figure}

Beyond the qualitative issues noted above, the GAL model is in good \emph{quantitative} agreement with NRG only near half filling ($\varepsilon=0$). Therefore, a phenomenological scaling, leading to the MGAL model, was introduced in Refs.~\cite{Kadlecova2019practical,Zonda2023generalized}. The improvement is evident in Fig.~\ref{fig:1d2lCheABS}(a), which shows the in-gap spectrum as a function of $\varepsilon$ for Josephson phase $\varphi=0.9\pi$. The ChE($L$) models, by contrast, require neither such corrections nor the ad-hoc parameter tuning often employed in ZBW studies. Short-chain models already reproduce the NRG results with high accuracy far from half-filling, as shown in Fig.~\ref{fig:1d2lCheABS}(b).

\begin{figure}[ht]
    \centering
   \includegraphics[width=1\linewidth]{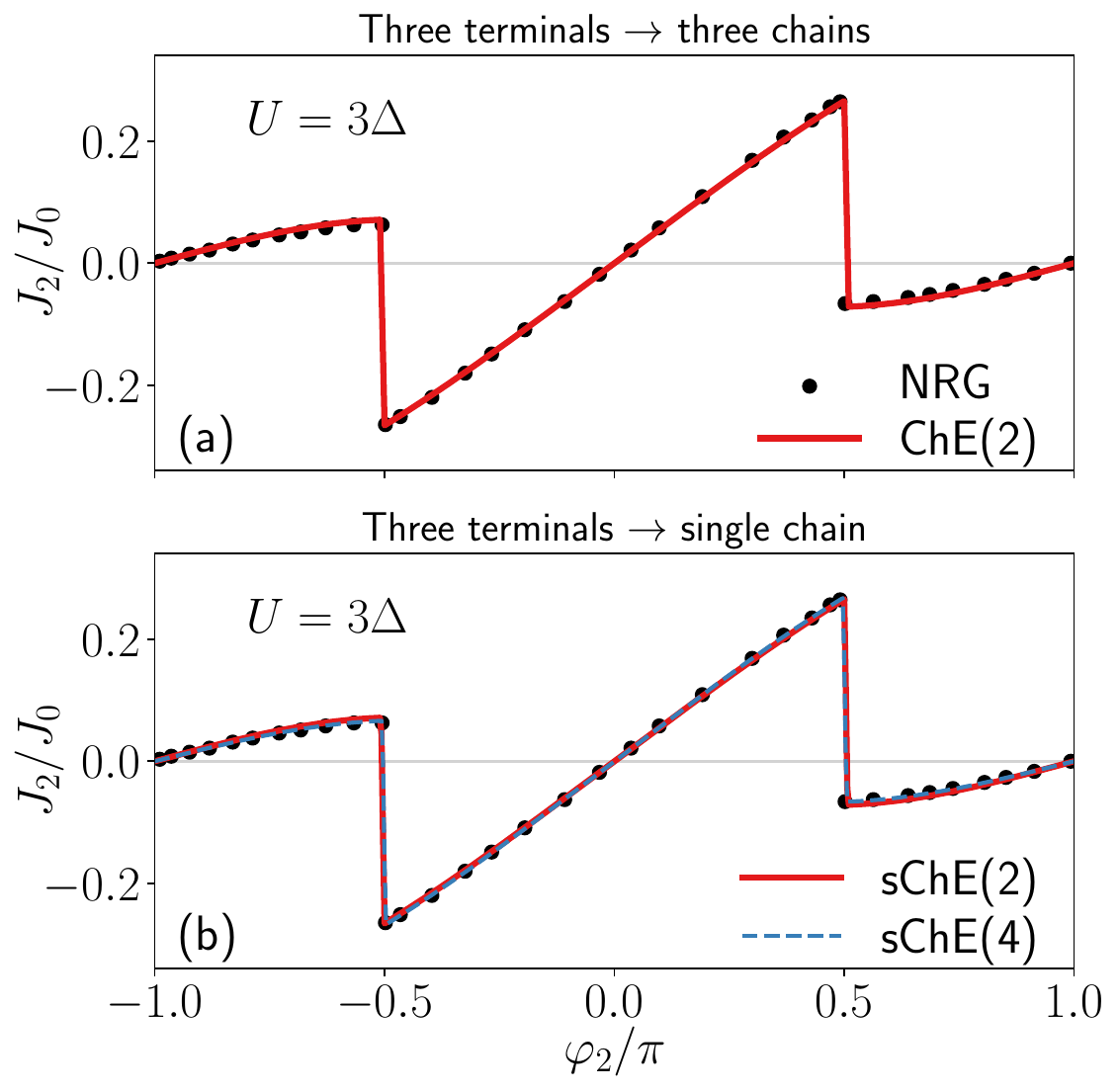}
    \caption{Supercurrent between the QD and the second lead in the three-terminal setup shown in Fig.~\ref{fig:bench}(c) with $\Gamma_1=0.45\Delta$, $\Gamma_2=0.4\Delta$, $\Gamma_3=0.15\Delta$ and $\varphi_1=\varphi_3=0$. Black circles denote NRG data obtained for a two-terminal setup and mapped to the three-terminal case via the transformation of Ref.~\cite{Zalom2023hidden}. (a) Comparison with ChE($L=2$) (red line) using three chains, each representing a single lead. (b) Comparison with sChE($L$) using a single chain to model all three leads, as described in Appendix~\ref{app:multileat}: red solid line for $L=2$ and blue dashed line for $L=4$. 
    \label{fig:1d3lcurr}
}
\end{figure}

To demonstrate the versatility of ChE models, we briefly discuss results for a three-terminal setup illustrated in Fig.~\ref{fig:bench}(c). Figure~\ref{fig:1d3lcurr} shows the supercurrent flowing between the QD and the second terminal for the system with $\Gamma_1=0.45\Delta$, $\Gamma_2=0.4\Delta$, $\Gamma_3=0.15\Delta$, $U=3\Delta$, $\varepsilon=0$, $\varphi_1=0$, and $\varphi_3=0$. The black circles show the NRG result (taken from Ref.~\cite{Zalom2023hidden}). These have been obtained by using the exact transformation introduced in Ref.~\cite{Zalom2023hidden}. This means that the raw data have been obtained for a two-terminal setup and then mapped to three terminals. In contrast, the ChE results in panel (a) are a direct calculation, where each lead was approximated by its own short chain of two sites. The results are in very good agreement.

Using the geometric factor of Ref.~\cite{Zalom2023hidden} (see Appendix~\ref{app:multileat}), multiple leads attached to the same dot can be mapped to a single chain. As shown in panel (b), the sChE representation matches the accuracy of the direct multi-chain calculation while yielding a much smaller effective model.

For this setup and for the cases in Figs.~\ref{fig:1d2lChe_Cur}--\ref{fig:1d2lCheABS}, the direct construction is therefore unnecessary. Nevertheless, it remains useful when the single-chain reduction does not apply--for example, when leads couple to multiple different dots, when metallic and SC leads coexist, when the SC gaps differ in magnitude, or under non-equilibrium driving. The direct formulation also enables hybrid strategies, e.g., to treat selected channels with NRG while describing additional leads by short ChE chains.

\paragraph{Driven systems:}
ChE models are not confined to equilibrium situations.  
To illustrate their scope, we consider here a junction subjected to a finite bias voltage.

\begin{figure}[ht]
    \centering
    \includegraphics[width=\linewidth]{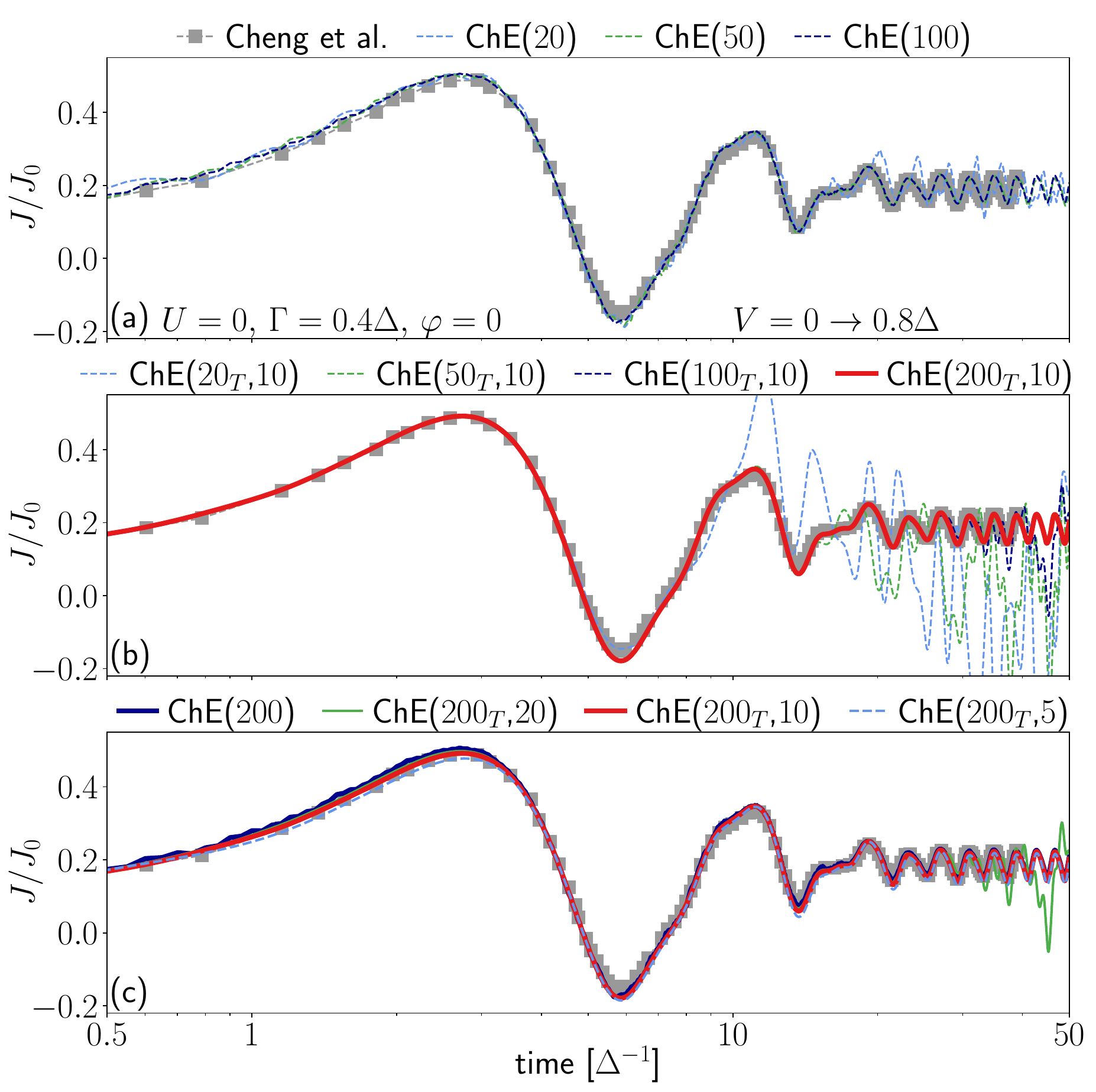}
    \caption{Single QD between two SC leads as illustrated in Fig.~\ref{fig:bench}(b).  
    Transient current after a quench from zero to a finite bias.  
    Grey squares: data extracted from Cheng \textit{et~al.}~\cite{Cheng2024quasiparticle}, obtained with the NEGF approach.  
    (a) Comparison with a tight-binding ChE($L$) chain.
    (b) Results of ChE($L_T,D_{\Delta}=10$) for several $L$s.  
    (c) Influence of the bandwidth on the transient dynamics.
    \label{fig:1d2lCheV}}
\end{figure}

Figure~\ref{fig:1d2lCheV} displays the transient current of a non-interacting junction ($U=0$) that is quenched from zero bias ($V=0$) to a finite bias $V=0.8\Delta$, where $V=\mu_1-\mu_2$, i.e., the difference between the chemical potentials of the leads. We use $\Gamma=0.4\Delta$, $\varphi=0$, and $\varepsilon=0$ because the same protocol was analysed by Cheng \emph{et al.} in Ref.~\cite{Cheng2024quasiparticle} with Keldysh NEGF; the gray squares represent their data taken from Fig.~7 of Ref.~\cite{Cheng2024quasiparticle}.

To tackle the quench, we follow the unitary transformation of Ref.~\cite{Cheng2024quasiparticle}.  
For the special case $\Gamma_1=\Gamma_2$ and $\mu_1=-\mu_2$, it transfers the static bias of the leads into time-dependent phases:
\begin{eqnarray}
    &&\mu_1 = \frac{V}{2} \rightarrow \mu_1 = 0,\qquad
    \mu_2 = -\frac{V}{2} \rightarrow \mu_2 = 0,\nonumber\\
    &&\varphi_1 = \frac{\varphi}{2} \rightarrow
    \varphi_1(t) = \frac{\varphi}{2} + Vt,\nonumber\\
    &&\varphi_2 = -\frac{\varphi}{2} \rightarrow
    \varphi_2(t) = -\frac{\varphi}{2} - Vt,
\end{eqnarray}
which leads to complex, time-dependent couplings
$\gamma_l(t)=\sqrt{h_0\Delta\Gamma_l}\,e^{i\varphi_l(t)/2}$. Because of $U=0$ ChE representation reduces to a tight-binding chain and, therefore, the evolution of chains comprising thousands of sites is straightforward, allowing us to examine relaxation processes over long time scales.

As Fig.~\ref{fig:1d2lCheV}(a) shows, a chain of only $L=50$ sites already reproduces the NEGF result up to the steady-state. Shorter chains, such as $L=20$, are inadequate because the signal reflected at the end of the chain reaches the dot before relaxation is completed. In this respect, the infinite-band ChE($L$) representation is superior to its truncated counterpart ChE($L_T,D_{\Delta}$), shown in Fig.~\ref{fig:1d2lCheV}(b), which requires longer chains. This stems from the larger hopping amplitudes $\tilde{h}_\ell$ in the truncated model as the propagation time between neighboring sites scales as $1/\tilde{h}_\ell$. Hence, as Fig.~\ref{fig:1d2lCheV}(c) illustrates, if finite $D_\Delta$ is required, it is advantageous to employ the smallest bandwidth permitted by the physical problem.

Despite the need for longer chains when analyzing long-time dynamics, the truncated representation can still be preferable in certain contexts. ChE($L$) possesses a distinct parameter set $h_\ell$ for every chain length, so short-time dynamics vary slightly with $L$ even when the initial state is identical, giving rise to the small oscillations visible in panel (a) even for short times. By contrast, the coefficients up to the site $L_T$ are identical for any chain length in ChE($L_T,D_{\Delta}$), because they are derived from the $L\!\to\!\infty$ limit.  
Consequently, the dynamics coincide for different $L$ until finite-size effects, namely the back-propagation of the signal reflected from the chain end, become relevant. In addition, spurious small oscillations are absent for ChE($L_T,D_{\Delta}$).

\subsection{Junction with a double QD in serial configuration}
Serial DQDs sandwiched between two SC leads have become an active field of study due to the potential to host qubits~\cite{Geier2024fermion,Steffensen2025YSR,Baran2021subgap} and their advanced tunability because, in principle, the parameters of each dot, in particular the level energies $\epsilon_j$, can be tuned independently~\cite{Saldana2018supercurrent,Saldana2020two,Zitko-2010-dqd,Lee-2010,Karrasch-2011}. This tunability was recently exploited in a proposal predicting an anomalous time-averaged Josephson current-phase relation, including complete rectification, when phase-shifted microwave gates are applied~\cite{Taberner2023anomalous}. For the latter, the results were obtained in the $\Delta\rightarrow\infty$ atomic limit or for a noninteracting system ($U_1=U_2=0$), and it is therefore not clear how they translate to more realistic conditions. Nonetheless, it was already shown that, in equilibrium, (G)AL does not suffer from the qualitative drawbacks observed for a single-QD junction~\cite{Zonda2023generalized}, suggesting that the non-equilibrium finite $U$ predictions may also remain qualitatively valid.
\begin{figure}[!ht]
    \centering
    \includegraphics[width=\linewidth]{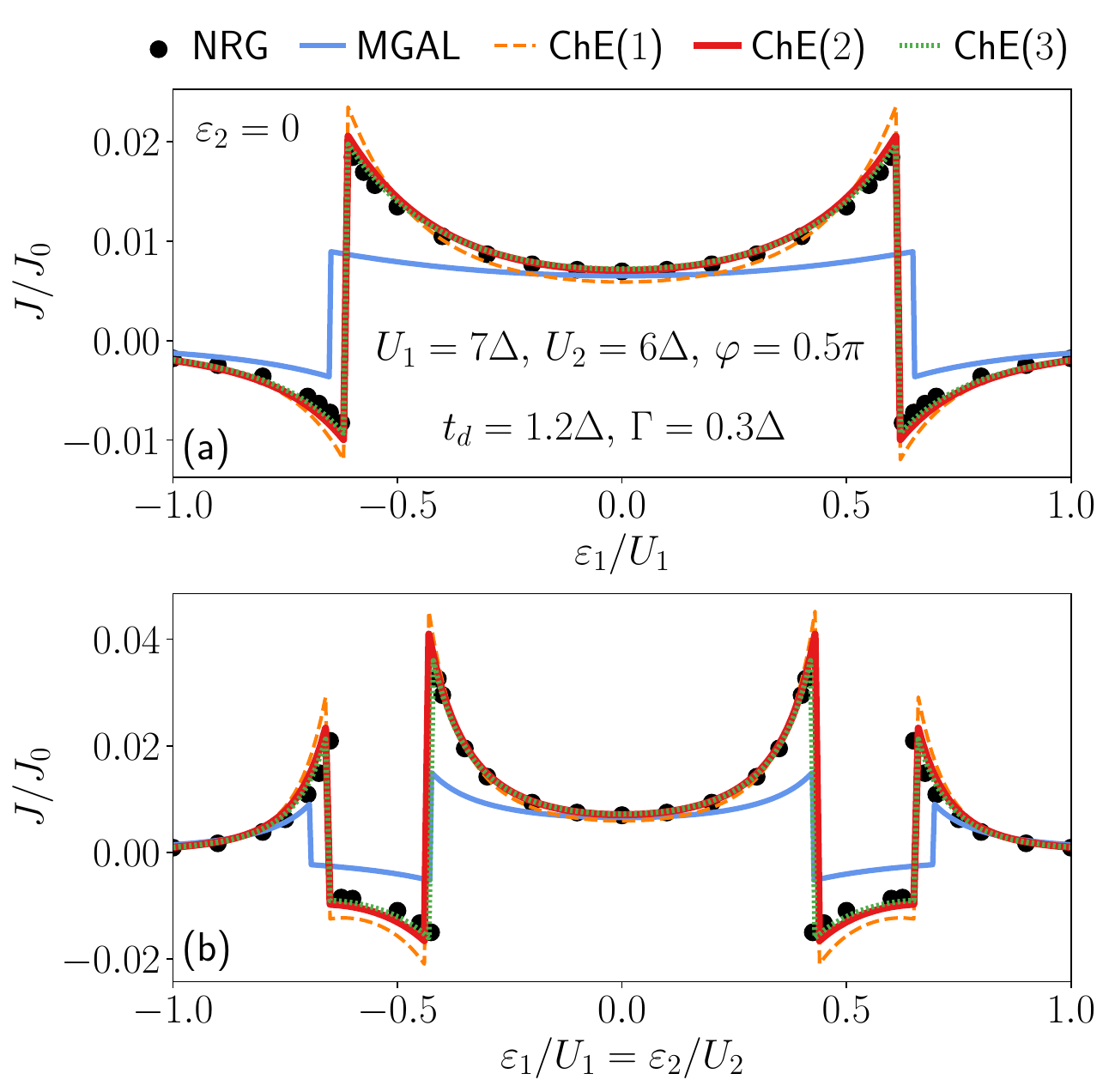}
    \caption{Serial DQD between two SC leads as illustrated in Fig.~\ref{fig:bench}(d). Josephson current for the experimental parameters of Ref.~\cite{Saldana2018supercurrent} as a function of $\varepsilon_1/U_1$ with $\varepsilon_2=0$ (a) and for $\varepsilon_1/U_1=\varepsilon_2/U_2$ (b). Black circles: NRG; blue line: MGAL~\cite{Zonda2023generalized}; orange, red, and green lines: ChE$(L)$ for $L=1$, $2$ and $3$, respectively. We use shifted energy levels $\varepsilon_j=\epsilon_j+U/2$.
    \label{fig:2d2lcurr}}
\end{figure}
\subsubsection{Equilibrium properties} 
Although (M)GAL remains quantitatively reliable across a wide parameter range, it fails in the strong-interaction limit relevant to many experiments. Figure~\ref{fig:2d2lcurr} probes this limit using the parameters of Ref.~\cite{Saldana2018supercurrent}: $U_1 = 7\Delta$, $U_2 = 6\Delta$, $t_d = 1.2\Delta$, $\Gamma = 0.3\Delta$, $\varphi = \pi/2$ and $W=0$. Panel (a) displays the equilibrium Josephson current $J$ versus $\varepsilon_1/U_1$ at fixed $\varepsilon_2 = 0$, whereas panel (b) follows $J$ along $\varepsilon_1/U_1 = \varepsilon_2/U_2$. While MGAL (blue line) locates the singlet-doublet phase boundary, signaled by the sign reversal of $J$, its magnitude deviates markedly from the NRG results. In contrast, the ChE model already improves the agreement at $L = 1$ (dashed orange line) and converges rapidly to NRG as $L$ increases.
\begin{figure}[ht]
\centering
\includegraphics[width=\linewidth]{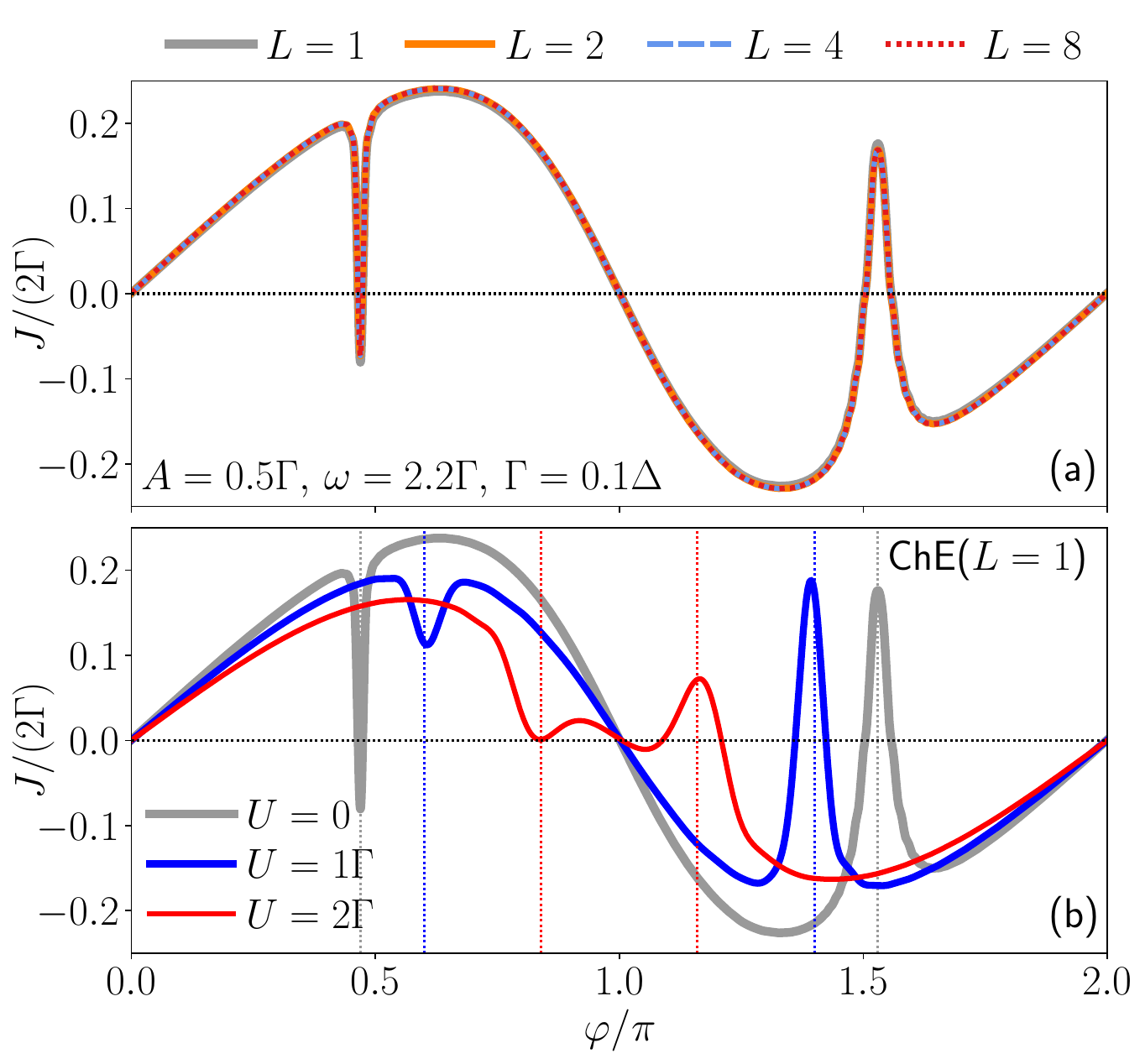}
\caption{Serial DQD between two SC leads as illustrated in Fig.~\ref{fig:bench}(d). Time-averaged Josephson current $J$ as a function of the Josephson phase $\varphi$ (current-phase relation). Parameters are taken from Ref.~\cite{Taberner2023anomalous}, except for $\Delta$, which is set to unity. To allow for an easier comparison with Ref.~\cite{Taberner2023anomalous}, we plot the data in units of $\Gamma$. The microwave drive frequency is set to $\omega = 2.2\Gamma$ with amplitude $A=0.1\Gamma$ and phase shift $\theta=0.5\pi$. (a) Non-interacting case with data obtained with ChE($L$) for $L = 1, 2, 4$, and $8$. (b) Impact of a finite Coulomb interaction $U$. Because the curves are extracted from direct real-time evolution with the initial state being the singlet ground state, only resonances from the singlet sector are visible. Their depth may depend on numerical details such as the sampling frequency.}
    \label{fig:2d2lmicro}
\end{figure} 

\subsubsection{Driven systems:} Figure~\ref{fig:2d2lmicro} revisits the phase-shifted microwave drive analyzed by Ortega-Taberner \textit{et al.}~\cite{Taberner2023anomalous} (cf.\ Fig.~10 therein). The dot levels are modulated as
\begin{equation}
\varepsilon_1(t) = \varepsilon_0 + A\cos(\omega t),\quad
\varepsilon_2(t) = \varepsilon_0 + A\cos(\omega t + \theta),
\end{equation}
but, in contrast to the original study, we retain a finite SC gap and set $\Gamma = 0.1\Delta$. Panel (a) shows the time-averaged Josephson current for $U = 0$ and several values of $L$, demonstrating that finite-$L$ effects are negligible in this regime as the lines lie practically on top of each other. This is because already ChE with $L=1$ gives a good estimate of the in-gap states.  Panel (b) introduces a finite on-site interaction $U$. Vertical lines of the corresponding color mark Josephson phase $\varphi$ where the singlet in-gap states meet integer multiples of the drive frequency, thereby identifying the origin of the observed anomalies. Clearly, these resonances persist even for $U=0.2\Delta$, although their position shifts and their width grow with increasing $U$. Our data were obtained by unitary time evolution from the singlet ground state at $\varepsilon_{1,2}(t) = \varepsilon_0$, therefore we probe only singlet-sector resonances. Coupling the ChE model to an effective metallic bath or using a mixed state at $t=0$, could lift this restriction, but it is left for future work.

\subsection{Double QD coupled to a common lead \label{sec:DQD1}}
\begin{figure}[ht]
    \centering    \includegraphics[width=\linewidth]{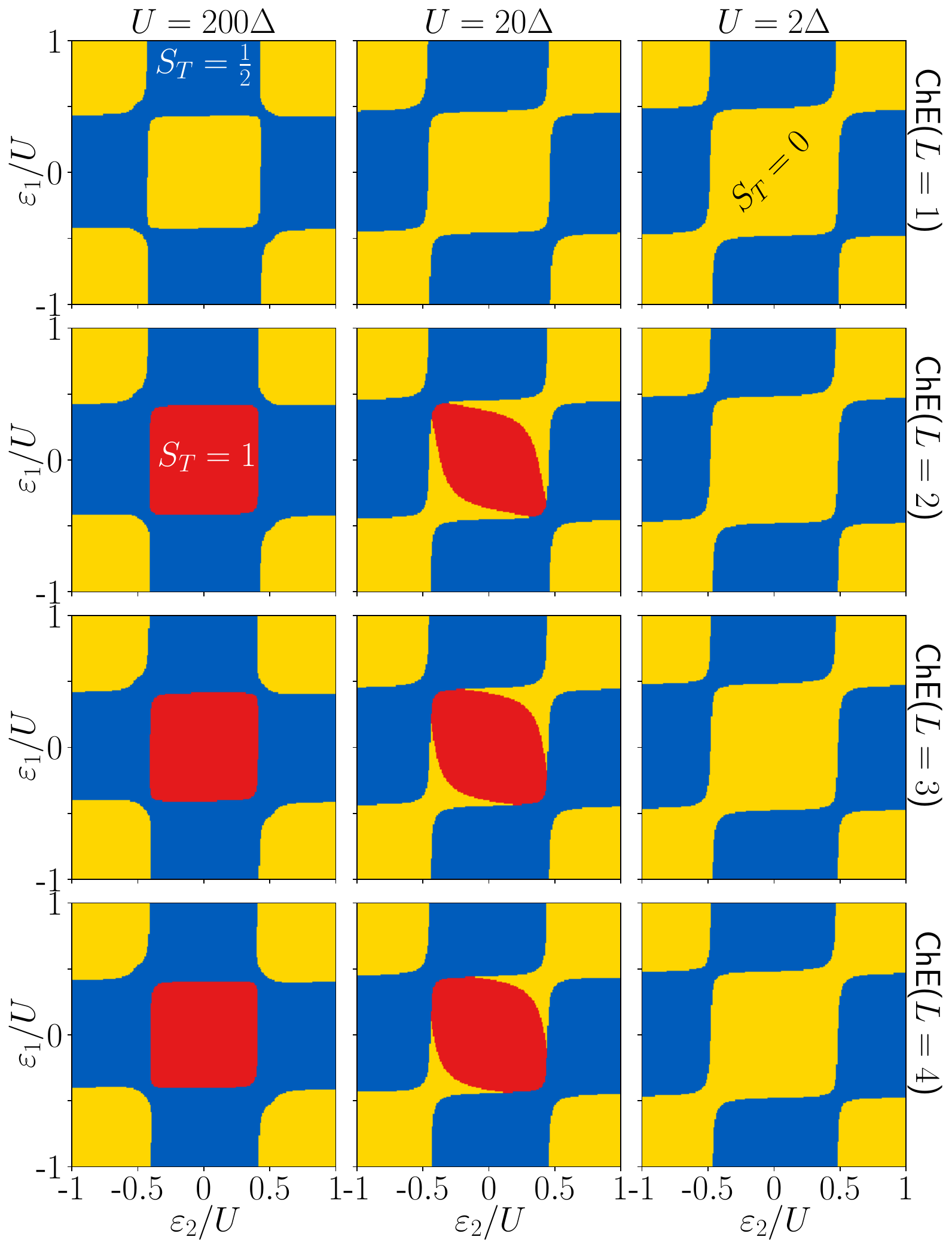}
    \caption{Phase diagrams away from half-filling for parallel DQD coupled to a single lead, illustrated in Fig.~\ref{fig:bench}(e) for $U=200\Delta$ (left column), $U=20\Delta$ (center column), and $U=2\Delta$ (right column). Each color represents a different value of total spin, yellow for $S_T=0$, blue for $S_T=1/2$, and red for $S_T=1$. The model parameters $\Gamma=0.05U$, $t_d=0$ and $W=0$ are taken from Ref.~\cite{zalom2024double}. First line shows results of ChE($L=1$) (ZBW), second of $L=2$ (eZBW), third $L=3$ and forth $L=4$, showing fast convergence.
    \label{fig:2d1lphd}}
\end{figure} 
As a final benchmark, we consider the parallel double-dot geometry sketched in Fig.~\ref{fig:bench}(e). Recent NRG analysis by Zalom \emph{et al.}~\cite{zalom2024double} revealed a parameter window in which the ground state of this device is a triplet. Neither the ZBW model nor (G)AL can capture this regime, as both allow only singlet or doublet ground states. The eZBW construction of Ref.~\cite{zalom2024double} remedies the problem qualitatively. The ChE framework delivers the eZBW parameters directly and, when necessary, enables systematic improvements beyond eZBW. Figure~\ref{fig:2d1lphd} shows the result of ChE results for $L=1,2,3,4$ and parameters of Fig.~2(e), (g), (h) in Ref.~\cite{zalom2024double}.
The colors represent different values of the total spin of the system, 
\begin{equation}
\hat{\bm{S}}^2_T\ket{\psi_\mathrm{GS}}=\left(\sum_{i}\hat{\bm{S}}_i\right)^2\!\!\!\ket{\psi_\mathrm{GS}}=S_T(S_T+1)\ket{\psi_\mathrm{GS}},
\end{equation}
where $i$ runs through all sites (QDs and chain sites) and $\ket{\psi_\mathrm{GS}}$ is the ground state. The yellow areas show singlet ($S_T=0$), blue doublet ($S_T=1/2$), and red triplet ($S_T=1$) ground state. Qualitatively, we observe the same behavior as discussed in Ref.~\cite{zalom2024double}, that is, $L=1$ (ZBW) is too simple to capture the triplet ground state. The extension to $L=2$ (eZBW) is already sufficient to solve the problem. Note that despite the strong interactions ($U=200\Delta$, $20\Delta$, $2\Delta$), ChE reproduces the NRG results with high fidelity. Although each panel in Fig.~\ref{fig:2d1lphd} consists of $221 \times221$ points, the diagrams were produced by ED on a standard PC. Because the parallel-dot configuration remains comparatively unexplored, we return to its ground-state phase diagram in Sec.~\ref{sec:DQD}.

\section{Results: Multiple quantum dots coupled to the same SC lead \label{sec:Results}}
\begin{figure}[ht!]
    \centering
    \includegraphics[width=1.01\linewidth]{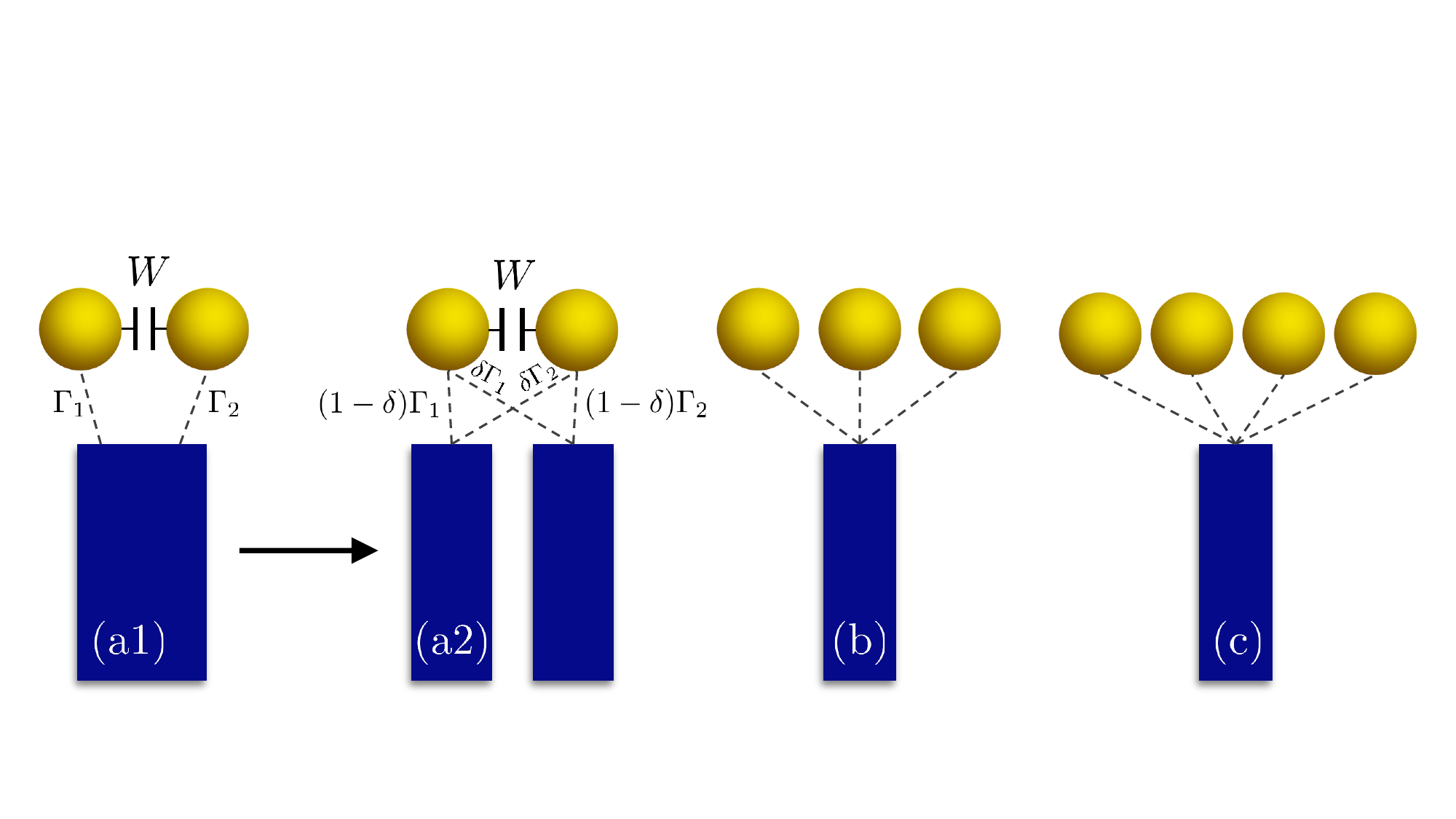}
    \caption{Illustration of the systems investigated in Sec.~\ref{sec:Results}. (a) Double QD coupled to a single SC lead. For $\zeta\neq1$ ($\delta\neq0.5$), this system maps onto two coupled chains. (b) Triple and (c) quadruple QDs with identical dots and couplings in the maximally correlated case $\zeta=1$.
    \label{fig:ilpDQ}}
\end{figure}
We use the ChE($L$) method supported by NRG data to investigate the ground state properties of multiple QDs coupled to the same lead, as illustrated in Fig.~\ref{fig:ilpDQ}.  We will first revisit the DQD from the previous section Sec.~\ref{sec:DQD1}, as it is a rich system with several open problems. Then we will address the ground state of triple and quadruple dots.

\subsection{Double QD\label{sec:DQD}}
As briefly outlined in Sec.~\ref{sec:DQD1} and examined in detail in Ref.~\cite{zalom2024double}, a DQD in the parallel configuration with $\zeta = 1$ already falls outside the reach of the most commonly employed effective descriptions, such as the ZBW and (G)AL. Extending the chain to larger $L$ resolves this limitation, yet, as we show below, short chains may still be inadequate for quantitative work in parts of the regions where $L=1$ fails completely. To streamline the analysis, we first neglect both the direct inter-dot hopping $t_d$ and the capacitance $W$ and focus on the half-filled case.
\begin{figure}[t!]
    \centering
    \includegraphics[width=1\linewidth]{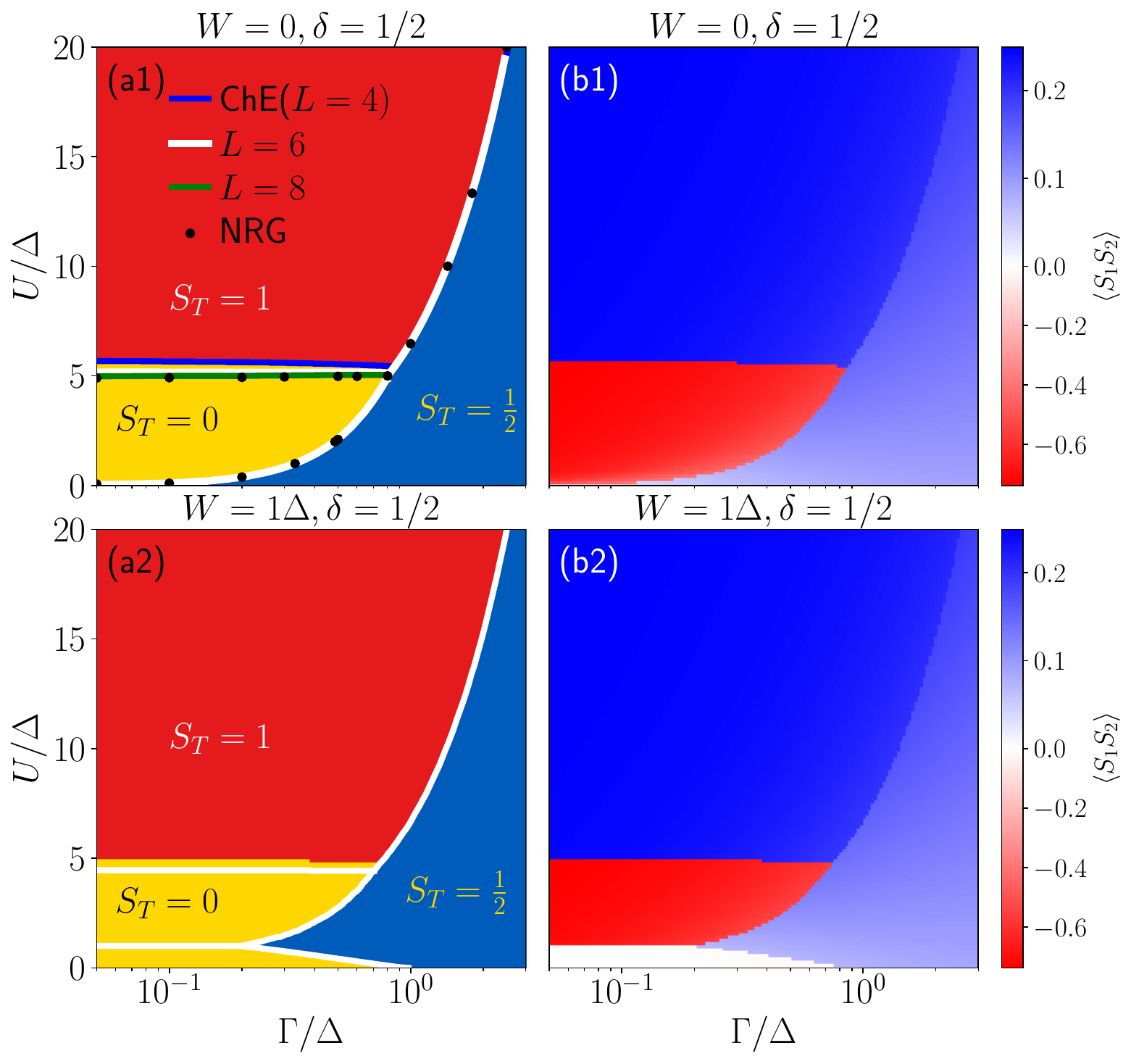}
    \caption{Phase diagrams for the DQD system sketched in Fig.~\ref{fig:ilpDQ}(a1). 
    (a1) Total spin $S_T$ in the $\Gamma-U$ plane for symmetric dots, $\Gamma_1=\Gamma_2\equiv\Gamma$ and $U_1=U_2\equiv U$, at half-filling, $W=0$, and in the maximally correlated limit $\zeta = 1$ ($\delta = 0.5$) calculated with ChE($L=4$). 
    (b1) Corresponding average spin correlation $\langle \hat{\mathbf{S}}_1 \!\cdot\! \hat{\mathbf{S}}_2 \rangle$. (a2) Same as (a1) but with a finite inter-dot capacitive coupling $W = \Delta$. 
    (b2) Mean spin correlation for the parameters of (a2). Phase boundaries (solid lines) computed with ChE($L=4,6,8$) and critical points (black circles) obtained with NRG were determined from the positions of the QPT in the in-gap spectrum. Maps have been obtained with ChE($L=4$).
    \label{fig:2dp1l_phd}
    }
\end{figure}

Figure~\ref{fig:2dp1l_phd}(a1) displays a color map of the total spin $S_T$ in the $\Gamma-U$ plane, obtained with ChE($L=4$) at half-filling. Yellow, blue, and red regions correspond to singlet, doublet, and triplet ground states, respectively. Panel (a2) shows the expectation value of spin correlations $\langle \hat{\mathbf{S}}_1 \!\cdot\! \hat{\mathbf{S}}_2 \rangle$, i.e., effective exchange, revealing that the singlet region is stabilized by an effective antiferromagnetic exchange, while the triplet region exhibits a ferromagnetic one. As noted previously in different contexts~\cite{plorin2024kondo,Li2025individual}, the doublet state likewise entails an effective ferromagnetic coupling, whose strength grows with both $\Gamma$ and $U$.

The colored lines in Fig.~\ref{fig:2dp1l_phd}(a1) mark the phase boundaries obtained with ChE($L$) for several chain lengths, and black circles denote the benchmark NRG results. The singlet-doublet and triplet-doublet boundaries converge rapidly with $L$, but the singlet-triplet line shifts appreciably toward lower $U$ as $L$ increases.
   
\begin{figure}[t!]
    \centering
   \includegraphics[width=1\linewidth]{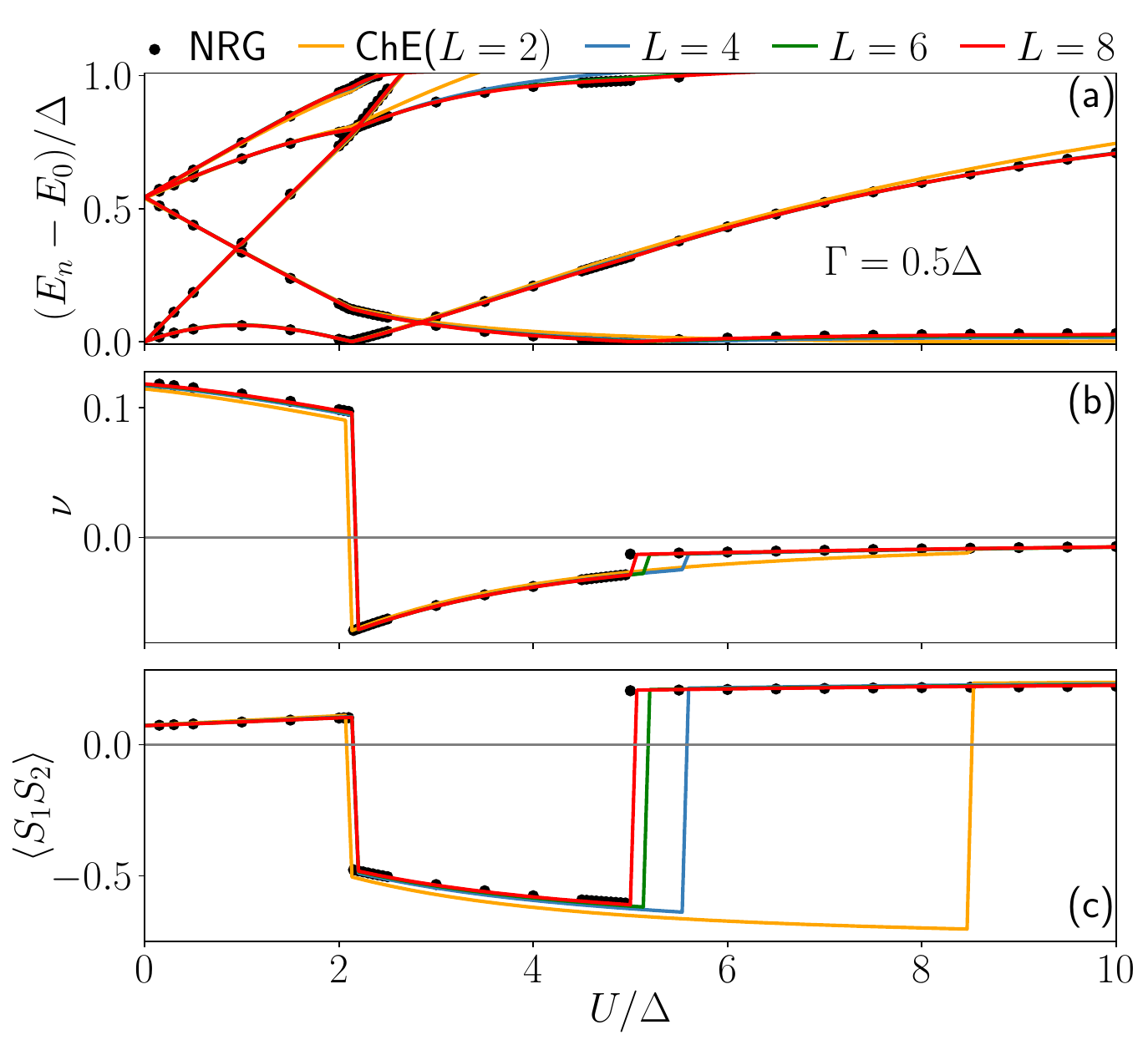}
    \caption{Parallel DQD coupled to a single lead illustrated in Fig.~\ref{fig:ilpDQ}(a1). (a) In-gap excitation energies, (b) induced pairing $\nu = \langle d^\dagger_{j\uparrow} d^\dagger_{j\downarrow} \rangle$, and (c) spin correlations as functions of interaction strength $U$ for $\Gamma = 0.5\Delta$ and $\zeta = 1$. Black circles denote NRG results, solid lines represent the corresponding ChE($L$) predictions.
    }
    \label{fig:2dp1l_Ucut}
\end{figure}

This slow convergence is highlighted in Fig.~\ref{fig:2dp1l_Ucut}: panel (a) shows the in-gap excitation energies, panel (b) the induced pairing on each dot, and panel (c) the effective inter-dot exchange, all evaluated at fixed $\Gamma = 0.5\Delta$. 
At this coupling rate, the system first undergoes the doublet-singlet transition, followed at larger $U$ by the singlet-triplet transition. 
Despite the intricate in-gap spectrum, short chains already reproduce nearly all NRG levels; the lone exception is the first excited state in the large-$U$ limit. 
For ChE($L=1$) (ZBW), this triplet state is missed entirely, as discussed in Sec.~\ref{sec:DQD1} and Ref.~\cite{zalom2024double}. 
Extending the chain to ChE($L=2$) removes that qualitative failure but still overestimates the critical $U$ of the singlet-triplet transition. 
Only when $L$ is sufficiently large, i.e., $L=8$, does the position of the QPT converge to the NRG benchmark, reflecting the small energy gap between the triplet ground state and the singlet first excited state at strong interactions.

\begin{figure}[ht]
    \centering
    \includegraphics[width=1\linewidth]{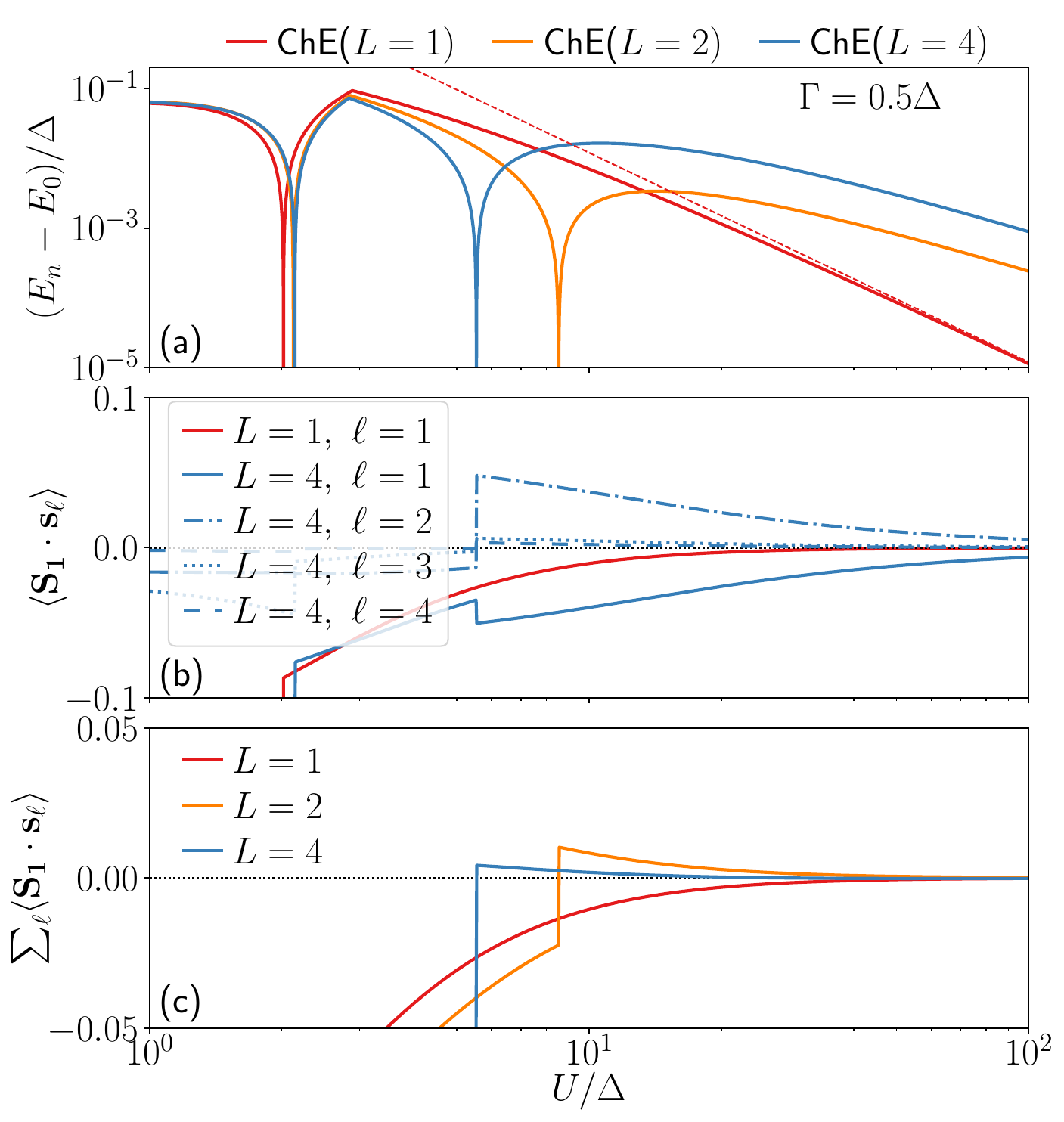}
    \caption{Parallel DQD coupled to a single lead illustrated in Fig.~\ref{fig:ilpDQ}(a1). 
    (a) First exited in-gap energy at $\Gamma=0.5\Delta$ for ChE($L=1,2,4$). 
    (b) Spin-spin correlations between the first QD and the $\ell$th lead site. 
    (c) Aggregated spin-spin correlations, showing that for $L>1$ the sum flips to positive values where the triplet emerges.  
    }
    \label{fig:2dp1l_Ucut_SS}
\end{figure}

Figure~\ref{fig:2dp1l_Ucut_SS}(a) plots the first-excited-state energy for several chain lengths $L$ on a log-log scale. For $L=1$, small $\Gamma$ and large $U$, an analytic expansion at half-filling shows that the lowest excitation is a triplet lying $\Delta E = 3\times16\Gamma^{2}/U^{3}$ above the singlet ground state (dashed red line). This positive splitting, generated by fourth-order tunneling processes in $\gamma$, yields an antiferromagnetic superexchange between the dots~\cite{Bacsi2023Exchange}. Removing the gap requires sixth-order virtual tunneling processes in $\widetilde{h}_{\ell}$ that appear in the ChE basis only for chains of length $L\ge2$.

The same trend is evident in the spin-spin correlators $\langle \hat{\mathbf{S}}_{j}\cdot\hat{\mathbf{s}}_{\ell}\rangle$ between a dot and chain site $\ell$ in Fig.~\ref{fig:2dp1l_Ucut_SS}(b). 
In the large-$U$ limit, the sum over $\ell$ approximates the effective Kondo exchange. For $L = 1$ the single term is strictly antiferromagnetic, so a triplet ground state cannot form~\cite{Zitko2006Multiple}. 
When $L > 1$, the sites with $\ell > 1$ can add ferromagnetic contributions; at large $U$ these are already strong enough for $L = 2$ to favor the triplet, and longer chains shift the crossover to smaller $U$, converging on the true critical value. As we will discuss in the following sections, this limitation of ZBW is rather general, as it also fails to predict the correct ground state for triple and quadruple QDs coupled to a common SC lead.

\begin{figure}[ht]
    \centering
   \includegraphics[width=1\linewidth]{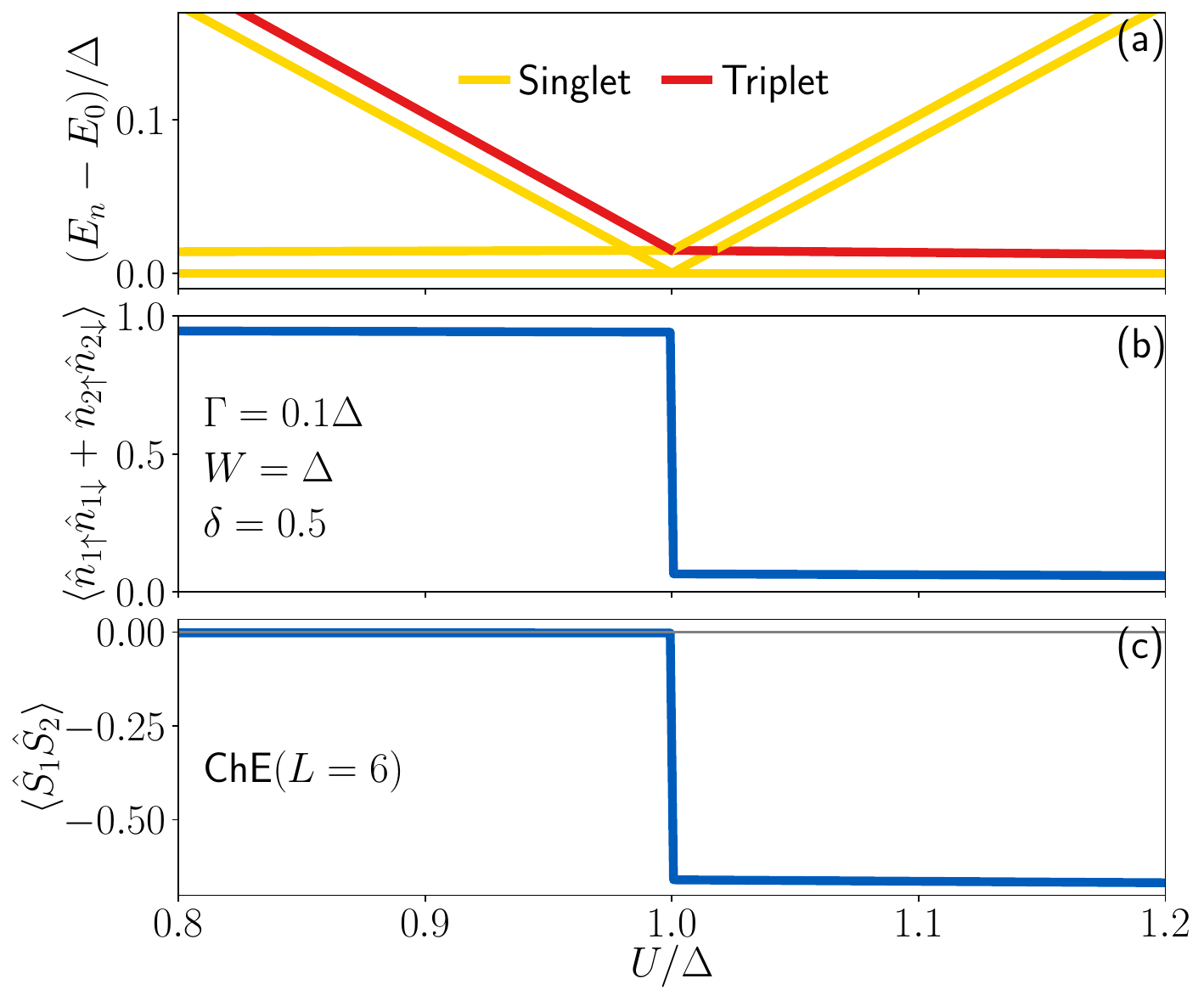}
    \caption{Evolution of low-energy properties across the singlet-singlet QPT in the DQD illustrated in Fig.~\ref{fig:ilpDQ}(a1), obtained with ChE($L=6$). (a) In-gap excitation energies, (b) double occupancy $\sum_j\langle n_{j\uparrow} n_{j\downarrow} \rangle$, and (c) inter-dot spin correlations, plotted as functions of $U$ at $\Gamma = 0.1\Delta$, $\zeta = 1$ ($\delta = 0.5$), and $W = \Delta$. Panel (a) shows the level crossing where the two singlet branches exchange order at the QPT. For $U < W$, the ground state is a singlet followed closely by the other singlet, whereas for $U > W$ the ground-state singlet is followed by a triplet. The singlet-triplet splitting is proportional to $\Gamma$.
}
    \label{fig:2dp1l_Ucut_W1}
\end{figure}

The phase diagram becomes even richer once the inter-dot capacitance is taken into account. Panels (a2) and (b2) of Fig.~\ref{fig:2dp1l_phd} show the same device with a finite $W = \Delta$. Under these conditions, the singlet sector splits into two distinct phases. In the admittedly non-physical limit $U < W$, the ground state is a charge singlet devoid of magnetic correlations. The detailed scan in Fig.~\ref{fig:2dp1l_Ucut_W1} confirms that the two singlets are separated by a QPT at $W = U$. For $W > U$, the system favors a charge-ordered singlet in which one dot is predominantly doubly occupied while the other remains empty, giving rise to pronounced double occupancy [panel (b)] and vanishing spin correlation [panel (c)]. We note that a singlet-singlet QPT with similar properties was also predicted in serial DQDs~\cite{Zitko2015numerical}.

\begin{figure}[ht]
    \centering
   \includegraphics[width=1\linewidth]{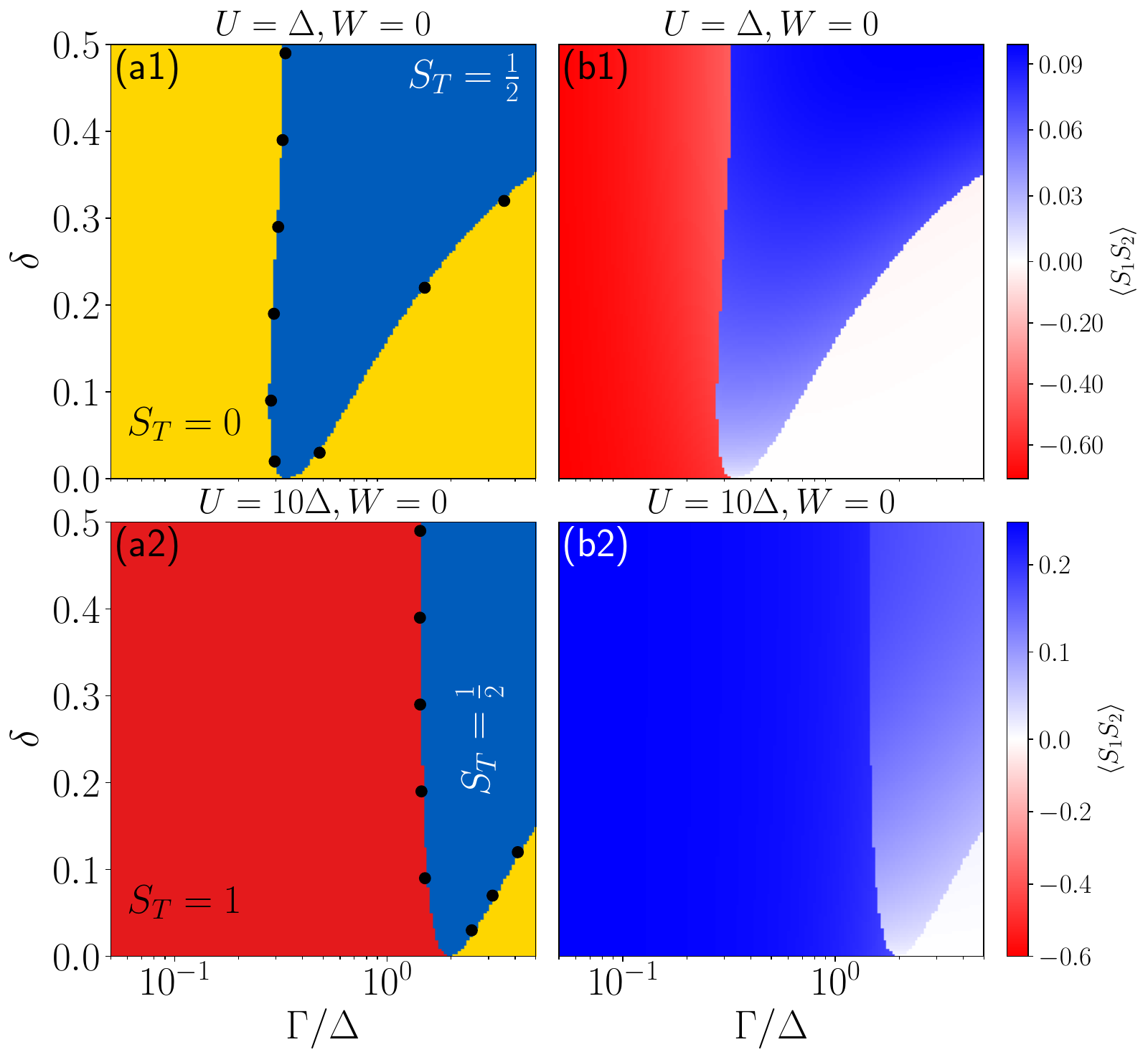}
    \caption{Phase diagrams for the DQD coupled to two SC chains, as illustrated in Fig.~\ref{fig:ilpDQ}(a2), evaluated by CheE($L=2$) and NRG (black circles) away from the maximally correlated point $\zeta = 1$, where two chains are needed. (a1) Total spin $S_T$ in the $\Gamma-\delta$ plane for symmetric dots, $\Gamma_1 = \Gamma_2 \equiv \Gamma$ and $U_1 = U_2 \equiv U = \Delta$, at half-filling with $W = 0$.  (b1) Corresponding average spin correlation $\langle \mathbf{S}_1 \!\cdot\! \mathbf{S}_2 \rangle$. (a2) Same as (a1) but for strong Coulomb interaction $U = 10\Delta$. (b2) Mean spin correlation for the parameters of (a2).
}
    \label{fig:2dp1l_phd_zeta}
\end{figure}

Two distinct singlet ground states arise at half-filling even at $W = 0$ once the system is detuned from the special point $\zeta = 1$ ($\delta = 0.5$); for $\delta \neq 0.5$ two chains must be retained [see the illustration in Fig.~\ref{fig:ilpDQ}(a2)]. Figure~\ref{fig:2dp1l_phd_zeta} shows examples of resulting phase diagrams. 
Panels (a1) and (a2) display the total spin in the $\Gamma-\delta$ plane for $U = \Delta$ and $U = 10\Delta$, respectively, while panels (b1) and (b2) show the corresponding effective exchange. 
The origin of the two singlet regions in (a1) and (a2) is most transparent in the $\delta \rightarrow 0$ limit, where the geometry crosses over to a serial configuration. In this limit, each dot behaves almost independently; its single-dot (i.e., at each half separately) ground state can be either a singlet or a doublet with the phase diagram shown in Fig.~\ref{fig:1d1lChe2}. The two-dot device can therefore form two nonequivalent composite ground states built from the tensor products $\text{singlet}\otimes\text{singlet}$ and $\text{doublet}\otimes\text{doublet}$~\cite{Zonda2023generalized}.

The $\text{singlet}\otimes\text{singlet}$ combination leads to a singlet composite ground state. However, the $\text{doublet}\otimes\text{doublet}$ gives, without any correlations between the dots, a singlet-triplet degenerate ground state. An infinitesimal $\delta$ will lift the degeneracy by promoting singlet at weak $U$ [Fig.~\ref{fig:2dp1l_phd_zeta} (a1)], or triplet at strong $U$ [Fig.~\ref{fig:2dp1l_phd_zeta}(a2)]. 

Note that the situation where both the ground state and first excited states are singlets is potentially interesting with respect to the recent proposal of YSR qubit~\cite{Steffensen2025YSR}. Similarly, the stability of the triplet state down to $\delta\approx 0$ leads to interesting conclusions. The two-chain setup discussed above can also be viewed as a serial DQD device [Fig.~\ref{fig:bench}] if $\delta$ is small. In this interpretation, direct inter-dot hopping is neglected ($t_d \approx 0$), while (small) cross couplings exist between the first lead and the second dot, and vice versa. 

\begin{figure}[ht]
    \centering
   \includegraphics[width=1\linewidth]{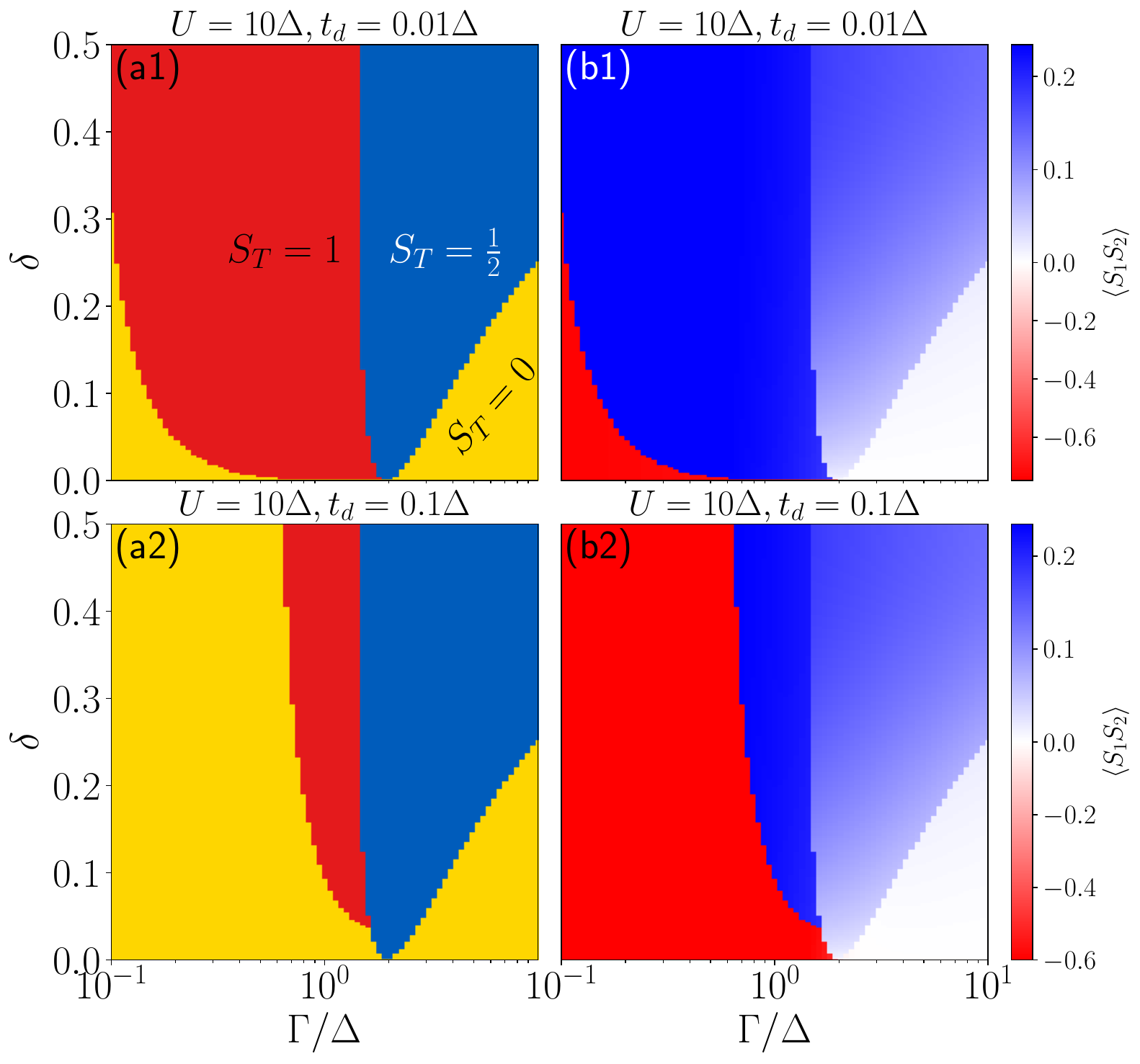}
    \caption{Phase diagrams for the DQD coupled to two SC leads, as illustrated in Fig.~\ref{fig:ilpDQ}(a2), evaluated by ChE($L=2$) away from the maximally correlated case $\zeta = 1$, where two chains are needed. 
    (a1) Total spin $S_T$ in the $\Gamma-\delta$ plane for symmetric dots, $\Gamma_1 = \Gamma_2 \equiv \Gamma$ and $U_1 = U_2 \equiv U = 10\Delta$, at half-filling with $t_d = 0.01\Delta$ and $W=0$. 
    (b1) Corresponding average spin correlation $\langle \mathbf{S}_1 \!\cdot\! \mathbf{S}_2 \rangle$. 
    Panel (a2) is the same as (a1), but for $t_d = 0.1\Delta$. 
    (b2) Mean spin correlations for the same parameters as in (a2).
}
    \label{fig:2dp2l_td}
\end{figure}

For a purely serial geometry ($\delta=0$) with identical dots ($U_1 = U_2$, $\varepsilon_1 = \varepsilon_2$, $\Gamma_1 = \Gamma_2$), the ground state is always a singlet because any finite $t_d$ splits the singlet-triplet pair by an exchange energy $\sim 4t_d^2/U$ \cite{Zitko2015numerical,Zonda2023generalized}. 
If, however, the cross couplings to the leads are allowed and $t_d$ stays small, a triplet ground state can emerge, as illustrated in Fig.~\ref{fig:2dp2l_td}. The finite $t_d$ clearly suppresses the triplet phase, which is most stable near the triplet-doublet boundary, where a larger effective $\Gamma$ helps to lock it in. It is important to note here that the half-filling is satisfied only approximately because the bipartite lattice condition is broken when both $\delta$ and $t_d$ are finite.

\subsection{Triple and Quadruple QD \label{sec:TQQD}}
\begin{figure}[ht]
    \centering
   \includegraphics[width=1\linewidth]{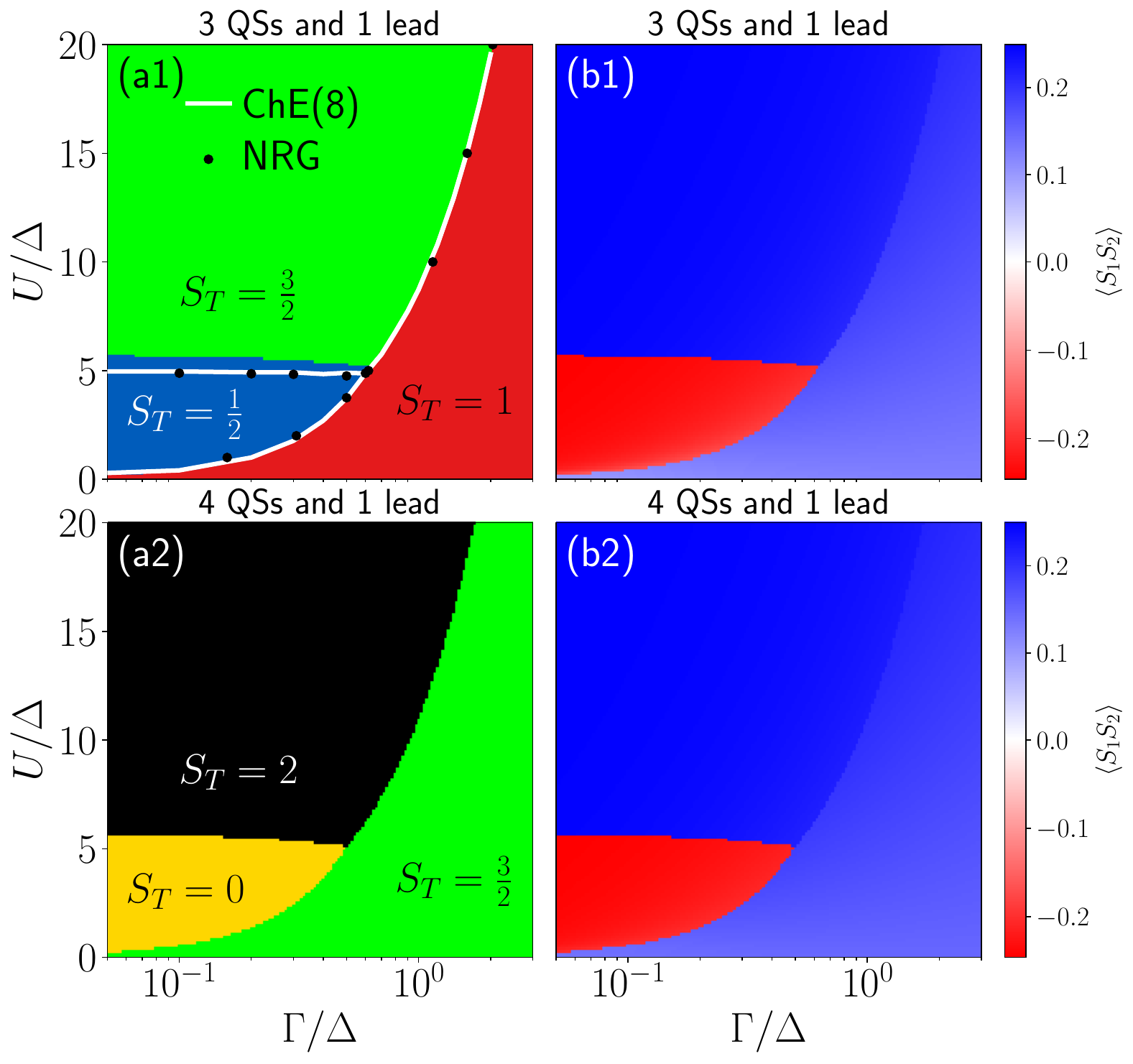}
    \caption{Phase diagrams for the triple (a) and quadruple (b) QD system with single SC lead as illustrated in Fig.~\ref{fig:ilpDQ}(b) and (c). 
    (a1) Total spin $S_T$ in the $\Gamma-U$ plane for three symmetric dots at half-filling, $W=0$, $t_d=0$, and in the maximally correlated limit $\zeta = 1$. (b1) Corresponding average spin correlation $\langle \hat{\mathbf{S}}_1 \!\cdot\! \hat{\mathbf{S}}_2 \rangle$. (a2) Same as (a1) but for four dots coupled to the same chain. 
    (b2) Mean spin correlation for the same parameters as in (a2). Phase boundaries (solid lines) computed with ChE($L$) and critical points (black circles) obtained with NRG were determined from the positions of the QPT in the in-gap spectrum. Maps have been obtained with ChE($L=4$) and white boundaries in (a1) with ChE($L=8$).
    \label{fig:3dp1l_phd}}
\end{figure}

Systems comprising three or more QDs have garnered growing interest, fueled by recent advances in the fabrication of ordered assemblies of magnetic atoms or molecules on SC surfaces~\cite{Korber-2018,Rutten2024-FeTrimers,Li2025individual,Li2025negative} as they represent tunable testbeds for studying various emerging phenomena. As an example, quadruple QDs are being explored for the interplay between Nagaoka ferromagnetism and the SC order~\cite{Siuda2025}. Here, we offer what may be viewed as an entry-level systematic exploration of such complex systems. For clarity, we restrict ourselves to the fully local $\zeta = 1$ limit, i.e., each dot is coupled to a single SC chain in both the ChE and NRG treatments. The resulting phase diagrams in the $\Gamma-U$ plane at half-filling, shown in Fig.~\ref{fig:3dp1l_phd}, resemble that of the DQD case both by the shape and number of ground state phases. Yet the individual phases differ markedly. 
For triple QD in panel (a1), the dominant regions are a triplet ground state (red) that takes place above critical $\Gamma$ and a quartet state ($S_T = 3/2$, green) stable above critical $U$, with a doublet phase (blue) wedged between them in small $\Gamma$ and weak to moderate $U$.
For quadruple QD in panel (a2), the respective dominant regions are a quartet (green) ground state above critical $\Gamma$ and a quintet state ($S_T = 2$, black) above critical $U$, and the third phase is a singlet (yellow). Still, the magnetic diagrams in panels (b1) and (b2), are equivalent.  

\begin{figure}[ht!]
    \centering
   \includegraphics[width=1\linewidth]{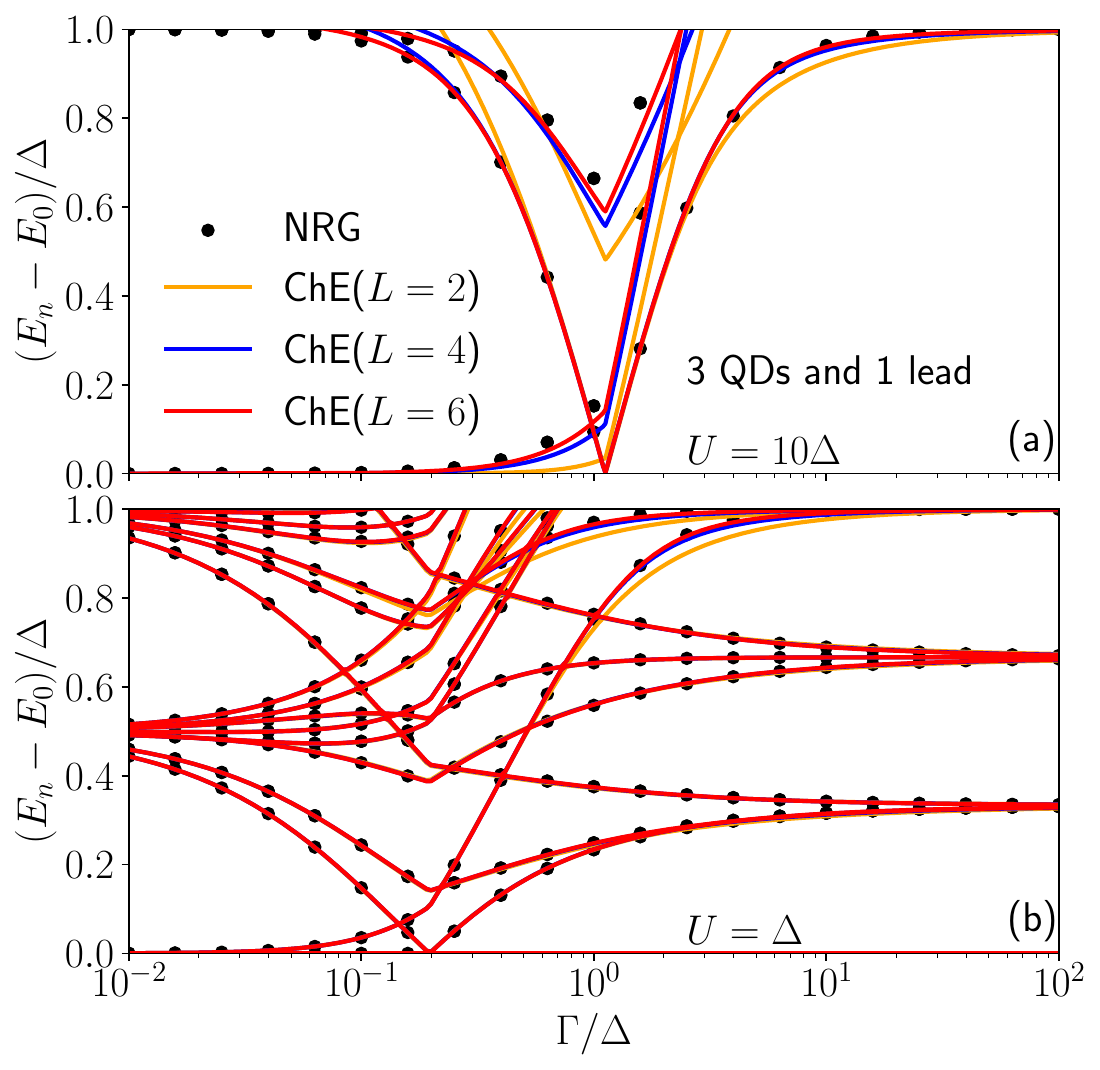}
    \caption{In-gap excitation energies for a triple QD coupled to a single SC lead as illustrated in Fig.~\ref{fig:ilpDQ}(b) at $U=10\Delta$ (a) and $U=\Delta$ (b) calculated with NRG (black circles) and ChE($L$) (solid lines) for different $L$.
}
    \label{fig:3dp1l_Gcut}
\end{figure}

Another common feature of the double-, triple-, and quadruple-dot systems is the stability of the phase boundary, which extends from $U = 0$, $\Gamma = 0$ to large values of $U$ and $\Gamma$. This boundary remains virtually unchanged as the chain length $L$ is increased.
Moreover, despite the intricate in-gap spectrum, a ChE description with only a few chain sites accurately reproduces the NRG excitation energies over several orders of magnitude in $\Gamma$, as demonstrated in Fig.~\ref{fig:3dp1l_Gcut}.  
For weak interaction, $U = \Delta$, two sites ($L = 2$) already suffice, and for $L \ge 4$ the ChE curves lie practically on top of the NRG results. The strong-coupling limit, $U = 10\Delta$, is qualitatively simpler, but attaining the same accuracy as NRG demands a longer chain.

\begin{figure}[ht!]
    \centering
   \includegraphics[width=1\linewidth]{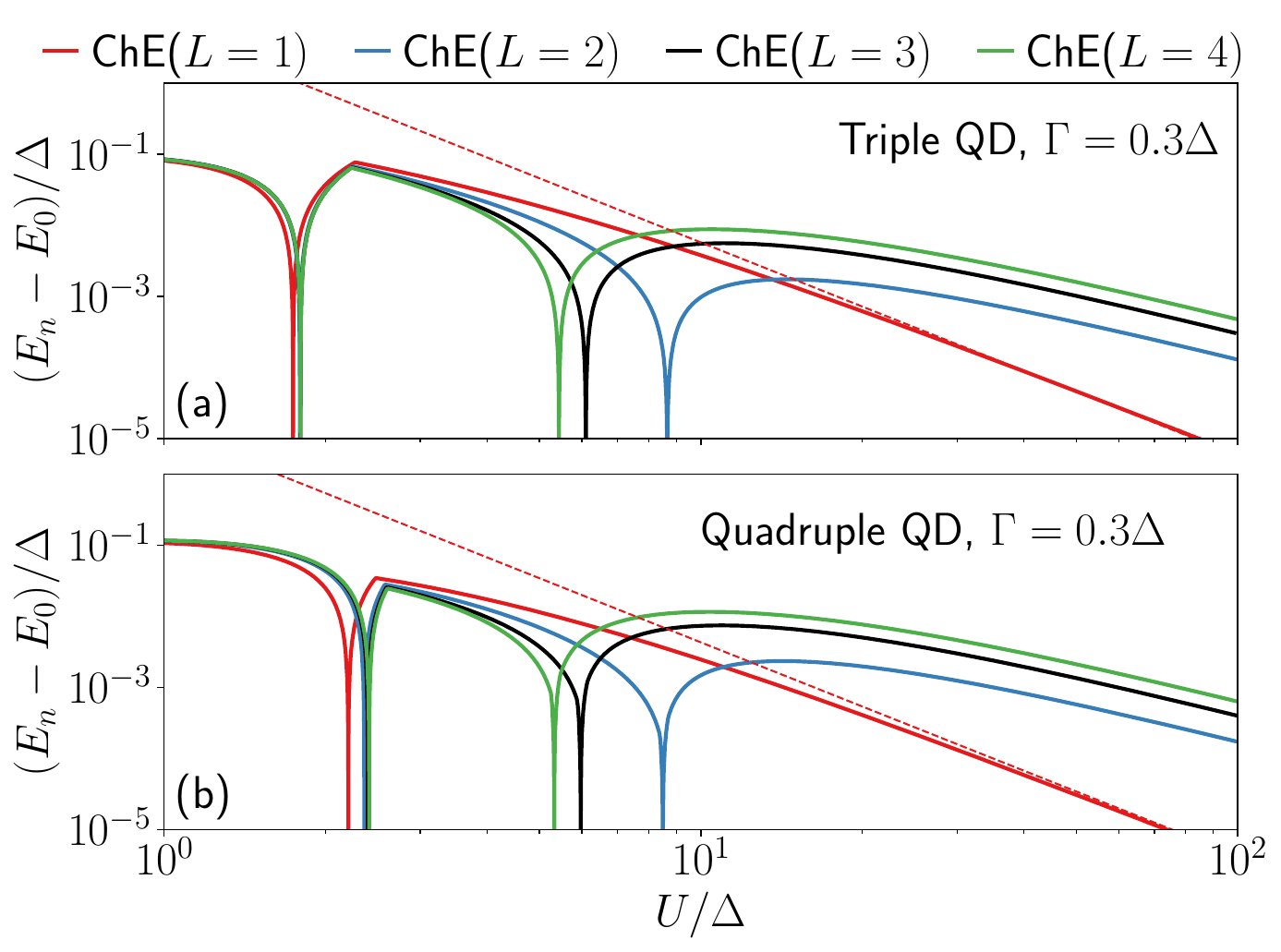}
    \caption{(a) First exited in-gap energy for $\Gamma=0.3\Delta$ for ChE($L=1,2,3,4$) for triple QD coupled to a single SC lead, illustrated in Fig.~\ref{fig:ilpDQ}(b). 
    (b) First exited in-gap energy for $\Gamma=0.3\Delta$ for ChE($L=1,2,3, 4$) for quadruple QD coupled to a single SC lead, illustrated in Fig.~\ref{fig:ilpDQ}(c). The dashed red lines show the strong $U$ and weak $\Gamma$ asymptotics described by $\Delta E = b\times16\Gamma^{2}/U^{3}$ with $b = 4$ for the triple QD and $b = 3$ for the quadruple QD.
}
    \label{fig:34dp1l_Ucut_SS}
\end{figure}
Finally, as in the DQD case, the ZBW approximation fails to capture the correct ground state in the small-$\Gamma$, large-$U$ regime, regardless of the specific multiplet involved. 
In this limit, the first excited state obtained with ChE($L = 1$) follows the same superexchange scaling, $\Delta E = b \times 16\Gamma^{2}/U^{3}$, with $b = 4$ for the triple dot (doublet-quartet transition) and $b = 3$ for the quadruple dot (singlet-quintet transition), as fitted to the numerical data in Fig.~\ref{fig:34dp1l_Ucut_SS}. 
This once again confirms that even for more complex systems, ZBW lacks the sixth-order virtual hopping processes in the lead needed to close the energy gap between the weak and strong $U$ ground state phases and correctly reproduce the phase structure. This explains its consistent failure across all three geometries.

Another clear feature is the pronounced shift in the critical value of $U$ that marks the transition from doublet to quartet in the triple dot, Fig.~\ref{fig:34dp1l_Ucut_SS}(a), and from singlet to quintet in the quadruple dot, Fig.~\ref{fig:34dp1l_Ucut_SS}(b), as the chain length $L$ increases. This is evident from the position of the sharp minima at larger values of $U$ in the difference between the first excited state and the ground state, corresponding to the crossing of the lowest eigen-energies. These minima shift significantly between $L = 2$ (blue), $L = 3$ (black) and $L = 4$ (green) in Fig.~\ref{fig:34dp1l_Ucut_SS}. In contrast, the minima at smaller values of $U$ remain stable for $L \ge 2$, indicating that the corresponding phase boundary has already converged.
\begin{figure}[ht!]
    \centering
    \includegraphics[width=\linewidth]{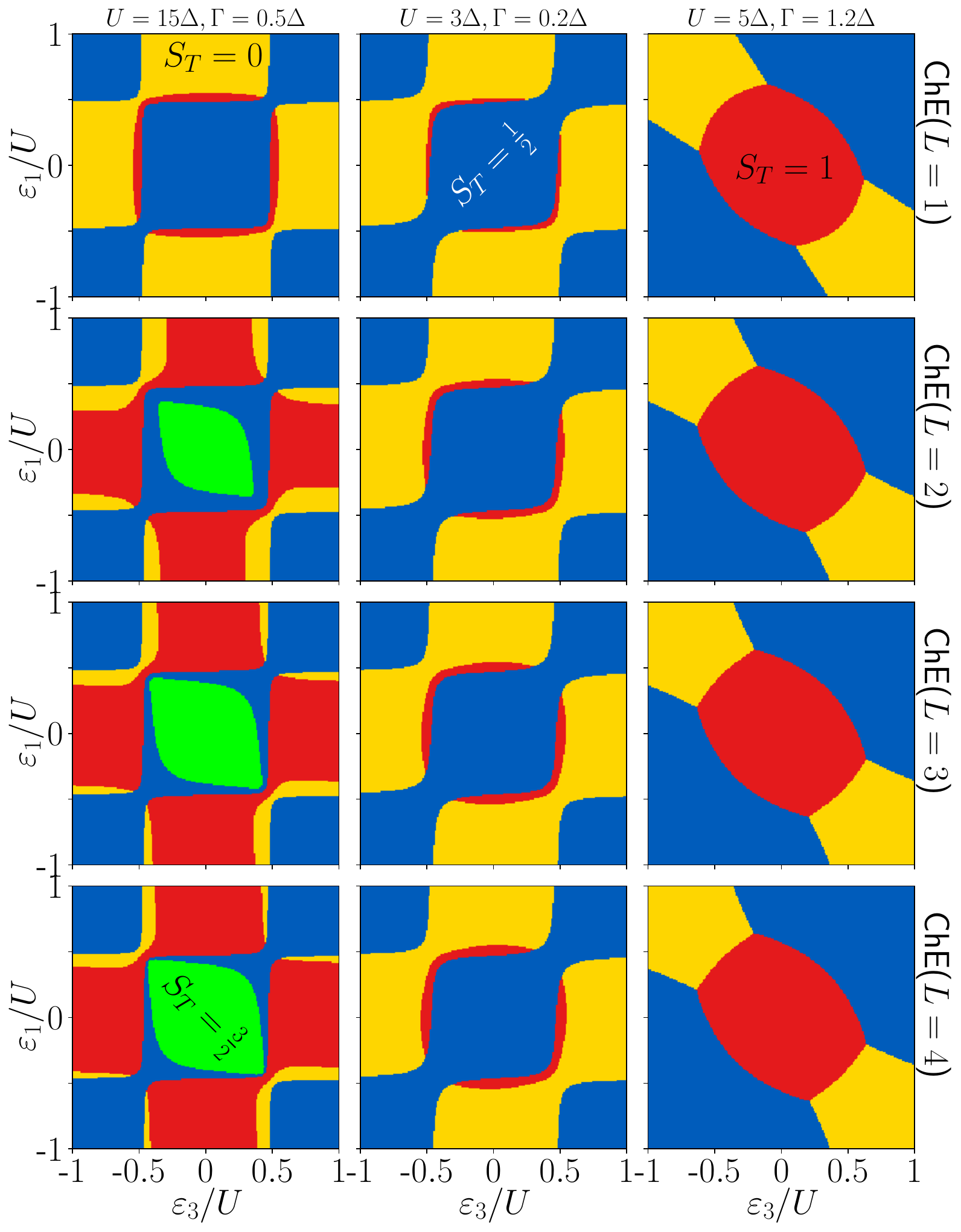}
    \caption{Phase diagrams for a triple QD coupled to a single SC lead illustrated in Fig.~\ref{fig:ilpDQ}(b), evaluated away from half-filling.  We fix $\varepsilon_2 = 0$ and vary $\varepsilon_1$ and $\varepsilon_3$.  Colors encode the total spin: $S_T = 0$ (yellow), $S_T = 1/2$ (blue), $S_T = 1$ (red) and $S_T = 3/2$ (green). The three columns correspond to different parameter sets: 
    (i) $U = 15\Delta$, $\Gamma = 0.5\Delta$;  
    (ii) $U = 3\Delta$, $\Gamma = 0.2\Delta$; and  
    (iii) $U = 5\Delta$, $\Gamma = 1.2\Delta$,  
    all with $t_d = 0$ and $W = 0$.  These choices place the half-filled point in each of the three ground-state phases identified in Fig.~\ref{fig:3dp1l_phd}(a). Rows display ChE results for increasing chain length: top, $L = 1$; middle, $L = 2$, and $L=3$; bottom, $L = 4$.}
    \label{fig:34d1laway}
\end{figure} 

The inability of the ZBW approximation to capture the correct ground state becomes even more pronounced away from half-filling. In this regime, the phase diagrams become significantly more complex. For a triple dot, this is illustrated in Fig.~\ref{fig:34d1laway}, where we fix $\varepsilon_2 = 0$ and vary $\varepsilon_1$ and $\varepsilon_3$. A noteworthy observation is that a singlet phase (yellow) -- absent at half-filling -- occupies a substantial portion of the diagram once particle-hole symmetry is broken.

For moderate interactions, such as $U = 3\Delta$, $\Gamma = 0.2\Delta$, and $U = 5\Delta$, $\Gamma = 1.2\Delta$, the ZBW phase boundaries agree well with those obtained from longer chains. In these regimes, ZBW proves reliable: the differences between $L = 1$ and $L = 2$ are minor, and larger values of $L$ introduce only negligible corrections to the $L=2$ case. However, in the strong-coupling limit $U = 15\Delta$, where the ground state at half filling is a quartet, the ZBW approximation fails even qualitatively. It completely misses the central $S_T = 3/2$ region (green) and significantly underestimates the extent of the surrounding triplet phase (red). While $L = 2$ provides qualitatively correct predictions, longer chains are necessary to accurately reproduce the full phase diagram.

\begin{figure}[ht!]
    \centering
   \includegraphics[width=\linewidth]{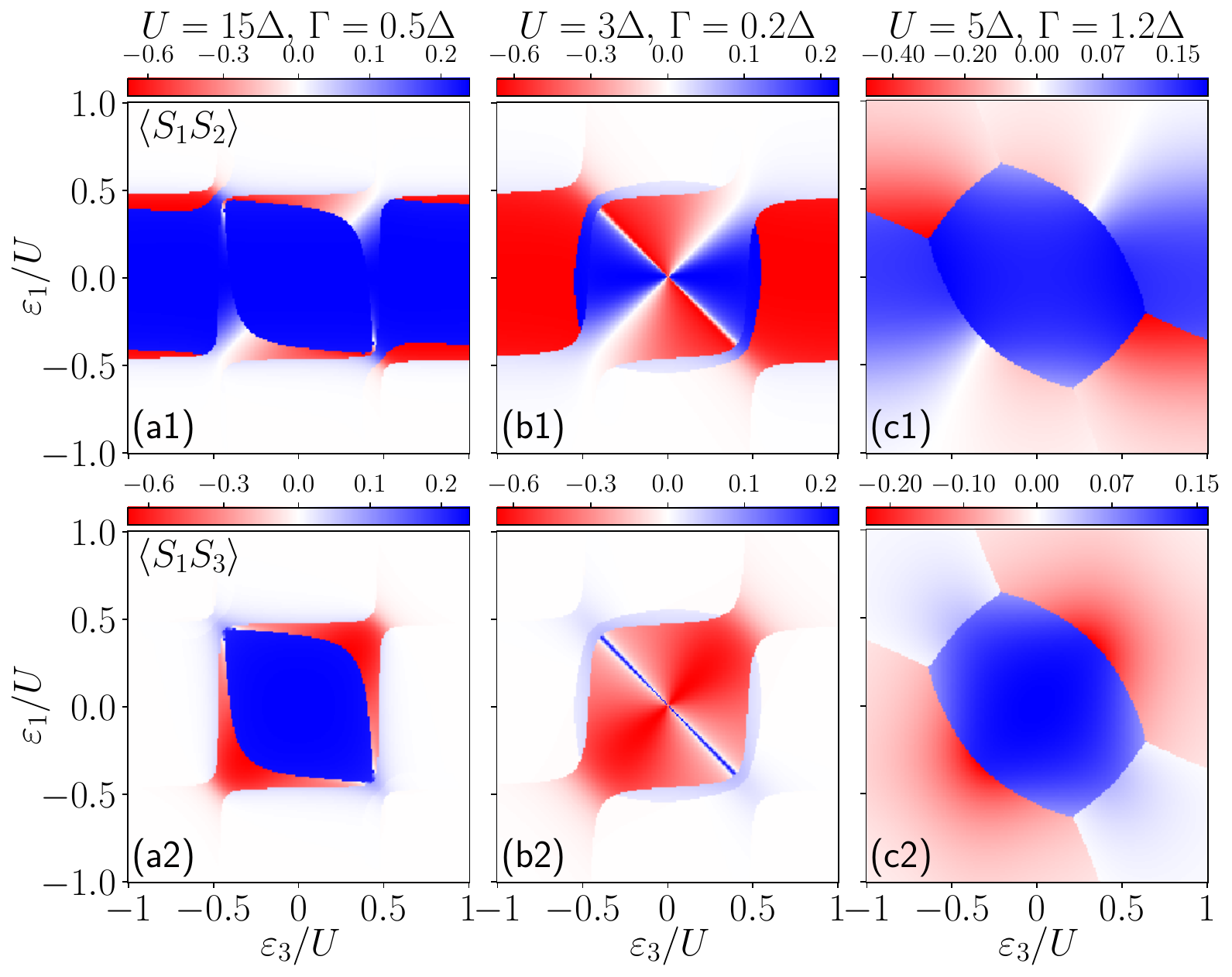}
    \caption{Triple QD coupled to a single SC lead illustrated in Fig.~\ref{fig:ilpDQ}(b). Maps of spin correlations  $\langle \bm{S}_1\!\cdot\!\bm{S}_2\rangle$ (first line) and $\langle\bm{S}_1\!\cdot\!\bm{S}_3\rangle$ (second line) for the same sets of parameters as phase diagrams in Fig.~\ref{fig:34d1laway}. Maps have been obtained using ChE($L=4$).
}
    \label{fig:3d1laway_mag}
\end{figure}
\subsubsection{Spin-spin correlations in triple QD}
The structure of the ground state diagram is, in fact, richer than suggested by the total spin alone.
Figure~\ref{fig:3d1laway_mag} shows spin-spin correlations, i.e., the effective magnetic exchange between the dots in the same setup as in Fig.~\ref{fig:34d1laway}. Specifically, it displays $\langle \bm{S}_1 \!\cdot\! \bm{S}_2 \rangle$ ($\langle \bm{S}_3 \!\cdot\! \bm{S}_2 \rangle$ is analogous) and $\langle \bm{S}_1 \!\cdot\! \bm{S}_3 \rangle$. These results reveal that identical total-spin ground states can exhibit very different magnetic correlations, not only between different pairs of dots (compare first and second line) but also for the same pair in different regions of the diagram.

A particularly interesting case occurs in the doublet ground state for $U = 3\Delta$ and $\Gamma = 0.2\Delta$. In this regime, spin correlations can vanish entirely in some regions of the diagram (e.g., the upper-left and lower-right corners), while in the central region -- near half-filling -- they exhibit a transition from strong ferromagnetic to strong antiferromagnetic character. The vanishing correlations arise from the QD fillings, as seen for $\varepsilon_3 = -\varepsilon_1$. In this case, the total filling is $3$, with the central dot occupied by one electron and the edge dots by approximately two and zero electrons, respectively. As a result, the spins on the edges are compensated, leading to zero spin correlations between the dots. The crossover in the central part of Fig.~\ref{fig:3d1laway_mag}, which forms a distinctive "propeller-like" pattern in panel (b1), has a more complex origin.

\begin{figure}[ht!]
    \centering
   \includegraphics[width=1\linewidth]{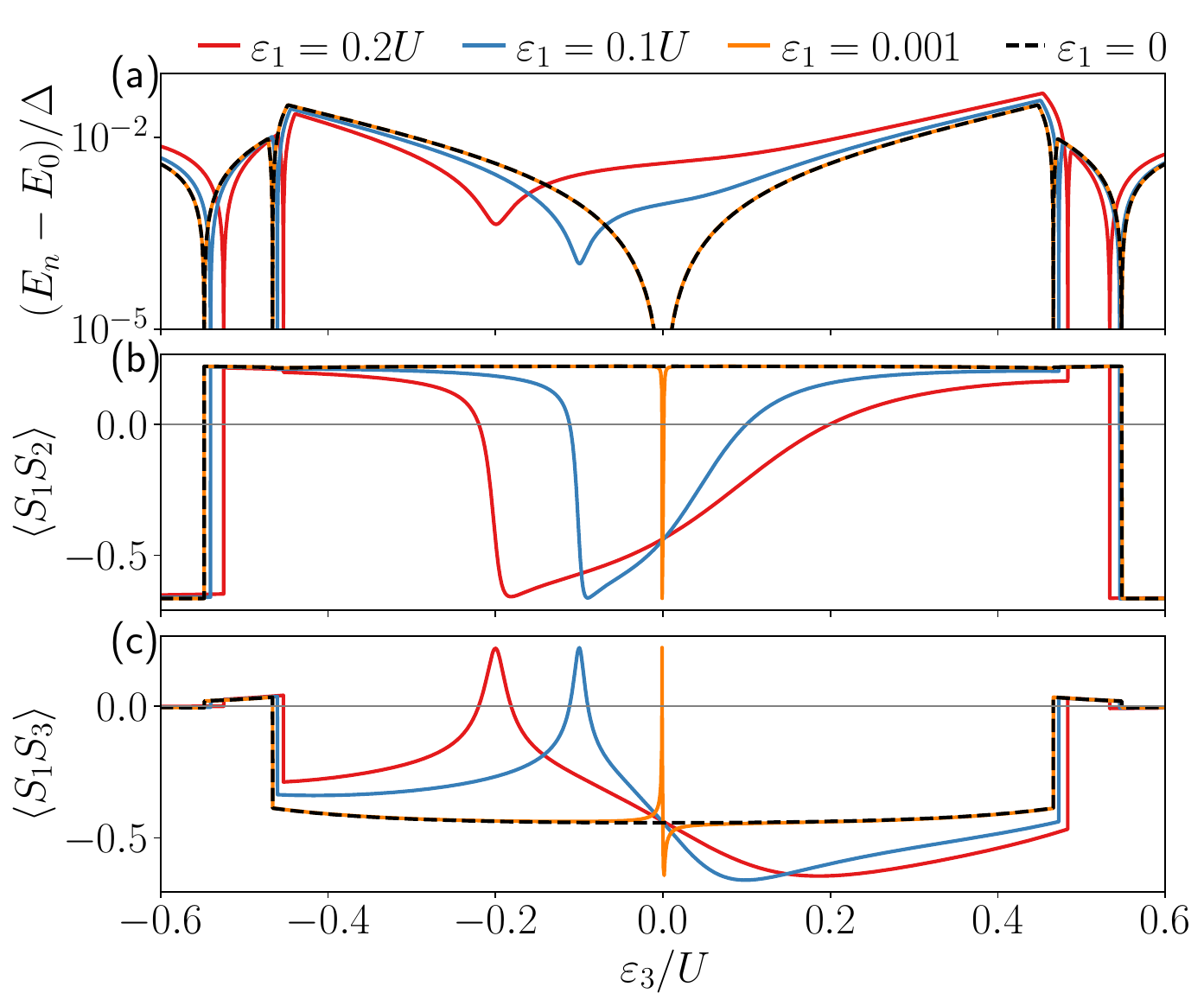}
    \caption{ (a) First exited in-gap energy as a function of $\varepsilon_3$ for triple QD with single SC lead illustrated in Fig.~\ref{fig:ilpDQ}(b), with $U=3\Delta$, $\Gamma=0.2\Delta$, $\varepsilon_2=0$ as in the panels (b1) and (b2) of Fig.~\ref{fig:3d1laway_mag} for  $\epsilon_1=0.2U$ (red), $0.1U$ (blue), $0.001U$ (orange) and $0$ (black).  (b) The respective spin-spin correlations between first and second dot $\langle \bm{S}_1 \!\cdot\! \bm{S}_2 \rangle$ and (c) between the first and third dot $\langle \bm{S}_1 \!\cdot\! \bm{S}_3 \rangle$. The system was modeled by ChE($L=4$).
}
    \label{fig:eps3cut}
\end{figure}

This origin lies in the structure of the doublet ground state. For symmetric dots, i.e., $\varepsilon_1 = \varepsilon_2 = \varepsilon_3$, the ground state is doubly degenerate. Introducing asymmetry in the level positions $\varepsilon_j$ lifts this degeneracy, turning one of the doublets into the first excited state. This is illustrated in Fig.~\ref{fig:eps3cut}(a). Lines of different colors show different cuts for $\varepsilon_2=0$ and four fixed values of $\varepsilon_1=0$ (black), $0.001U$ (orange), $0.1U$ (blue) and $0.2U$ (red) and varying $\varepsilon_3$. The sharp dips at the edges ($|\varepsilon_3|>0.4U$) correspond to QPTs: singlet to triplet, and triplet to doublet. 
However, the only central minimum that approaches zero is the one for equal $\varepsilon_1=\varepsilon_2=0$, where at $\varepsilon_3=0$ (black) two doublets meet. Nevertheless, even for this case, we have not identified a true QPT. The minima for $\varepsilon_3\neq \varepsilon_2$ positioned at $\varepsilon_3=-\varepsilon_1$ are accompanied by a sharp dip in spin correlations $\langle \bm{S}_1 \!\cdot\! \bm{S}_2 \rangle$ [orange line in panel (b)] for small $\varepsilon_1$ that becomes broader with increasing  $\varepsilon_1$. Analogously, the spin correlations $\langle \bm{S}_1 \!\cdot\! \bm{S}_3 \rangle$ shows an overshoot-undershoot type of behavior [orange line in panel (c)] that gets broadened to a still relatively sharp peak at $\varepsilon_3=- \varepsilon_1$ and broad dip at $\varepsilon_3=\varepsilon_1$ with increasing $\varepsilon_1$. These broadened shapes lead to regions with different effective exchange.

\begin{figure}[ht!]
    \centering
   \includegraphics[width=1\linewidth]{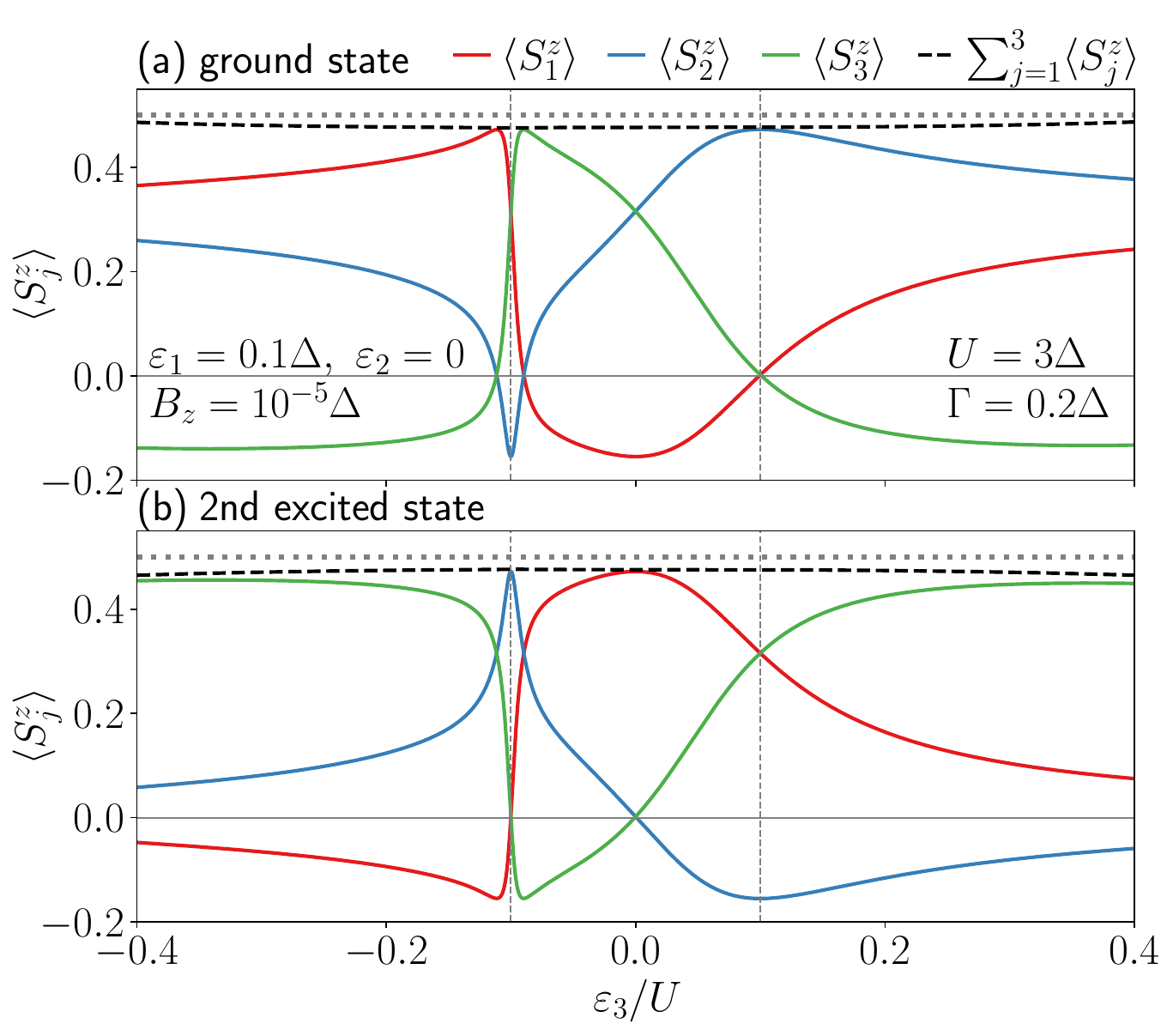}
    \caption{Triple QD coupled to a single SC lead illustrated in Fig.~\ref{fig:ilpDQ}(b) with $U=3\Delta$, $\Gamma=0.2\Delta$, $\varepsilon_1=0.1\Delta$ and $\varepsilon_2=0$ in a weak magnetic field $B_z = 10^{-5}\Delta$, which splits the doublet states. (a) Mean spin in the $z$ direction for the ground state at each dot (red, blue, and green lines) and their sum (black dashed line). The horizontal gray solid line marks zero, while the dashed gray line indicates $1/2$. Vertical dotted lines mark $\varepsilon_3=\varepsilon_1$ and  $\varepsilon_3=-\varepsilon_1$. (b) Same as in (a), but for the second-excited state, i.e., the lower component of the split second doublet.     
}
    \label{fig:eps3cutSZ}
\end{figure}

To clarify this behavior, we applied a small magnetic field in the $z$ direction. Its effect is described by the Hamiltonian.
\begin{equation}
H_B=-B_z\sum_{j=1}^3\left( n_{j\uparrow}-n_{j\downarrow}\right),
\end{equation}
which further splits the ground state and first excited doublets of the $B_z=0$ solution. Figure~\ref{fig:eps3cutSZ} shows the average $z$ component of the spin on each dot (red, blue, and green lines) and their sum (black dashed line) as a function of $\varepsilon_3$ for $\varepsilon_1=0.1\Delta$ and $B_z=10^{-5}\Delta$, for both the ground state (a) and the second excited state (b). The first excited state (not shown here) corresponds to the ground state with reversed spin polarities on each dot. In both panels (a) and (b), the sum of $S^z$ is close to $1/2$, indicating that only a small fraction of the spin resides in the lead. As in Fig.~\ref{fig:eps3cut}, two special points, $\varepsilon_3=\pm\varepsilon_1$, stand out: for $\varepsilon_3=\varepsilon_1$, $\langle S^z_2\rangle\approx 1/2$ while spins on the other two dots vanish, giving zero inter-dot correlations; for $\varepsilon_3=-\varepsilon_1$, a sharp negative peak in $\langle S^z_2\rangle$ is balanced by large positive values of $\langle S^z_2\rangle$ and $\langle S^z_3\rangle$. In the second excited state, this behavior is reversed.

These two points divide the central region into three regimes. For $\varepsilon_3<-\varepsilon_1$, the ordering is \textcolor{red}{$\uparrow$}\textcolor{blue}{$\uparrow$}\textcolor{mygreen}{$\downarrow$}, with effective ferromagnetic coupling between dots 1 and 2, and antiferromagnetic coupling between dot 3 and the other two. For $-\varepsilon_1<\varepsilon_3<\varepsilon_1$, the pattern is \textcolor{red}{$\downarrow$}\textcolor{blue}{$\uparrow$}\textcolor{mygreen}{$\uparrow$}; for $\varepsilon_3>\varepsilon_1$, it returns to \textcolor{red}{$\uparrow$}\textcolor{blue}{$\uparrow$}\textcolor{mygreen}{$\downarrow$} but with reversed magnitudes of $\langle S^z_1\rangle$ and $\langle S^z_3\rangle$. Since the central region has width $2\varepsilon_1$, the transition sharpens as $\varepsilon_1$ decreases, with $\langle S^z_1\rangle$ and $\langle S^z_3\rangle$ converging. Even at $\varepsilon_1=0$, with only one special point, the ground and excited states do not exchange character as $\varepsilon_3$ crosses zero, so no true QPT is observed.       

\begin{figure}[ht!]
    \centering
    \includegraphics[width=\linewidth]{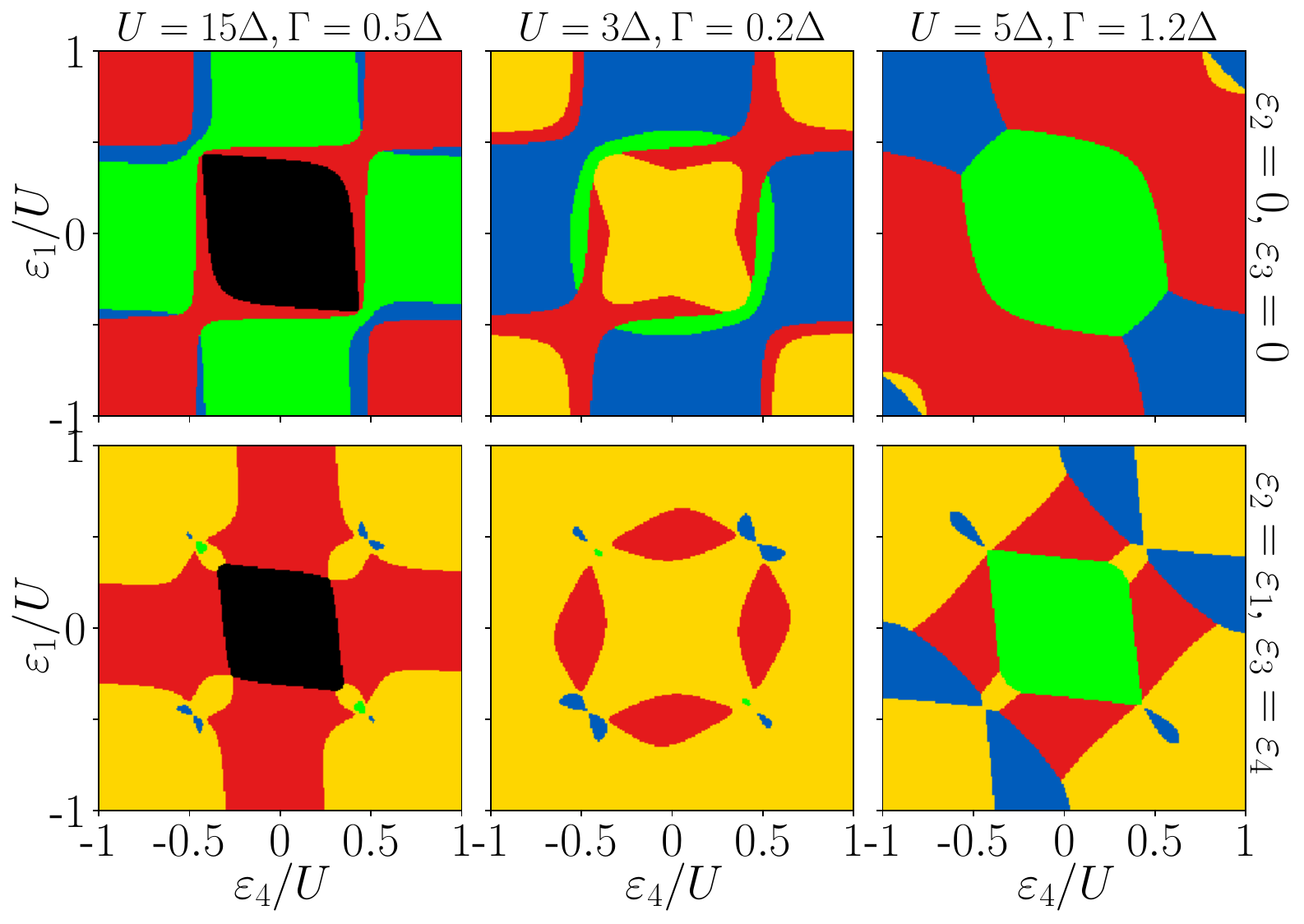}
    \caption{Phase diagrams for a quadruple QD coupled to a single SC lead, illustrated in Fig.~\ref{fig:ilpDQ}(c), evaluated away from half-filling using ChE($L = 4$). In the first line we fix $\varepsilon_2 = \varepsilon_3=0$ and vary $\varepsilon_1$ and $\varepsilon_4$.
    The second line shows results where $\varepsilon_1=\varepsilon_2$ and $\varepsilon_3=\varepsilon_4$.
    Colors encode the total spin: yellow for $S_T = 0$, blue for $S_T = 1/2$, red for $S_T = 1$, green for $S_T = 3/2$ and black for $S_T = 2$. The three columns correspond to the parameter sets  
    (i) $U = 15\Delta$, $\Gamma = 0.5\Delta$;  
    (ii) $U = 3\Delta$, $\Gamma = 0.2\Delta$; and  
    (iii) $U = 5\Delta$, $\Gamma = 1.2\Delta$,  
    all with $t_d = 0$ and $W = 0$.}
    \label{fig:phd4away}
\end{figure}

\subsubsection{Quadruple QD}
The final results, shown in Fig.~\ref{fig:phd4away}, present phase diagrams away from half filling for a quadruple dot using the same values of $U$ and $\Gamma$ studied in the double- and triple-dot cases. The first row shows cases where only the energy levels of the first and fourth dots are varied, with the other two kept at half-filling condition, i.e., $\varepsilon_2 = \varepsilon_3 = 0$. The second row explores a scenario in which tuning gates are applied symmetrically to pairs of neighboring dots, i.e., $\varepsilon_1 = \varepsilon_2$ and $\varepsilon_3 = \varepsilon_4$.

The resulting diagrams form a kaleidoscope of phases, with colors representing different total spin states. In addition to broad regions of singlet, doublet, triplet, quartet, and quintet phases (the latter seen in the leftmost panels), small, isolated pockets, not observed for double and triple dots, also appear in the diagrams. These regions are stable with respect to increasing chain length $L$ (not shown here), indicating that they are robust features of the model. This illustrates how the addition of a single QD can generate qualitatively new behavior, consistent with the "more is different" concept~\cite{Anderson1972more}. It also highlights a key advantage of ChE over heavy numerical methods such as NRG or QMC. While detailed and reliable maps such as Fig.~\ref{fig:34d1laway} can be obtained on standard PCs using ChE, achieving comparable results with NRG or QMC would require computational clusters and hundreds of node hours.

It is worth emphasizing that, for triple and quadruple dots coupled to a superconductor, this work represents an entry-level study. Realistic systems are far more complex, even within the already simplified framework of a generalized SC-AIM. Finite direct hopping and capacitive coupling between dots can be significant. The hybridization parameters $V_{il}$ are, in general, complex. The spatial arrangement of the dots -- for example, their relative positions on the SC substrate in an STM study -- also plays a crucial role~\cite{Li2025individual,Li2025negative}. Furthermore, effects such as Kondo screening, RKKY interaction, and superexchange can interfere in subtle ways. Exact numerical methods that account for all these ingredients remain prohibitively expensive for broad parameter scans. This underscores the need for approximate approaches that are nonetheless controlled, reliable, and scalable -- such as the chain expansion scheme presented in this work     

\section{Conclusions \label{sec:Conc}}
We have presented a chain-expansion scheme that builds low-energy effective models for SC heterostructures by expanding the tunneling self-energy into a finite tight-binding chain. The construction is flexible: different \emph{Pad\'e} orders give chains of different lengths that can be chosen to match the task at hand.

Short \emph{Pad\'e} chains containing only 2-4 sites already capture equilibrium properties in the typical parameter ranges. Because these Hamiltonians remain small, straightforward ED is fast enough to scan broad phase spaces, even for complicated multi-dot layouts. Increasing the chain length $L$ improves the accuracy in a controlled way, and they work equally well out of equilibrium, where they agree with NEGF calculations.

Using ChE($L$), we mapped the ground-state phase diagrams of double, triple, and quadruple dots coupled to a single SC lead. We demonstrated that at the half-filling, these phase diagrams have similar overall shapes but consist of different ground-state phases. Particularly interesting is the regime of large $U$ and small $\Gamma$. Here, the ZBW limit, a heavily used effective model, fails to give the correct ground state. Longer chains correctly capture these phases, but lengths up to $L=4$ are typically needed to pinpoint the position of the relevant QPT. 

For the double dot, we show that capacitive coupling can drive a singlet-singlet QPT. We also varied the strength of the lead-mediated inter-dot pairing. Depending on the other parameters, a relatively weak pairing can stabilize triplet ground states even when direct hopping between dots is present. 

Moving away from half-filling makes the phase diagrams more complex, especially for a larger number of dots. We show that although the ZBW approximation can still be reliable in some regimes, it often predicts qualitatively incorrect ground states.

We presented ChE not just as a model but also as a general strategy. Like NRG or the recently proposed surrogate models, it replaces the lead by a finite or semi-infinite chain. Its main advantages, as presented here, are that the chains are built using \emph{Pad\'e} approximants; therefore, they are the optimal low-energy representations to a given \emph{Pad\'e} order. We supplied an algorithm for arbitrary bandwidth and simple analytic formulas for WBL and the infinite-chain limit. Very short chains give quantitative results with ED; longer chains are still manageable by (t)DMRG or Monte Carlo methods, letting one tackle multi-terminal and time-dependent problems. ChE chains can also be combined with other approaches, such as Wilson chains, where the main channel is treated by NRG and a side-coupled short ChE chain captures additional channels. In addition, the accuracy of ChE schemes makes them suitable substitutes for other, more demanding impurity solvers and can eventually be used to treat lattice models like the SC periodic Anderson model within the dynamical mean-field theory~\cite{Luitz2010weak,Pokorny-2021-PAM}. Finally, the only convergence parameter in the presented scheme is the chain length $L$, providing a controllable alternative to methods that require non-trivial parameter fitting and advanced discretization schemes.

In this way, ChE offers not just another method, but a practical and broadly adaptable framework for advancing the study of strongly correlated superconducting systems.


\begin{acknowledgments}
The authors acknowledge support from the Czech Science Foundation through project No.~23-05263K. This work was also supported by the Ministry of Education, Youth and Sports of the Czech Republic through the e-INFRA CZ (ID:90254), the Czech Republic-Germany Mobility programme (ID:8J25DE001), and EU COST action CA21144 SUPERQUMAP. We thank Artur Slobodeniuk, Peter Zalom, Tom\'a\v{s} Novotn\'y, and Wolfgang Belzig for many productive discussions. 
\end{acknowledgments}
\appendix

\section{Tunneling self-energy of ChE model as a continued fraction \label{app:CF}}

To obtain the dot Green's function for the ChE models from its all-sites form in Eq.~\eqref{eq:ChEGF}, which is equivalent to the full SC-AIM analogue of Eq.~\eqref{eq:G0}, we must integrate out the states that belong to the chain. A convenient way is to eliminate them iteratively using a block-matrix identity~\cite{Lu-2002}
\begin{align}
    &\begin{pmatrix}
        \mathbf{A} & \mathbf{c} \\
        \mathbf{d} & \mathbf{B}
    \end{pmatrix}^{-1}=\nonumber\\
    &\begin{pmatrix}
        (\mathbf{A} -\mathbf{c} \mathbf{B}^{-1}\mathbf{d})^{-1} & -\mathbf{A}^{-1}\mathbf{c}(\mathbf{B} -\mathbf{d}\mathbf{A}^{-1}\mathbf{c})^{-1} \\
        -\mathbf{B}^{-1}\mathbf{d}(\mathbf{A}-\mathbf{c}\mathbf{B}^{-1}\mathbf{d})^{-1} & (\mathbf{B} -\mathbf{d}\mathbf{A}^{-1}\mathbf{c})^{-1}
    \end{pmatrix},
    \label{eq:block_matrix_approx}
\end{align}
Due to the pentadiagonal structure, we can formally apply the above identity successively to the $2\times 2$ $\mathbf{A}$-blocks, starting from the upper left corner of the matrix in Eq.~\eqref{eq:ChEGF}, one block at a time. After this formal forward sweep, we get a ladder of auxiliary matrices $\mathcal{A}$ that can be calculated in reverse order
by taking advantage of the identity 
\begin{equation}
\sigma_z
\begin{pmatrix}
i \omega_n & \wD \\
\wD & i \omega_n
\end{pmatrix}^{-1}\!\!\!\!\!\!
\sigma_z
=
\frac{-1}{\omega_n^2 +\wD^2}
\begin{pmatrix}
i \omega_n & \wD \\
\wD & i \omega_n
\end{pmatrix}\equiv\frac{-1}{\omega_n^2 +\wD^2}\mathbb{A},
\end{equation}
where $\sigma_z$ is the third Pauli matrix. The series then reads 
\begin{widetext}
\begin{align*}
    \mathcal{A}_{L-1}^{-1} &=\left(\mathbb{A} - \wh^2_{L-1}\sigma_z \mathbb{A}^{-1}\sigma_z \right)^{-1} =
    \left(\mathbb{A} +\wh^2_{L-1} (\wD^2+\omega_n^2)^{-1} \mathbb{A}\right)^{-1}=(\wD^2+\omega_n^2)(\wD^2+\wh^2_{L-1}+\omega_n^2)^{-1}\mathbb{A}^{-1},\\
    \mathcal{A}_{L-2}^{-1} &= \left(\mathbb{A} - \wh^2_{L-2} \sigma_z{A}_{L-1}^{-1}\sigma_z\right)^{-1}=\left(1+\wh^2_{L-2}(\wD^2+\wh^2_{L-1}+\omega_n^2)^{-1}\right)^{-1}\mathbb{A}^{-1},\\
     \mathcal{A}_{L-3}^{-1} &= \left(\mathbb{A} - \wh^2_{L-3} \sigma_z{A}_{L-2}^{-1}\sigma_z\right)^{-1}=(\wD^2+\omega_n^2)\left(\wD^2+\omega_n^2 + \wh^2_{L-3}\left(1+\wh^2_{L-2}(\wD^2+\wh^2_{L-1}+\omega_n^2)^{-1}\right)^{-1} \right)^{-1}\mathbb{A}^{-1},\\
\mathcal{A}_{L-4}^{-1} &= \left(1+\wh^2_{L-4}\left(\wD^2+\omega_n^2 + \wh^2_{L-3}\left(1+\wh^2_{L-2}(\wD^2+\wh^2_{L-1}+\omega_n^2)^{-1}\right)^{-1} \right)^{-1}\right)^{-1}\mathbb{A}^{-1},\\
\vdots.
\end{align*}
\end{widetext}
There are two important observations here. The first is that the matrix structure is preserved solely by $\mathbb{A}$. The second is that the prefactor is clearly a continuous fraction. When rolled up to the QD, the prefactor reads
\begin{widetext}
\begin{equation}
    \widetilde{\cal P}(\omega_n;\wD,\gamma,\{\wh_\ell\})=\cfrac{\gamma^2}{\wD^2+\omega_n^2+\cfrac{\wh^2_1}{1+\cfrac{\wh^2_2}{\wD^2+\omega_n^2+\cfrac{\wh^2_3}{1+\cfrac{\wh^2_4}{\wD^2+\omega_n^2+\cfrac{\wh^2_5}{1+\cfrac{\wh^2_6}{\vdots}}}}}}}.
\end{equation}
\end{widetext}
The continued fraction terminates with $(\wD^2+\omega_n^2+\wh^2_{L-1})$ for a chain with an even number of sites and with $(1+\wh^2_{L-1}/(\wD^2+\omega_n^2))$ in the case of an odd number of sites. 
This leads to the non-interacting dot Green's function that reads
\begin{widetext}
\begin{equation}
    G^\text{ChE(L)}_{U=0} =
    \begin{pmatrix}
    i \omega_n\left[1+\widetilde{\cal P}\left(\omega_n;\wD,\gamma,\{\wh_\ell\}\right)\right]-\epsilon 
    & \wD \widetilde{\cal P}\left(\omega_n;\wD,\gamma,\{\wh_\ell\}\right) \\
    \wD \widetilde{\cal P}\left(\omega_n;\wD,\gamma,\{\wh_\ell\}\right) 
    & i \omega_{n}\left[1+\widetilde{\cal P}\left(\omega_n;\wD,\gamma_n,\{\wh_\ell\}\right)\right]+\epsilon \\
    \end{pmatrix}^{-1}.
    \label{eq:G0ChE1_2}
\end{equation}
\end{widetext}
After the rescaling introduced in Eq.~\eqref{eq:resc}, repeated here for convenience  
\begin{equation}
    \gamma = \sqrt{\frac{\Gamma}{\Delta}h_0}\Delta,\qquad
    \wh_\ell = \sqrt{h_\ell}\Delta,
    \label{eq:appresc}
\end{equation}
the comparison of Eq.~\eqref{eq:G0ChE1_2} with Eq.~\eqref{eq:G0} sets $\wD=\Delta$ and 
gives the dot Green's function Eq.~\eqref{eq:G0ChE1}
where ${\cal P}_L(\frac{\omega_n}{\Delta};\{h_\ell\})$ is a finite continued fraction 
\begin{widetext}
\begin{align}
    {\cal P}\left(\frac{\omega_n}{\Delta};\{h_\ell\}\right)&=\cfrac{h_0}{1+\frac{\omega^2_n}{\Delta^2}+\cfrac{h_1}{1+\cfrac{h_2}{1+\frac{\omega_n^2}{\Delta^2}+\cfrac{h_3}{1+\cfrac{h_4}{1+\frac{\omega_n^2}{\Delta^2}+\cfrac{h_5}{1+\cfrac{h_6}{\vdots}}}}}}}.
    \label{eq:ConFrac}
\end{align}
\end{widetext}
Therefore, it can be expressed using the matrix product form as ${\cal P}_L(x;\{h_\ell\})=P_L/Q_L$, where 
\begin{equation}
\begin{pmatrix}
P_L \\
Q_L 
\end{pmatrix}=
\prod_{\ell=0}^{L-1}
\begin{pmatrix}
0 & h_\ell \\
1 & \left(1 +\left[(\ell+1)\hspace{-0.2cm}\mod 2\right]x^2\right) \\
\end{pmatrix}
\begin{pmatrix}
0 \\
1 
\end{pmatrix},
\label{eq:PMP}
\end{equation}
leading to the rational functions (Eqs.~\eqref{eq:PadeEven} and~\eqref{eq:PadeOdd} in the main text)
\begin{equation}
    {\cal P}_L(x;\{h_\ell\})=\frac{\sum_{j=0}^{L/2-1}p_j(\{h_\ell\}) x^{2j}}
    {\sum_{j=0}^{L/2}q_j(\{h_\ell\}) x^{2j}}
    \label{eq:PadeEvenAp}
\end{equation}
for even $L$ and
\begin{equation}
    {\cal P}_L(x;\{h_\ell\})=\frac{1}{1+x^2}\frac{\sum_{j=0}^{(L-1)/2}p_j(\{h_\ell\}) x^{2j}}
    {\sum_{j=0}^{(L-1)/2}q_j(\{h_\ell\}) x^{2j}}
    \label{eq:PadeOddAp}
\end{equation}
for odd $L$.
\begin{figure}[ht!]
    \centering
    \includegraphics[width=1.\linewidth]{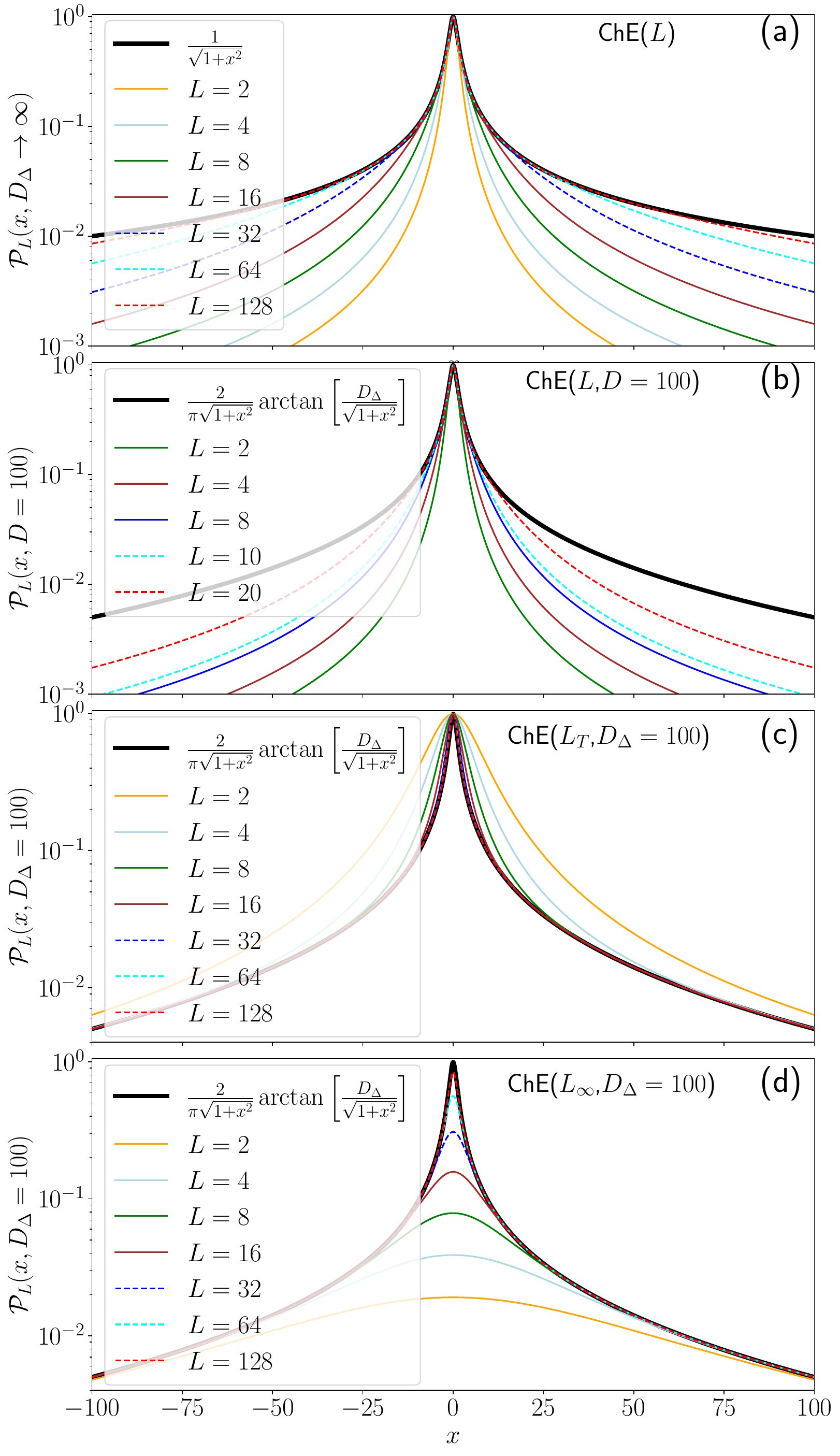}
    \caption{Comparison of the approximate function ${\cal P}_L(x,{h_\ell})$ with the full hybridization function $\tGD$ for $\Delta = 1$.
    (a) WBL, where the coefficients ${h_\ell}$ take the simple analytic form given in Eq.\eqref{eq:coefWBL}.
    (b) Finite-bandwidth case with $D_\Delta = 100$, where ${h_\ell}$ are obtained using the method described in the main text and Appendix~\ref{app:CF}, based on matching the \emph{Pad\'e} approximant to the full tunneling self-energy.
    (c) Truncated expansion of the infinite-chain model, where the last coefficient is evaluated using Eq.~\eqref{eq:trancation}, with $D_\Delta = 100$.
    (d) Truncated infinite-chain expansion for $D_\Delta = 100$, without applying the correction to the final coefficient.  
    \label{fig:rozvoj_app}}
\end{figure}

The calculation of $h_\ell$ therefore differs slightly between chains with even and odd numbers of sites $L$. In what follows, we use $\Delta$ as an energy unit. For \emph{even} $L$ the needed \emph{Pad\'e} expansion reads 
\begin{align}
    \frac{1}{\sqrt{1 + x^2}}&\frac{2}{\pi}
    \arctan\left[\frac{D}{\sqrt{1+x^2}}\right]\approx
    \frac{\sum_{j=0}^{L/2-1}p_jx^{2j}}
    {\sum_{j=0}^{L/2}q_jx^{2j}},\label{eq:evenPade}
\end{align}
for \emph{odd} $L$ we use the \emph{Pad\'e} approximant for the expression in the curly brackets 
\begin{equation}
\begin{aligned}
&\frac{1}{1 + x^2}\left\{\frac{2}{\pi}\sqrt{1 + x^2}
\arctan\left[\frac{D}{\sqrt{1+x^2}}\right]\right\}\\
&\approx
\frac{1}{(1+x^2)}\frac{\sum_{j=0}^{(L-1)/2}p_jx^{2j}}
{\sum_{j=0}^{(L-1)/2}q_{j}x^{2j}},\label{eq:oddPade}
\end{aligned}
\end{equation}
We now transform these expressions to a standard rational polynomials using: 
\begin{equation}
\begin{aligned}
    \overline{p}_k &= \sum_{j=k}^{N-1}\binom{j}{k} (-1)^{j-k} p_j,\quad\wip_k=\overline{p}_k/\overline{q}_{N},\\
    \overline{q}_k &= \sum_{j=k}^{N}\binom{j}{k} (-1)^{j-k} q_j,\quad\wiq_k =\overline{q}_k/\overline{q}_{N}
    \label{eq:even_pq}
\end{aligned}
\end{equation}
with $N=L/2$ for even $L$ and
\begin{equation}
\begin{aligned}
    \overline{p}_k &= \sum_{j=k}^{N-1}\binom{j}{k} (-1)^{j-k} p_j,\quad\wip_k=\overline{p}_k/\overline{q}_{N},\\
    \overline{q}_{k+1} &= \sum_{j=k}^{N-1}\binom{j}{k} (-1)^{j-k} q_j,\quad\wiq_k=\overline{q}_k/\overline{q}_N \label{eq:pol}
\end{aligned}
\end{equation}
with $N=(L-1)/2+1$ and  $\overline{q}_0=0$ for odd $L$.  By applying the substitution $z=1+x^2$ we get the standard rational polynomials, which allows us to use some continued fraction identities
\begin{align}
{\cal R}(z)&= 
\frac{\sum_{k=0}^{N-1}\widetilde{p}_{k}z^{k}}
{\sum_{k=0}^{N}\widetilde{q}_{k}z^{k}}.
\end{align}
The rest of the solution is the same for odd and even $L$, because the above manipulation allows us to rewrite the fraction Eq.~\eqref{eq:ConFrac} as 
\begin{equation}
    \frac{h_0}{z }\lowplus\frac{h_1}{1} \lowplus \frac{h_2}{z} \lowplus\frac{h_3}{1} \lowplus \frac{h_4}{z} \dots
\end{equation} 
The coefficients ${h_\ell}$ are calculated by the method outlined in Ref.~\cite{wall2018analytic} using the odd ($M_j^o$) and even ($M^e_j$) (not related to odd and even $L$) principal minors, i.e., the determinants of the sub-matrices
\begin{equation}
\label{eq:matrix}   
\begin{NiceMatrix}[]  
  M^o_1 & M^e_1 & M^o_2 & M^e_2 & M^o_3 & M^e_3 &\dots \\
  \wip_{N-1} & \wip_{N-2} & \wip_{N-3} & \wip_{N-4} & \wip_{N-5} & \wip_{N-6} &\dots \\
  \wiq_{N\phantom{-0}} & \wiq_{N-1} & \wiq_{N-2} & \wiq_{N-3} & \wiq_{N-4} & \wiq_{N-5} &\dots \\
  0_{\phantom{N+0}} & \wip_{N-1} & \wip_{N-2} & \wip_{N-3} & \wip_{N-4} & \wip_{N-5} & \dots \\
  0_{\phantom{N+0}} & \wiq_{N\phantom{+0}} & \wiq_{N-1} & \wiq_{N-2} & \wiq_{N-3} & \wiq_{N-4} & \dots \\
  0_{\phantom{N+0}} & 0_{\phantom{N+0}} &  \wip_{N-1}  &  \wip_{N-2}  &  \wip_{N-3}  &  \wip_{N-4}  &\dots \\
  0_{\phantom{N+0}} & 0_{\phantom{N+0}} & \wiq_{N\phantom{-0}} & \wiq_{N-1} & \wiq_{N-2} & \wiq_{N-3} & \dots \\
  \vdots &  \vdots &  \vdots &  \vdots &  \vdots &  \vdots & \ddots \\
  \CodeAfter
        \tikz \draw[thick]
            (2-1.north east) -- (2-1.south east)   
            (2-1.south west) -- (2-1.south east); 
        \tikz \draw[thick]
            (2-2.north east) -- (3-2.south east)   
            (3-1.south west) -- (3-2.south east); 
        \tikz \draw[thick]
            (2-3.north east) -- (4-3.south east)   
            (4-1.south west) -- (4-3.south east);
        \tikz \draw[thick]
            (2-4.north east) -- (5-4.south east)   
            (5-1.south west) -- (5-4.south east);
        \tikz \draw[thick]
            (2-5.north east) -- (6-5.south east)   
            (6-1.south west) -- (6-5.south east);
        \tikz \draw[thick]
            (2-6.north east) -- (7-6.south east)   
            (7-1.south west) -- (7-6.south east);
\end{NiceMatrix}
\end{equation}
where $\wip_j$ and $\wiq_j$ with $j<0$ are zero and we also need $M^o_0=\wiq_{N}=1$ and $M^e_0=1$. The coefficients $h_\ell$ are then
\begin{equation}
h_0=\frac{M_1^o}{M_0^o},\,\,h_{2j-1} = \frac{M^o_{j-1}M^e_j}{M^o_jM^e_{j-1}}
,\,\,h_{2j} =\frac{M^o_{j+1}M^e_{j-1}}{M^o_jM^e_{j}},
\label{eq:hcoefM}
\end{equation}
where $j>0$. Examples for ChE($L=1$,$D_\Delta$) (ZBW) and  ChE($L=2$,$D_\Delta$) (eZBW) can be found in the main text. For WBL ($D_\Delta\rightarrow\infty$) this leads to Eq.~\eqref{eq:coefWBL}.

\subsubsection{Examples:} To illustrate the procedure for determining the coefficients of ChE($L$,$D_\Delta$), we present two simple step-by-step examples for $L=3$ with finite band width and $L=4$ for the WBL.

We begin with the less intuitive case of odd $L$, represented here by $L=3$ and $D_\Delta=10$. For odd $L$, we use the \emph{Pad\'e} $[(L-1)/(L-1)]$ form, as shown in Eq.~\eqref{eq:even_pq}, applied only to the term in the curly brackets, rather than to the entire expression. For $L=3$ we need \emph{Pad\'e} $[2/2]$ which leads to 
\begin{align}
    &\frac{1}{1+x^2} \biggl\{ \sqrt{1+x^2} \frac{2}{\pi}
    \arctan\left[\frac{10}{\sqrt{1+x^2}}\right] \biggr\} \approx \nonumber\\ 
    &\approx \frac{1}{1+x^2}\frac{0.936549 + 0.704351 x^{2}}{1. + 0.285722 x^{2}}.
\end{align}
Therefore, we identify $p_0 \doteq 0.936549 $, $p_1 \doteq 0.704351$, $q_0 = 1$, and $q_1 \doteq 0.285722$. Using the coefficient transformation formulas for odd-site chains, we find the normalized coefficients
\begin{align}
    &\wip_0 = (p_0 - p_1)/q_1,\qquad \wip_1 = p_1/q_1,\\
    &\wiq_0 = 0,\qquad \wiq_1 =(q_0 - q_1)/q_1,\qquad \wiq_2 = 1, \nonumber
\end{align}
which allows us to construct the matrix~\eqref{eq:matrix},
\begin{align}
    \begin{pmatrix}
        \wip_1 & \wip_0 & 0 \\
        \wiq_2 & \wiq_1 & \wiq_0 \\
        0 & \wip_1 & \wip_0
    \end{pmatrix}.
\end{align}
The required minors (subdeterminants), therefore, read
\begin{align}
    M_0^e &=M_0^o= 1, \nonumber\\
    M_1^o & = \wip_1, \nonumber\\
    M_1^e &= \wip_1\wiq_1 - \wip_0\wiq_2,\nonumber\\
    M_2^o &=  \wip_0 M_1^e.
\end{align}
Using Eq.~\eqref{eq:hcoefM} we finally get the $h$ coefficient values as
\begin{align}
    h_0 &= \wip_1 \doteq 2.46516,\nonumber\\
    h_1 &= (\wip_1\wiq_1 - \wip_0\wiq_2)/\wip_1 \doteq 2.17024,\\
    h_2 &= \wip_0/\wip_1 \doteq 0.329662.\nonumber
\end{align}

For the even $L$ and WBL we calculate the \emph{Pad\'e} approximant $[L-2/L]$ of the self energy function $1/\sqrt{1 + x^2}$. Taking as an example the case $L=4$, and therefore \emph{Pad\'e} approximant$ [2,4]$, we get
\begin{equation}
    \frac{1}{\sqrt{1 + x^2}} \approx \frac{1 + \frac{1}{2} x^2}{1 + x^2 + \frac{1}{8} x^4},
\end{equation}
which gives coefficients $p_0 = 1, p_1 = \frac{1}{2}, q_0 = 1, q_1 = 1$, and $q_2 = \frac{1}{8}$. Using the coefficient transformation
formulas for even-site chains, we find the normalized coefficients
\begin{align}
    \widetilde{p}_{0} &= 4, \qquad\widetilde{p}_{1} = 4,\nonumber\\
    \widetilde{q}_{0} &= 1, \qquad\widetilde{q}_{1} = 6,\qquad\widetilde{q}_{2} = 1.
\end{align}
Now we substitute the values of $\widetilde{p}_{k}$ and $\widetilde{q}_{k}$ into the matrix ~\eqref{eq:matrix} and calculate the needed minors
\begin{align}
    M^o_0 &= 1 ,\, \, 
    M^e_0 = 1,\nonumber\\
    M^o_1 &= 4 ,\, \, 
    M^e_1 = 20,\\
    M^o_2 &= 64,\, \, 
    M^e_2 = 64.\nonumber
\end{align}
Finally, we use the Eq.~\eqref{eq:hcoefM} to calculate coefficients $h_l$ obtaining
\begin{align}
    h_0 = 4,\quad h_1 = 5,\quad h_2 = 4/5,\quad h_3 = 1/5.
\end{align} 
These values are in agreement with the Eq.~\eqref{eq:coefWBL} for $L = 4$.

\subsubsection{Infinite $L$ limit:} 
For very long chains ($L\gg1$), it is advantageous to use the result for the infinite $L$, which follows directly from the continued fraction expansion of the $\arctan$ function, 
\begin{equation}
\begin{aligned}
    &\frac{2}{\pi}\frac{1}{\sqrt{1+x^2}}\arctan\left[\frac{D_\Delta}{\sqrt{1+x^2}}\right]=\\
    &\cfrac{2D_\Delta/\pi}{1(1+x^2)+\cfrac{(1D_\Delta)^2}{3+\cfrac{(2D_\Delta)^2}{5(1+x^2)+\cfrac{(3D_\Delta)^2}{7 +\cfrac{(4D_\Delta)^2}{9(1+x^2)+\cfrac{(5D_\Delta)^2}{\vdots}}}}}}
\end{aligned}
\end{equation}
after straightforward rearrangement of this continued fraction to the form of Eq.~\eqref{eq:ConFrac} we can directly read the coefficients for $L\rightarrow\infty$:
\begin{equation}
    h_0=\frac{2}{\pi}D_{\Delta},\qquad
    h_\ell=\frac{\ell^2 D_{\Delta}^2}{(2\ell+1)(2\ell-1)}.
\end{equation}
For finite lattices, we truncate the series by assuming $x=0$ for all terms with $h_{\ell>L-1}$ and evaluate the resulting infinite continued fraction
\begin{align}
h_{L-1}=
\cfrac{\frac{(L-1)^2D_{\Delta}^2}{4(L-1)^2-1}}{1+\cfrac{\frac{L^2D_{\Delta}^2}{4L^2-1}}{1+\cfrac{\frac{(L+1)^2D_{\Delta}^2}{4(L+1)^2-1}}{1 +\cfrac{\frac{(L+2)^2D_{\Delta}^2}{4(L+2)^2-1}}{1+\cfrac{\frac{(L+3)^2D_{\Delta}^2}{4(L+3)^2-1}}{\vdots}}}}}
\label{eq:trancation}
\end{align}
to get $h_{L-1}$ that improves the approximation for small $x$.
Note that the coefficients of ChE($L$,$D_\Delta$) approach the special limits of ChE($L$) and ChE($L_\infty$,$D_\Delta$) as expected. 
\vspace{0.2cm}

\section{Systems with multiple leads coupled to the same QD \label{app:multileat}}
Representing each SC lead by its own chain is straightforward and works even for unequal leads, i.e., with different superconducting gaps or chemical potentials, or leads coupled to multiple dots. However, when multiple leads are coupled to the same QD and not to any other dot, all such leads can be represented by a single chain. This construction relies on the geometric factor $\bm{\chi}_j$ introduced in Ref.~\cite{Zalom2023hidden} and its generalization to multi-dot systems discussed in Appendix VI of that work. In these cases, the tunneling self-energies are block diagonal with $2\times 2$ Nambu blocks
\begin{align}
\Sigma_j(\omega_n)=
    \begin{pmatrix}
    i\omega_n\frac{1}{\Delta}, & 
    \bm{\chi}_j\\
    \bm{\chi}_j^* & 
    i\omega_n\frac{1}{\Delta}
    \end{pmatrix}\Gamma_{jT}\tGD
    \label{eq:SigmaChi},
\end{align}
which describes the local coupling of dot $j$ to all its leads. Here, the complex geometric factor is
\begin{align}
	\bm{\chi}_j 
	&=
	\sum_{\ell} 
	\gamma_{j\ell} e^{i\varphi_{j\ell}}
\end{align}
where $\Gamma_{jT}=\sum_{\ell}\Gamma_{j\ell}$ is the total coupling constant to dot $j$ and $\gamma_{j\ell}=\Gamma_{j\ell}/\Gamma_{jT}$. Writing $\bm{\chi}_j=\chi_j e^{i\Phi_j}$, the amplitude $\chi_j\in\langle 0,1\rangle$ and phase $\Phi_j$ characterize the net pairing. By gauge invariance, all $\varphi_{j\ell}$ can be shifted by a constant so that one $\Phi_j$ (e.g., for $j=1$) is set to zero, making $\bm{\chi}_1=\chi_1$ real \cite{Zalom2023hidden}.
\begin{figure}[ht]
    \centering
   \includegraphics[width=1\linewidth]{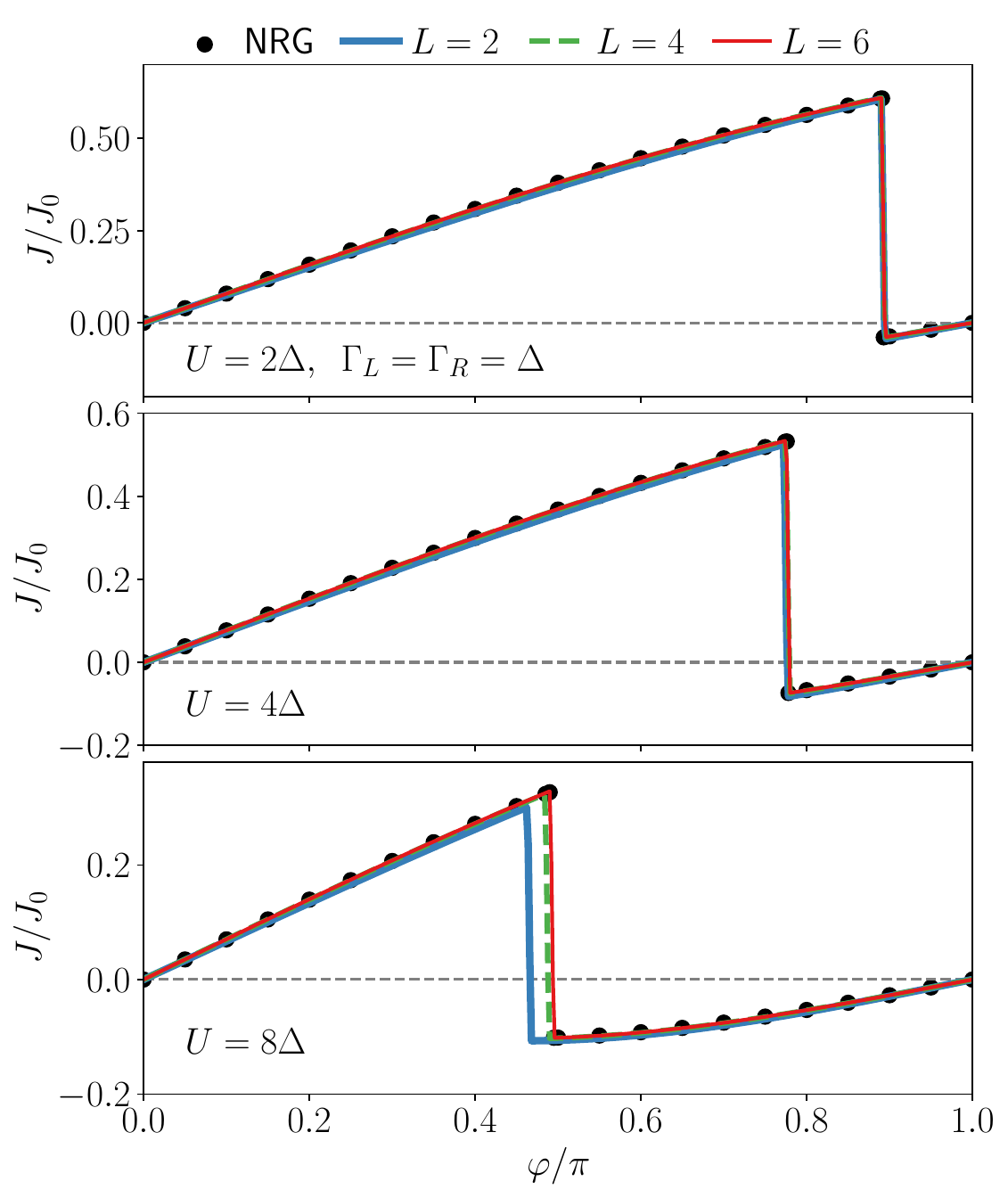}
    \caption{Single QD between two SC leads as illustrated in Fig.~\ref{fig:bench}(b). Comparison of the supercurrent as a function of the Josephson phase for different $U$ at half-filling ($\varepsilon=0$) between NRG (black circles) and single chain effective model sChE($L$). 
} 
    \label{fig:chidep}
\end{figure}

A single-chain expansion (sChE) effective model that reproduces the self-energy~\eqref{eq:SigmaChi} for block $j$ can be set as
\begin{align}
\mathcal{H}^{\chi_j}&=-\sqrt{h_0\Gamma_{jT}\Delta}\sum_{\sigma} \left(e^{i\Phi_j/2} d_{j\sigma}^\dagger c^\pdag_{1j\sigma} +  \mathrm{H.c.}\right)\nonumber\\
    &-\Delta\sum_{\ell=1}^{L-1}\sum_\sigma \left(\sqrt{h^\pdag_{\ell}} c_{\ell j\sigma}^\dagger c^\pdag_{\ell+1,j\sigma} +  \mathrm{H.c.}\right)\\ 
    &-\chi_j\Delta\sum_{\ell=1}^L\left(c_{\ell j\uparrow}^{\dagger}c_{\ell j\downarrow}^\dagger + \mathrm{H.c.}\right),\nonumber
    \label{eq:ChEM}
\end{align}
with coefficients $h_\ell$ obtained by the same procedure as in the main text or by the algorithm in Appendix~\ref{app:CF} (with minor modifications, e.g., the substitution $z=\chi_j^2+x^2$ which also changes the transformation factor $(-1)^{j-k}$ to $(-\chi^2)^{j-k}$ in the Eq.~\ref{eq:even_pq}.

For convenience, explicit $h_\ell$ for the smallest even chain lengths in the WBL are listed in Table~\ref{tab:1}. These formulas are useful because the supercurrent through lead $\ell$ can be computed by the Hellmann--Feynman theorem, which for zero temperature and zero chemical potential in the leads reads
\begin{equation}
J_{\ell} = \frac{2e}{\hbar}\frac{\partial E}{\partial \varphi_\ell}
\end{equation}
where $E$ is the ground state energy.

Figure~\ref{fig:chidep}, which uses the same NRG data as Fig.~\ref{fig:1d2lChe_Cur}, shows that a single chain suffices to capture the supercurrent, including the correct placement of current reversals associated with the underlying QPT. An even more interesting case is illustrated in the Fig.~\ref{fig:1d3lcurr}, which shows a three-terminal setup. 

\begin{table*}[t]
\centering
\caption{Chain-expansion coefficients for a block coupled to multiple leads, parameterized by the total coupling $\Gamma_{T}$ and the geometric factor $\bm{\chi}=\chi e^{i\Phi}$. We list only the even $L$ up to $L=10$.}
\begin{tabular}{cccccccc}
\toprule
$L=$ & $2$ & $4$ & $6$ & $8$ & 10 \\
\midrule
$h_0$ &2&4& 6 & 8& 10\\
$h_1$ & $2-\chi^2$ &  6-$\chi^2$ &$\frac{38-3\chi^2}{3}$&  $22-\chi^2$ & $34-\chi^2$\\
$h_2$  & & $\frac{4}{6-\chi^2}$ & $\frac{224}{3(38-3\chi^2)}$ & $\frac{84}{22-\chi^2}$ &$\frac{1056}{5(34-\chi^2)}$\\ 
$h_3$  & & $\frac{8-8\chi^2+\chi^4}{6-\chi^2}$ & $\frac{3(172 - 116\chi^2+7\chi^4)}{7(38-3\chi^2)}$ &  $\frac{424-200\chi^2+7\chi^4}{7(22-\chi^2)}$ & $\frac{644 - 220 \chi^2 + 5 \chi^4}{5(34-\chi^2)}$\\ 
$h_4$  & & &$\frac{4(38-3\chi^2)}{7(172-116\chi^2+7\chi^4)}$ & $\frac{176(22-\chi^2)}{21(424-200\chi^2+7\chi^4)}$ & $\frac{52 (34 - \chi^2)}{3 (644 - 220 \chi^2 + 5 \chi^4)}$ \\
$h_5$  & & & $\frac{7(32-48\chi^2+18\chi^4-\chi^6)}{172-116\chi^2+7\chi^4}$  & $\frac{7(3440 - 3992 \chi^2 + 1014 \chi^4 - 33 \chi^6)}{33(424 - 200 \chi^2 + 7\chi^4)}$ &$\frac{5 (9248 - 8624 \chi^2 + 1554 \chi^4 - 33 \chi^6)}{33 (644 - 
   220 \chi^2 + 5 \chi^4)}$\\
$h_6$  & & & & $\frac{4(424-200\chi^2+7\chi^4)}{11(3440-3992\chi^2 + 1014\chi^4-33\chi^6)}$ & $\frac{320 (644 - 220 \chi^2 + 5 \chi^4)}{143 (9248 - 8624 \chi^2 + 
   1554 \chi^4 -33 \chi^6)}$\\
$h_7$  & & & & $\frac{33(128 - 256 \chi^2 + 160\chi^4 - 32 \chi^6 + \chi^8)}{3440-3992\chi^2 + 1014\chi^4-33\chi^6}$ &$\frac{3 (311232 - 509056 \chi^2 + 244156 \chi^4 - 34892 \chi^6 + 
   715 \chi^8)}{65 (9248 - 8624 \chi^2 + 1554 \chi^4 - 
   33 \chi^6)}$\\
$h_8$  & & & & & $\frac{4 (9248 - 8624 \chi^2 + 1554 \chi^4 - 33 \chi^6)}{
 5 (311232 - 509056 \chi^2 + 244156 \chi^4 - 34892 \chi^6 + 
    715 \chi^8)}$\\
$h_9$  & & & & & $\frac{715 (512 - 1280 \chi^2 + 1120 \chi^4 - 400 \chi^6 + 
   50 \chi^8 - \chi^{10})}{311232 - 509056 \chi^2 + 
 244156 \chi^4 - 34892 \chi^6 + 715 \chi^8}$\\
\bottomrule\label{tab:1}
\end{tabular}
\end{table*}

\section{Non-interacting Green's function of a DQD\label{app:double dot}}

Consider the Hamiltonian of a parallel DQD connected to one SC lead. Following the procedure from the main text, we can write the Matsubara Green's function in the Nambu basis
$\mathscr{D}^T=(d^\pdag_{1\uparrow},d^\dag_{1\downarrow},d^\pdag_{2\uparrow},d^\dag_{2\downarrow},
c^\pdag_{\mathbf{k}},c^\dag_{-\mathbf{k}})$,
\begin{widetext}
\begin{equation}
    G_0(i\omega_n,\mathbf{k})^{-1} =
    \begin{pmatrix}
    i\omega_n - \epsilon_{1} & 0 & -t_d & 0 & -V_{1\mathbf{k}} & 0 \\
    0 & i\omega_n + \epsilon_{1} & 0 & t_d & 0 & V_{1\mathbf{k}} \\
    -t_d & 0 & i\omega_n - \epsilon_{2} & 0 & -V_{2\mathbf{k}} & 0 \\
    0 & t_d & 0 & i\omega_n + \epsilon_{2} & 0 & V_{2\mathbf{k}} \\
    -V_{1\mathbf{k}} & 0 & -V_{2\mathbf{k}} & 0 & i\omega_n - \epsilon_{\mathbf{k}} & \Delta \\
    0 & V_{1\mathbf{k}} & 0 & V_{2\mathbf{k}} & \Delta & i\omega_n + \epsilon_{\mathbf{k}}
    \end{pmatrix},
\end{equation}
\end{widetext}
where we assume real $V_{j\mathbf{k}}$.
We are interested in the impurity Green's function $G^{d}_0$. Following a standard block-matrix inversion scheme~\eqref{eq:block_matrix_approx} we obtain a 4$\times$4 matrix,

\begin{widetext}
\begin{equation}
\begin{aligned}
    \relax [G^d_0(i\omega_n)]^{-1} & =
    \begin{pmatrix}
    i\omega_n - \epsilon_{1} & 0 & -t_d & 0 \\
    0 & i\omega_n + \epsilon_{1} & 0 & t_d  \\
    -t_d & 0 & i\omega_n - \epsilon_{2} & 0  \\
    0 & t_d & 0 & i\omega_n + \epsilon_{2} &  \\
    \end{pmatrix} \\
    &-\sum_\mathbf{k}\frac{1}{\omega_n^2+\epsilon_{\mathbf{k}}^2+\Delta^2}
    \begin{pmatrix}
    (i\omega_n+\epsilon_{\mathbf{k}})V_{1\mathbf{k}}^2 
     & \Delta V_{1\mathbf{k}}^2 
     & (i\omega_n+\epsilon_{\mathbf{k}}) V_{1\mathbf{k}} V_{2\mathbf{k}} 
     & \Delta V_{1\mathbf{k}} V_{2\mathbf{k}} \\
    \Delta V_{1\mathbf{k}}^{2} 
     & (i\omega_n-\epsilon_{\mathbf{k}})V_{1\mathbf{k}}^2
     & \Delta V_{1\mathbf{k}}V_{2\mathbf{k}} 
     & (i\omega_n-\epsilon_{\mathbf{k}})V_{1\mathbf{k}}V_{2\mathbf{k}} \\
    (i\omega_n+\epsilon_{\mathbf{k}})V_{1\mathbf{k}}V_{2\mathbf{k}}
     & \Delta V_{1\mathbf{k}}V_{2\mathbf{k}}
     & (i\omega_n+\epsilon_{\mathbf{k}})V_{2\mathbf{k}}^2 
     & \Delta V_{2\mathbf{k}}^2 \\
    \Delta V_{1\mathbf{k}}V_{2\mathbf{k}} 
     & (i\omega_n-\epsilon_{\mathbf{k}})V_{1\mathbf{k}}V_{2\mathbf{k}}
     & \Delta V_{2\mathbf{k}} 
     & (i\omega_n-\epsilon_{\mathbf{k}})V_{2\mathbf{k}}^2
    \end{pmatrix}.
\end{aligned}
\end{equation}
\end{widetext}
Assuming a constant DOS with half-bandwidth $D$, $\rho(\epsilon) = \Theta(|D-\epsilon|)/(2D)$, 
We rewrite the momentum summations as integrals,
\begin{equation}
    \sum_\textbf{k} F(\epsilon_\textbf{k}) = \frac{1}{2D}\int_{-D}^D d\epsilon F(\epsilon).
\end{equation}
This allows us to define real tunneling rates,
\begin{equation}
    \Gamma_{ij}(\epsilon) = \pi\sum_{\mathbf{k}} V_{i\mathbf{k}} V_{j\mathbf{k}}
    \delta(\epsilon-\epsilon_\mathbf{k}) = \Gamma_{ij}\Theta(D - |\epsilon|)
\end{equation}
which are constant in WBL, $\Gamma_{ij}(\epsilon)=\Gamma_{ij}$ for $D\rightarrow\infty$. We denote
\begin{equation}
    \widetilde{\Gamma}_{ij}(\omega_n)=\frac{\Gamma_{ij}}{\Delta}\tGD,
\end{equation}
where $\tGD$ is given by Eq.~\eqref{eq:GomegaD}. The inverse of the impurity Green's function then reads
\begin{widetext}
\begin{equation}
\begin{aligned}
    &[G^d_0(i\omega_n)]^{-1} = \\
    &\begin{pmatrix}
    i\omega_n[1+\widetilde{\Gamma}_{11}(\omega_n)] -\epsilon_{1} 
    & -\Delta \widetilde{\Gamma}_{11}(\omega_n) 
    & i\omega_n \widetilde{\Gamma}_{21}(\omega_n) - t_d
    & -\Delta \widetilde{\Gamma}_{21}(\omega_n)
    \\
    -\Delta \widetilde{\Gamma}_{11}(\omega_n) 
    & i\omega_n[1+\widetilde{\Gamma}_{11}(\omega_n)] + \epsilon_{1}
    & -\Delta \widetilde{\Gamma}_{21}(\omega_n)
    & i\omega_n \widetilde{\Gamma}_{21}(\omega_n) + t_d
    \\
    i\omega_n \widetilde{\Gamma}_{12}(\omega_n) - t_d
    & -\Delta \widetilde{\Gamma}_{12}(\omega_n)
    & i\omega_n[1+\widetilde{\Gamma}_{22}(\omega_n)] - \epsilon_{2}
    & -\Delta \widetilde{\Gamma}_{22}(\omega_n)
    \\
    -\Delta \widetilde{\Gamma}_{12}(\omega_n)
    & i\omega_n \widetilde{\Gamma}_{12}(\omega_n) + t_d
    & -\Delta \widetilde{\Gamma}_{22}(\omega_n)
    & i\omega_n[1+\widetilde{\Gamma}_{22}(\omega_n)] + \epsilon_{2}
    \end{pmatrix}.
\end{aligned}
\end{equation}
\end{widetext}

\section{Technical details for numerical calculations \label{app:RGmethods}}
Most of the results presented in the main text using the ChE method were obtained via ED on standard PCs, employing various linear algebra routines from the \texttt{SciPy} package \cite{2020SciPy-NMeth}. For longer chains, we used DMRG techniques implemented in the \texttt{TenPy} \cite{tenpy2024} and \texttt{iTensor} \cite{itensor} libraries. 
The online examples published in Ref.~\cite{ChECode} make use of the \texttt{NetKet} library~\cite{netket2:2019,netket3:2022}, owing to its convenient representation of fermionic Hamiltonians and operators.

All NRG results were obtained using the open-source \texttt{NRG Ljubljana} package \cite{Zitko2021nrg}. For single-channel problems (one chain), we used a logarithmic discretization parameter $\Lambda = 2$, and for two-channel problems (two chains), we used $\Lambda = 4$. Unless stated otherwise, the typical half-bandwidth was set to $D = 100\Delta$, which effectively suppresses band-edge-related effects.   

\input{paper_main.bbl}

\end{document}

%% file: paper_main.bbl
%